\documentclass[a4paper,11pt]{article}
\pdfoutput=1 % if your are submitting a pdflatex (i.e. if you have
\usepackage{jcappub} % for details on the use of the package, please
\usepackage[T1]{fontenc} % if needed

\usepackage{graphicx}% Include figure files
\usepackage{dcolumn}% Align table columns on decimal point
\usepackage{hyperref}
\hypersetup{colorlinks}
\usepackage{color}
\usepackage{tikz}
\usepackage[caption=false]{subfig}
\def\beq{\begin{eqnarray}}
\def\eeq{\end{eqnarray}}

\newcommand{\av}[1]{\langle{#1\rangle}} 

\usepackage{bm}
\let\vec\bm
\newcommand{\Mpch}{h^{-1}\mathrm{Mpc}}
\newcommand{\hMpc}{h\,\mathrm{Mpc}^{-1}}

\newcommand{\A}{\mathcal{A}}
\newcommand{\B}{\mathcal{B}}
\newcommand{\mA}{\mathcal{A}}
\newcommand{\mB}{\mathcal{B}}

\newcommand{\mK}{\mathcal{K}}
\newcommand{\mC}{\mathcal{C}}

\newcommand{\dpt}[1]{\delta^{(#1)}}

\newcommand{\dptM}[1]{\delta_M^{(#1)}}
\newcommand{\dptgR}[1]{\delta_{g,R}^{(#1)}}
\newcommand{\dptg}[1]{\delta_{g\,}^{(#1)}}
\newcommand{\delD}{(2\pi)^3\delta_D}
\newcommand{\vx}{\vec x}
\newcommand{\vr}{\vec r}
\newcommand{\vk}{\vec k}
\newcommand{\vn}{\vec n}
\newcommand{\vp}{\vec p}
\newcommand{\vq}{\vec q}
\newcommand{\ikk}{\int_{\vec k_{12}=\vec k}}

\definecolor{darkgreen}{RGB}{0,120,0}
\definecolor{brown}{RGB}{120,60,0}

\newcommand{\resub}[1]{#1}%\textcolor{darkgreen}{#1}}
\numberwithin{equation}{section}

\title{\boldmath Modeling the Marked Spectrum of Matter and Biased Tracers in Real- and Redshift-Space}

\author[a,b,1]{Oliver H.\,E. Philcox,\note{Corresponding author.}}
\author[c,d]{Alejandro Aviles,}
\author[e]{Elena Massara}

\affiliation[a]{Department of Astrophysical Sciences, Princeton University,\\ Princeton, NJ 08540, USA}
\affiliation[b]{School of Natural Sciences, Institute for Advanced Study, 1 Einstein Drive,\\ Princeton, NJ 08540, USA}
\affiliation[c]{Consejo Nacional de Ciencia y Tecnolog\'ia, Av. Insurgentes Sur 1582,\\
Colonia Cr\'edito Constructor, Del. Benito Ju\'arez, 03940, Ciudad de M\'exico, M\'exico}
\affiliation[d]{Departamento de F\'isica, Instituto Nacional de Investigaciones Nucleares,\\
Apartado Postal 18-1027, Col. Escand\'on, Ciudad de M\'exico, 11801, M\'exico}

\affiliation[e]{Waterloo Centre for Astrophysics, University of Waterloo, 200 University Ave W,\\
Waterloo, ON N2L 3G1, Canada}

\emailAdd{ohep2@cantab.ac.uk}
\emailAdd{avilescervantes@gmail.com}
\emailAdd{emassara@uwaterloo.ca}

\abstract{We present the one-loop perturbation theory for the power spectrum of the marked density field of matter and biased tracers in real- and redshift-space. The statistic has been shown to yield impressive constraints on cosmological parameters; to exploit this, we require an accurate and computationally inexpensive theoretical model. Comparison with $N$-body simulations demonstrates that linear theory fails on all scales, but inclusion of one-loop Effective Field Theory terms gives a substantial improvement, with $\sim 5\%$ accuracy at $z = 1$. The expansion is less convergent in redshift-space (achieving $\sim 10\%$ accuracy), but there are significant improvements for biased tracers due to the freedom in the bias coefficients. The large-scale theory contains non-negligible contributions from all perturbative orders; we suggest a reorganization of the theory that contains all terms relevant on large-scales, discussing both its explicit form at one-loop and structure at infinite-loop. This motivates a low-$k$ correction term, leading to a model that is sub-percent accurate on large scales, albeit with the inclusion of two (three) free coefficients in real- (redshift-)space. We further consider the effects of massive neutrinos, showing that beyond-EdS corrections to the perturbative kernels are negligible in practice. It remains to see whether the purported gains in cosmological parameters remain valid for biased tracers and can be captured by the theoretical model.}

\begin{document}
\maketitle
\flushbottom

\section{Introduction}\label{sec:intro}

The overdensity field, $\delta(\vx)$, is the fundamental building block of large-scale structure analyses. It is the source of many observables: galaxy densities, weak lensing, and cosmic shear, to name but a few. For galaxy surveys, the principal method of extracting information from $\delta$ is through the two-point correlator of the galaxy overdensity, $\xi_g(\vr) \equiv \av{\delta_g(\vx)\delta_g(\vx+\vr)}$, or its Fourier space counterpart, $P_g(\vk)$. If the Universe may be considered Gaussian, such statistics encapsulate all the cosmological information, and, for the past decades, analyses have focused on measuring such quantities to high precision. In the present epoch, the assumption of Gaussianity is not valid however; structures form on a vast range of scales, and the two-point correlator is no longer all-encompassing. To proceed, there are two avenues: (1) analyze statistics beyond the power spectrum, encompassing the non-Gaussian information; or (2) consider the two-point correlator of a different, transformed, density field.

For the first option, the lowest-order extension lies in the \textit{three-point} correlator, or bispectrum, though there exists information beyond even this \citep{2001ApJ...553...14V,2016JCAP...06..052B,2020arXiv200902290G}. The bispectrum has been considered in a number of works \citep[e.g.,][]{2001ApJ...546..652S,2006PhRvD..74b3522S,2020JCAP...03..040H,2017MNRAS.468.1070S,2015JCAP...05..007B}, though its use for cosmological parameter inference still presents considerable challenges, with particular difficulties arising from its estimation  \citep{2015MNRAS.454.4142S,2015PhRvD..91d3530S,2017MNRAS.472.2436W,2020MNRAS.492.1214P,2020arXiv200501739P} and high-dimensionality \citep{2020arXiv200903311P}. It has thus been little adopted, though Refs.\,\citep{2017MNRAS.465.1757G,2018MNRAS.478.4500P,2017MNRAS.469.1738S} are notable recent exceptions. Statistics not based on $n$-point functions can also be used, including counts-in-cells \citep[e.g.,][]{1980lssu.book.....P} and void-statistics \citep[e.g.,][]{2019BAAS...51c..40P}. A different approach is to analyze the galaxy density field directly without use of summary statistics; whilst this field is still in its infancy it is nonetheless promising \citep{2019PhRvD.100d3514S,2020JCAP...04..042C,2020arXiv200714988C,2020JCAP...11..008S}.

Regarding the second case, transformations of the density field have been shown to produce promising results by transforming higher-point information into the two-point correlators. Examples include Gaussianized fields \citep{1992MNRAS.254..315W,2011ApJ...731..116N,2011ApJ...742...91N}, log-normal transformations \citep{2009ApJ...698L..90N,2011ApJ...735...32W} and reconstructed density fields \citep{2007ApJ...664..675E}, with the latter having been used in many analyses \citep[e.g.,][]{2017MNRAS.464.3409B,2020JCAP...05..032P,2020MNRAS.498.2492G}. An example of significant recent interest is the \textit{marked} density field, which, in its most basic form, is simply a weighted density field \citep{doi:10.1002/mana.19841160115,2005MNRAS.364..796S}, where the weights can represent galaxy properties \citep{2005astro.ph.11773S,2006MNRAS.369...68S,2000ApJ...545....6B,2002A&A...387..778G} or be density-based \citep{2009MNRAS.395.2381W,2016JCAP...11..057W}. Assuming a local-overdensity weighting, it has been shown to produce strong constraints on cosmology, particularly the neutrino mass \citep{2020arXiv200111024M}, and modified gravity models \citep{2016JCAP...11..057W,2018PhRvD..97b3535V,2018MNRAS.478.3627A,2018MNRAS.479.4824H,2020JCAP...01..006A}, due to the ability to upweight low-density, unvirialized, regions and the transfer of higher-order moments into the two-point function \citep{2020PhRvD.102d3516P}.  

Modeling such quantities is non-trivial. Most involve non-linear transformations of the density field (be it for matter or biased tracers), which must be expanded as a Taylor series to facilitate perturbative analyses. This poses difficulties since the density field is an inherently non-linear quantity, with series convergence only guaranteed if the field is sufficiently smoothed. The overdensity-weighted marked field satisfies this criterion however, since the weighting, and hence Taylor expansion, depends only on the \textit{smoothed} density field whose variance is parametrically controlled. There exists extensive literature pertaining to modeling the unsmoothed density field, $\delta$, most notably using the Effective Field Theory of Large Scale Structure (hereafter EFT) \citep{2012JHEP...09..082C,2012JCAP...07..051B}, which has been applied to a variety of statistics both for matter and biased tracers, in real- and redshift-space \citep[e.g.,][]{2014arXiv1409.1225S,2016arXiv161009321P,2015JCAP...09..029A}. Here, we build upon the treatment of Ref.\,\citep{2020PhRvD.102d3516P}, which derived the one-loop EFT of the marked density field, $\delta_M$, for matter in real-space.

This primary goal of this work is to further develop the EFT of the marked power spectrum, $M(\vk)$; in particular, to create a full description of the statistic valid to one-loop order in perturbations for general tracers in real- or redshift-space. Schematically, this is straightforward;
\beq
    M(\vk) \propto M_{11}(\vk) + M_{22}(\vk)+2M_{13}(\vk) + M_\mathrm{ct}(\vk) + M_\mathrm{shot}(\vk),
\eeq
where the `11' term is the linear contribution, `22' and `13' terms are from one-loop perturbation theory, `ct' encodes the effects of small-scale physics on large scale modes, and the `shot' piece encapsulates stochastic contributions. The theoretical model is not manifestly more complex when accounting for bias or redshift-space distortions (RSD); since $\delta_M$ is a functional of the underlying density field, we need simply replace the perturbative kernels of matter arising in the above expression with those of biased tracers. Computation of these is non-trivial however, and thoroughly investigated in this work. Furthermore, we extend the statistic to include lowest-order treatment of neutrino effects following the work of Ref.\,\citep{2020JCAP...10..034A}.

As noted in Ref.\,\citep{2020PhRvD.102d3516P}, the marked power spectrum is difficult to model perturbatively, since even the contribution proportional to the linear power spectrum (which dominates on large scales), depends on all loop orders, including gravitational kernels usually found only in the low-$k$ limit of higher-point statistics. This arises since the marked field is not well modeled by the first terms in its linear density field expansion, even though the Taylor series are strictly convergent. This is particularly apparent for the mark functions found to be optimal for parameter estimation in Ref.\,\citep{2020arXiv200111024M}. A significant section of this work is devoted to addressing this complication, and we introduce a \textit{reorganized} linear theory, formally encompassing all contributions important on large scales, motivating its functional form at infinite-loop order;
\beq
    \left.M(\vk)\right|_\mathrm{large\,\,scale} = \left[(a_0+a_1W_R(k))+(a_2+a_3W_R(k))\mu^2\right]^2P_L(k) + \left[c_0+c_1W_R(k)\right]
\eeq
\eqref{eq: Mr0-inf-loop}, where $a_i$ and $c_i$ are free coefficients (with $a_2=a_3=0$ in real-space), $W_R$ a smoothing window and $P_L$ the linear power spectrum. Comparison of the theoretical models to $N$-body simulations demonstrates the validity of our theory, and such studies pave the way towards applications to data, which will allow one to reap the significant cosmological rewards available.

This work begins by outlining the one-loop model for the marked spectrum in Sec.\,\ref{sec: theory}, before we discuss its low-$k$ limit and reorganized form in Sec.\,\ref{sec: reorg}. Comparison to $N$-body simulations, both for matter and biased tracers, is shown in Sec.\,\ref{sec: data} before we summarize in Sec.\,\ref{sec: summary}. Appendices \ref{appen: integral-simp}\,\&\,\ref{appen: low-k-limits} include supplementary material relating to Secs.\,\ref{sec: theory}\,\&\,\ref{sec: reorg}, whilst Appendices \ref{appen: massive-nu}\,\&\,\ref{appen: rot-integ} discuss application to massive neutrino cosmologies and present the derivation of a useful integral relation.

\section{Theory Model for the Marked Spectrum}\label{sec: theory}
Below, we present a brief derivation of the theory model for the marked density field, first starting from its definition in Sec.\,\ref{subsec: theory-basics}, before introducing the one-loop theory and tracer-specific behavior in Secs.\,\ref{subsec: theory-spt}\,\&\,\ref{subsec: theory-uv}. Much of these sections parallels that of Ref.\,\citep{2020PhRvD.102d3516P}, though in greater generality. 

\subsection{The Marked Density Field}\label{subsec: theory-basics}
We begin with the definition of the marked density field, 
\beq\label{eq: rhoM-def}
    \rho_M(\vec x) = m(\vec x)n_g(\vec x) = m(\vec x)\bar{n}_g\left[1+\delta_g(\vec x)\right],
\eeq
where $\bar{n}_g=\av{n_g(\vec x)}$ is the average density and $\delta_g(\vec x)$ is the usual overdensity field. Though we denote the field by the subscript $g$, it is fully general and can apply to \textit{any} quantity; biased tracers or matter, real- or redshift-space. As in Refs.\,\citep{2016JCAP...11..057W,2020JCAP...01..006A,2020arXiv200111024M,2020PhRvD.102d3516P}, we define the mark, $m(\vec x)$ via
\beq\label{eq: mark-def}
    m(\vec x) = \left(\frac{1+\delta_s}{1+\delta_s+\delta_{g,R}(\vec x)}\right)^p,
\eeq
where $\delta_{g,R}$ is the overdensity field smoothed on scale $R$, and the exponent $p$, the offset $\delta_s$ and the smoothing $R$ are model hyperparameters. As for the unmarked field, it is simpler to consider the marked \textit{overdensity}, given by
\beq\label{eq: deltaMdef}
    \delta_M(\vec x) \equiv \frac{\rho_M(\vec x) - \av{\rho_M}}{\av{\rho_M}} = \frac{1}{\bar{m}}m(\vec x)\left[1+\delta_g(\vec x)\right] - 1,
\eeq
where $\bar{m}$ is the density-weighted mean mark, $\av{n_g(\vec x)m(\vec x)}/\bar{n}_g$, which can be measured from simulations or data.

For an analytic treatment, we proceed by expanding $\delta_M(\vec x)$ as a Taylor series in the non-linear fields $\delta_g$ and $\delta_{g,R}$, using the expansion
\beq\label{eq: Mexpan}
    m(\vec x) = \sum_{n = 0}^\infty (-1)^n\frac{p(p+1)...(p+n-1)}{n!(1+\delta_s)^n}\delta_{g,R}^n(\vec x) \equiv \sum_{n=0}^\infty (-1)^nC_n \delta_{g,R}^n(\vec x),
\eeq
where the coefficients $C_n \equiv \binom{p+n-1}{p-1}(1+\delta_s)^{-n}$ encode all dependence on the mark parameters $p$ and $\delta_s$. Note that this decomposition, and all succeeding formulae, can be applied to any mark depending only on $\delta_{g,R}$ with suitably defined Taylor coefficients $\{C_n\}$. In our example, the condition for a convergent Taylor expansion is given by
\beq\label{eq: taylor-condition}
    \sigma_{g,RR}(z) < 1+\delta_s,
\eeq
where $\sigma^2_{g,RR}$ is the variance of $\delta_{g,R}$. Convergence is thus guaranteed by sufficiently large $R$ and $\delta_s$.\footnote{Technically, we require $R$ to be at least as large as the non-linear scale $k_\mathrm{NL}^{-1}$ if $\delta_s = 0$, with a weaker condition required by increasing $\delta_s$.} For biased tracers in real-space, assuming $R$ to be sufficiently large, we can assume $\delta_{g,R}(z) \sim b_1(z)D(z)$ for linear bias $b_1$ and growth factor $D(z)$, such that the condition becomes
\beq
    b_1(z)D(z)\sigma_{RR}(0) < 1+\delta_s,
\eeq
where $\sigma_{RR}(0)$ is the variance of the matter field on scale $R$ at redshift zero. For most biased tracers, $b_1(z) > 1$, thus this is a stricter bound than for matter and, additionally, is not necessarily ameliorated by moving to higher redshift. Furthermore, for (angle-averaged) biased tracers in redshift-space the condition becomes stricter still;
\beq\label{eq: taylor-condition-rsd}
    \left(b^2_1(z)+\frac{2}{3}f(z)b_1(z)+\frac{1}{5}f^2(z)\right)^{1/2}D(z)\sigma_{RR}(0) < 1+\delta_s,
\eeq
for growth rate $f(z)$ (which is scale-independent for $\Lambda$CDM cosmologies without massive neutrinos).

\subsection{Eulerian Perturbation Theory}\label{subsec: theory-spt}
To construct an Eulerian perturbation theory (EPT; also Standard PT) for the marked field, we proceed by expanding $\delta_M$ in powers of the linear density field $\dpt{1}$. As an intermediary step, we expand $\delta_g$ and $\delta_{g,R}$ perturbatively as
\beq
    \delta_g(\vec x) = \sum_{n = 0}^\infty \dptg{n}(\vec x), \,\quad \delta_{g,R}(\vec x) = \sum_{n = 0}^\infty \dptgR{n}(\vec x),
\eeq
where the superscript $(n)$ indicates that the field is $n$-th order in $\dpt{1}$. The relation between smoothed and unsmoothed fields is simply given in Fourier space; $\delta_{g,R}(\vec k) = W_R(k)\delta_g(\vec k)$ where $W_R$ is some (isotropic) smoothing window, usually assumed to be Gaussian (but see Ref.\,\citep{2020PhRvD.102d3516P} for a discussion of this choice). This trivially extends to the perturbative solutions: $\dptgR{n}(\vec k) = W_R(k)\dptg{n}(\vec k)$. Inserting these into \eqref{eq: deltaMdef}, coupled with the Taylor expansion of \eqref{eq: Mexpan}, gives the following series
\beq
    \bar{m}\left[1+\delta_M(\vec x)\right] &=& \sum_{n = 0}^\infty (-1)^nC_n\delta^n_{g,R}(\vec x)\left[1+\delta_g(\vec x)\right] \\\nonumber
    &=& \sum_{n = 0}^\infty (-1)^nC_n\prod_{i=1}^n \left[\sum_{m_i = 0}^\infty \dptgR{m_i}(\vec x)\right]\times \left[1+\sum_{j=1}^\infty\dptg{j}(\vec x)\right]\\\nonumber
    &\equiv& \left[1+\sum_{n = 1}^\infty\dptM{n}(\vec x)\right].
\eeq
In this work, we consider the one-loop theory of the two-point $\delta_M$ correlator, thus will use only the first three terms, given by
\beq\label{eq: dptM-def}
    \dptM{1}(\vx) &=& \left[C_0\dptg{1} - C_1\dptgR{1}\right](\vx)\\\nonumber
    \dptM{2}(\vx) &=& \left[C_0\dptg{2} - C_1\dptgR{2}-C_1\dptgR{1}\dptg{1} + C_2\dptgR{1}\dptgR{1}\right](\vx)\\\nonumber
    \dptM{3}(\vx) &=& \left[C_0\dptg{3}-C_1\dptgR{3}-C_1\dptgR{2}\dptg{1}-C_1\dptgR{1}\dptg{2}+2C_2\dptgR{1}\dptgR{2}\right.\\\nonumber
    &&\left.\quad+C_2\dptgR{1}\dptgR{1}\dptg{1}-C_3\dptgR{1}\dptgR{1}\dptgR{1}\right](\vx).
\eeq
%In this paper, we work only in Fourier space, thus we can drop the $\vec x$-independent zero-lag terms, \textit{i.e.} $\delta_X^{(0)}$ \alej{Why $\delta_X^{(0)}$? this guy is zero by definition, and is not a zero-lag}. Inserting \alej{...?}

To proceed, we require expressions for $\dptg{n}$ in terms of the linear spectrum $\dpt{1}$. Assuming Einstein-de-Sitter (EdS) kernels %\footnote{\alej{I dont think is necessary to specify a cosmology here. And in such a case it would be better to say EdS kernels, instead of cosmology.}\oliver{I think we have to specify EdS kernels since only then is the time part separable. But I agree: kernels > cosmology}} 
and switching to Fourier space,\footnote{In this paper, we define the Fourier and inverse Fourier transforms as
\beq
    X(\vec k) &=& \int d\vec x\,e^{-i\vec k\cdot\vec x}X(\vec x),  \qquad
    X(\vec x) = \int \frac{d\vec k}{(2\pi)^3}e^{i\vec k\cdot\vec x}X(\vec k)\nonumber
\eeq
leading to the definition of the Dirac function $\delta_D$
\beq
    \int d\vec x\,e^{i(\vec k_1-\vec k_2)\cdot\vec x} = (2\pi)^3\delta_D(\vec k_1-\vec k_2).\nonumber
\eeq
The correlation function and power spectrum of the density field are defined as 
\beq
    \xi(\vec r) = \av{\delta(\vec x)\delta(\vec x+\vec r)},\qquad (2\pi)^3\delta_D(\vec k+\vec k')P(\vec k) = 
    \av{\delta(\vec k)\delta(\vec k')}\nonumber
\eeq
with the power spectrum as the Fourier transform of the correlation function.} these take the standard form
\beq
    \dptg{n}(\vec k) \equiv \int_{\vk_1...\vk_n}Z_n(\vk_1,...,\vk_n)\dpt{1}(\vk_1)...\dpt{1}(\vk_n)\times (2\pi)^3\delta_D\left(\vk_1+...+\vk_n-\vk\right),
\eeq
where the $Z_n$ kernels can be found in e.g., Refs.\,\citep{2002PhR...367....1B,2020JCAP...05..042I}, and we adopt the shorthand $\int_{\vk_1..\vk_n} \equiv (2\pi)^{-3n}\int d\vk_1...\int d\vk_n$. Note that we have still assumed nothing about the form of the input field $\delta_g$, thus the expressions in this section are relevant to any tracer with complexities such as redshift-space distortions (RSD) only appearing in the $\{Z_n\}$ kernels.\footnote{Note that we mark-transform the \textit{redshift-space} field rather than applying RSD to the marked field. This is correct since the redshift-space field is the observable quantity.} 

Inserting this definition into \eqref{eq: dptM-def} gives analogous kernels, $\{H_n\}$, for the marked field, \textit{i.e.}
\beq\label{eq: Hdef}
    \dptM{n}(\vec k) &\equiv& \int_{\vk_1...\vk_n}H_n(\vk_1,...,\vk_n)\dpt{1}(\vk_1)...\dpt{1}(\vk_n)\\\nonumber
    &&\qquad \times (2\pi)^3\delta_D\left(\vk_1+...+\vk_n-\vk\right)\\\nonumber
    H_1(\vk) &=& C_{\delta_M}(k)Z_1(\vk)\\\nonumber
    H_2(\vk_1,\vk_2) &=& C_{\delta_M}(k) Z_2(\vk_1,\vk_2) + C_{\delta^2_M}(k_1,k_2)Z_1(\vk_1)Z_1(\vk_2)\\\nonumber
    H_3(\vk_1,\vk_2,\vk_3) &=& C_{\delta_M}(k) Z_3(\vk_1,\vk_2,\vk_3) + 2 C_{\delta^2_M}(k_1,k_{23})  Z_1(\vk_1) Z_2(\vk_2,\vk_3)\\\nonumber
    &&\,+C_{\delta^3_M}(k_1,k_2,k_3)Z_1(\vk_1)Z_1(\vk_2)Z_1(\vk_3),
\eeq
where $\vk=\sum_i \vk_i$, $\vk_{ij} = \vk_i+\vk_j$, and $k = |\vec k|$, and the $H_3$ kernel should properly be symmetrized over its arguments. For convenience, we have introduced the following functions;
\beq\label{eq: CdeltaMdef}
    C_{\delta_M}(k) &=& - C_1 W_R(k)+C_0 \\\nonumber
    C_{\delta^2_M}(k_1,k_2) &=& C_2W_R(k_1)W_R(k_2) - \frac{1}{2}C_1\left[W_R(k_1)+W_R(k_2)\right]\\\nonumber
    C_{\delta^3_M}(k_1,k_2,k_3) &=&  - C_3 W_R(k_1) W_R(k_2)W_R(k_3) \\\nonumber
    &&\,+  \frac{1}{3} C_2 \left[W_R(k_2)W_R(k_3) + W_R(k_3)W_R(k_1) + W_R(k_1)W_R(k_2)\right],
\eeq
which are simply the coefficients of an expansion of $\delta_M(\vk)$ in powers of the full $\delta_g(\vk)$ field, \textit{i.e.};
\beq
    \bar{m}\,\delta_M(\vk) \equiv \sum_{n=1}^\infty\int_{\vk_1...\vk_n}C_{\delta_M^n}(k_1,...,k_n)\delta_g(\vk_1)...\delta_g(\vk_n)\times (2\pi)^3\delta_D\left(\vk_1+...+\vk_n-\vk\right).
\eeq

From these expansions, we can easily compute summary statistics, such as the marked power spectrum, by using Wick's theorem to evaluate products of $\dpt{1}$ fields. Just as for the unmarked case, we obtain
\beq\label{eq: Mk-def}
    \bar{m}^2M(\vec k) &=& \bar{m}^2\left|\delta_M(\vec k)\right|^2 \equiv M_{11}(\vk) + 2M_{13}(\vk) + M_{22}(\vk)\\\nonumber
    M_{11}(\vk) &=& H_1^2(\vk)P_L(k)\\\nonumber
    M_{13}(\vk) &=& 3H_1(\vk)P_L(k)\int_{\vp}H_3(\vp,-\vp,\vk)P_L(\vp)\\\nonumber
    M_{22}(\vk) &=& 2\int_{\vp}\left|H_2(\vp,\vk-\vp)\right|^2P_L(\vp)P_L(\vk-\vp),
\eeq
where $P_L(k)$ is the linear power spectrum, \textit{i.e.} $\av{\dpt{1}(\vk)\dpt{1}(-\vk)}$. The linear spectrum $M_{11}$ is simply equal to that of the unmarked field, $P_{11} = Z_1^2(\vk)P_L(k)$ damped by the function $C_{\delta_M}^2(k)$. Notably, this damping prefactor appears in \textit{any} $H_n$ term proportional to $Z_n$, and sources a marked spectrum contribution proportional to $C_{\delta_M}^2(k)P(\vk)$ for non-linear power spectrum $P(\vk)$. This has important consequences for evaluation of the theory: third order contributions to $\delta_g$ appear only in the term proportional to $P(\vk)$, thus for the remaining terms we can work to second order in $\delta_g$, \textit{i.e.} consider only $Z_1$ and $Z_2$. Throughout this work, we will compute terms proportional to $P(\vk)$ using the FFTLog algorithm \citep{2018JCAP...04..030S}, via the publicly available \texttt{CLASS-PT} package \citep{2020PhRvD.102f3533C},\footnote{\href{https://github.com/michalychforever/CLASS-PT}{github.com/michalychforever/CLASS-PT}} including the inbuilt infra-red resummation procedure \citep{2015JCAP...02..013S}.

In Appendix \ref{appen: integral-simp}, we consider simplifications of the above expressions, given also the considerations in the following subsections. This allows the theoretical model to be straightforwardly computed, involving at most two-dimensional numerical integrals. Considering redshift-space, our expressions are somewhat more involved than those in Ref.\,\citep{2020PhRvD.102d3516P}, however, the one-loop terms can all be written as even polynomials in $\mu$ (the angle cosine between $\vk$ and the line-of-sight (LoS) vector $\hat{\vec n}$). For comparison to data, it is usually more convenient to express the function as a set of Legendre multipoles, $M_\ell(k)$, \textit{i.e.},
\beq
    M(k,\mu) \equiv \sum_n \tilde{M}_n(k)\mu^{2n} \equiv \sum_\ell M_\ell(k)L_\ell(\mu),
\eeq
and the even Legendre moments $M_\ell$ are given in terms of $\tilde{M}_n$ by
\beq
    M_\ell(k) &=& \sum_n \tilde{M}_n(k)\frac{2\ell+1}{2}\int_{-1}^1 L_\ell(\mu)\mu^{2n}d\mu \\\nonumber
    &=& \begin{cases}\sum_{n=\ell/2}^\infty \tilde{M}_n(k) \frac{(2\ell+1)}{2^{2n+1}}\frac{\sqrt{\pi}\,\Gamma(1+2n)}{\Gamma(1+n-\ell/2)\Gamma(n+\ell/2+3/2)} & \text{even }\ell\\
    0 & \text{else }\end{cases}
\eeq
\citep[Eq.\,7.126]{2007tisp.book.....G}, where $\{L_\ell(\mu)\}$ are the Legendre polynomials and $\Gamma$ is the Gamma function. To compute $M_\ell(k)$, we thus only need expressions for $M(\vk)$ as an expansion in powers of $\mu^2$.

\subsection{Biases, Counterterms and Shot-Noise}\label{subsec: theory-uv}

To evaluate \eqref{eq: Mk-def}, we require the $Z_n$ kernels, and thus, for biased tracers, a bias expansion. Here we follow Ref.\,\citep{2020JCAP...05..042I} and utilize the third-order expansion
\beq\label{eq: bias-expan}
    \delta_g(\vx) = b_1\delta(\vx) + \frac{b_2}{2}\left[\delta^2\right](\vx) + b_{\mathcal{G}_2}\left[\mathcal{G}_2\right](\vx) + b_{\Gamma_3}\left[\Gamma_3\right](\vx),
\eeq
which contains all terms relevant to the one-loop power spectrum,\footnote{At one-loop order, additional terms such as $\delta^3$ and $\mathcal{G}_3$ contain no new shapes ($k$-dependencies) and can thus be absorbed into the bias parameters of lower-order contributions.} with $\mathcal{G}_2$ being the Galileon tidal operator. In the following, we set $b_{\Gamma_3}$ to zero, since it was found to be highly degenerate with $b_{\mathcal{G}_2}$ in the BOSS analysis of Ref.\,\citep{2020JCAP...05..042I}, which considered similar volumes to the simulations used in this work. This is strictly an expansion in terms of \textit{renormalized} operators \citep{2006PhRvD..74j3512M,2014JCAP...08..056A}, which, at second order, are related to the usual fields via
\beq
   [\delta^2](\vec x) = \delta^2(\vec x) - \sigma^2,\qquad [\mathcal{G}_2](\vx) = \mathcal{G}_2(\vx).
\eeq
$\sigma^2$ is the variance of the unsmoothed field $\delta(\vx)$, \textit{i.e.},
\beq\label{eq: sig2-def}
    \sigma^2 = \int_{\vp} P_L(p),
\eeq
which strictly depends on the ultraviolet (UV) momentum cut-off of the theory.\footnote{We may safely ignore the third order contributions to these expressions since they renormalize only the (well-known) $P(\vk)$ terms and are thus already incluided in the standard power spectrum models.} When considering only the unmarked power spectrum, the $\sigma^2$ term in $[\delta^2]$ can be neglected, since it gives only a zero-lag contribution; here, greater caution is needed since the marked theory contains products of operators evaluated at the same location. In Fourier space, its inclusion leads to the redefinition
\beq
    \delta_g(\vec k) \rightarrow \delta_g(\vec k) - \frac{b_2}{2}\sigma^2\delD(\vk),
\eeq
or, for the marked field
\beq\label{eq: bias-renorm-cont}
    \dptM{2}(\vk) &\rightarrow& \dptM{2}(\vk) - \frac{b_2}{2}\sigma^2C_{\delta_M}(k)\delD(\vk)\\\nonumber
     \dptM{3}(\vk) &\rightarrow& \dptM{3}(\vk) - b_2\sigma^2C_{\delta^2_M}(k,0)\dpt{1}(\vk).
\eeq
As discussed below, properly including the $\dptM{3}(\vk)$ contribution is crucial for the UV-safety of the one-loop $M(\vk)$ theory. 

Further discussion is needed regarding the effects of small-scale physics on the $M(\vk)$ model. As shown originally in Refs.\,\citep{2012JHEP...09..082C,2012JCAP...07..051B}, standard perturbative treatments are incomplete since they (a) are not manifestly UV-convergent, (b) ignore terms arising from imperfections in the cosmological fluid, and (c) rely on an ill-posed expansion of the unsmoothed density field. The Effective Field Theory of Large Scale Structure (EFT) ameliorates these concerns, by deriving the theory from the smoothed imperfect fluid equations. For unmarked matter at one-loop order, this gives only a single change to the perturbative expansion of $\delta(\vk)$: the introduction of a third-order %\footnote{\alej{Minor point: I don't see the counterterm as third order. This is given by a double expansion, one in $1/k_{NL}$ and the other in $1/\Lambda$ (the smoothing EFT scale). In fact, the expression for $c^2_s$ is zero-order in fluctuations, as in \cite{2012JHEP...09..082C}.}\oliver{I disagree. The counterterm is required to correct the UV behavior of the third order $P_{13}$ part, and is only needed if one expands $\delta$ to third order. Further, the whole point of EFT is to remove the dependence of the theory on $\Lambda$ which otherwise appears in the loop integrals. Its time dependence is often approximated as $D^2(z)$ for the third-order reason.}}
\textit{counterterm} $\dpt{ct}(\vk) = -c_s^2k^2\dpt{1}(\vk)$, which both captures the UV-dependence of $P_{13}(\vk)$, and accounts for the backreaction of short-scale physics on long-wavelength modes. Here, the amplitude $c_s^2$ (known as the effective speed-of-sound) is a free parameter that should be predicted from observations, real or simulated.

Ref.\,\citep{2020PhRvD.102d3516P} showed that this $c_s^2$ counterterm was the only one needed to capture the UV divergences of the one-loop $M(\vk)$ theory in real-space; equivalently, all other terms are manifestly convergent for hard loop momenta $p\gg k$ due to the presence of smoothing windows that depend on the physical scale $R$ used in the definition of the mark. Since the expansion of $\delta_M(\vk)$ contains at maximum one unsmoothed $\delta(\vk)$ field, this is a general result, true at arbitrary loop order; all possible UV divergences must arise from terms in $P(\vk)$, thus the marked theory requires no additional counterterms relative to the unmarked theory.\footnote{This implicitly assumes that $R^{-1}$ is small compared to the cut-off scale $\Lambda$ of the EFT.} Note that we assume $R$  Furthermore, since the $P(\vk)$ terms always enter in the combination $\bar{m}^2M(\vk)\supset C_{\delta_M}^2(k)P(\vk)$, the relevant counterterm in $M(\vk)$ is 
\beq
    M_{ct}^\mathrm{real}(\vk) = -2c_s^2 k^2 C_{\delta_M}^2(k)P_L(k)
\eeq
in real-space, where $c_s^2$ strictly depends on redshift.

For biased tracers in redshift-space the counterterm structure of $P(\vk)\equiv P(k,\mu)$ is somewhat more complex. Here, we follow Refs.\,\citep{2014arXiv1409.1225S,2016arXiv161009321P,2020JCAP...05..042I}, and use
\beq
    M_{ct,\ell}(k) = -2c_{\ell}^2 k^2 C_{\delta_M}^2(k)P_L(k),
\eeq
for the multipole counterterms. Note that we use free coefficients for each $\ell$ to encapsulate additional effects including higher-derivative bias (affecting $\ell = 0$) and fingers-of-God (FoG), both of which inherit the $k^2$ scaling at leading order. A full treatment of the UV behavior of the unbiased real-space loop integrals was performed in Ref.\,\citep{2020PhRvD.102d3516P}; for the more general case, it suffices to say that the only additional divergences are of the form:
\beq
    M_{13}(\vk) &\supset& 2C_{\delta_M}(k)C_{\delta^2_M}(k,0)Z_1^2(\vec k)P_L(k)\int_{\vec p}\frac{b_2}{2}P_L(p)\\\nonumber
    M_{22}(\vk) &\supset& \frac{b_2^2}{2}C_{\delta_M}^2(k)\int_{\vec p} P_L(p)P_L(|\vk-\vp|) = \frac{b_2^2}{2}C_{\delta_M}^2(k)\int_{\vec p} \left[P_L(p)P_L(|\vk-\vp|) - P_L^2(p)\right]\\\nonumber
    &&\qquad\qquad\qquad\qquad\qquad\qquad\qquad +  \frac{b_2^2}{2}C_{\delta_M}^2(k)\int_{\vec p} P_L^2(p).
\eeq
%\citep{2006PhRvD..74j3512M}. 
The first of these vanishes when including the properly renormalized bias operators \eqref{eq: bias-renorm-cont}, and the second contains a divergent part  (explicitly given in the final line) also present in $P(\vk)$. %\footnote{\alej{It should be, as in \cite{2006PhRvD..74j3512M}, $M_{22}(\vk) \supset \frac{b_2^2}{2}C_{\delta_M}^2(k)\int_{\vec p} P_L(p)P_L(|\vk-\vp|)=\frac{b_2^2}{2}C_{\delta_M}^2(k)\int_{\vec p} P_L(p)(P_L(|\vk-\vp|)-P_L(p))  +  \frac{b_2^2}{2}C_{\delta_M}^2(k)\int_{\vec p} P^2_L(p)$ and the second term absorbed by the shot-noise. But it is not wrong per-se as it is written.}\oliver{Included!}} 
Assuming the following form for the stochastic part of the power spectrum (\textit{i.e.} that uncorrelated with the matter field $\delta$),
\beq
    M_{\mathrm{stoch},\ell}(k) = C_{\delta_M}^2(k)\delta^K_{\ell 0}\times P_\mathrm{shot},
\eeq
for constant $P_\mathrm{shot}$ and Kronecker delta $\delta^K$, the UV sensitive part of $M_{22}(\vk)$ is fully absorbed. In practice, we find that a $k$-independent stochastic contribution of the form $M_{\mathrm{stoch},\ell}(k) = \delta^K_{\ell 0}P_\mathrm{shot}$ (as suggested by Poisson statistics) gives a better fit to the data and can absorb the majority of the above divergence. This will be assumed henceforth. Finally, following Ref.\,\citep{2020JCAP...05..042I}, we include an additional higher-order counterterm
\beq
    M_{ct,\mathrm{NLO}}(\vk) = \tilde{c}\,(kf\mu )^4\times  C_{\delta_M}^2(k)Z_1^2(\vk)P_L(k),
\eeq
where $f$ is the growth rate; this accounts for the next order contribution of FoG.\footnote{Recall that any FoG kernel is a function of $(k\mu\sigma_v)^2$ for some velocity dispersion $\sigma_v^2$; our approach is to take the terms in its Taylor expansion allowing the coefficients to be free; this is more general than assuming some functional form.}

\resub{One final issue of note concerns \textit{infra-red resummation}. Any Eulerian PT necessarily incurs an error by perturbatively expanding long-wavelength (IR) displacements that are not guaranteed to be small. This is contrary to the approach of Lagrangian PT and leads to insufficiently damped oscillations in the power spectrum, or, in configuration-space, an erroneous enhancement of the BAO peak \citep{2015JCAP...02..013S}. For the power spectrum of matter at leading and next-to-leading order, this can be accounted for by performing the following resummation:
\beq\label{eq: IR-res}
    P^\mathrm{IR-res}_\mathrm{LO}(\vk) &=& P_{L}^{nw}(\vk) + e^{-k^2\Sigma^2(\mu)}P_{L}^{w}(\vk)\\\nonumber
    P^\mathrm{IR-res}_\mathrm{NLO}(\vk) &=& P_L^{nw}(\vk)+P^{nw}_\mathrm{1-loop}(\vk) + e^{-k^2\Sigma^2(\mu)}\left[\left(1+k^2\Sigma^2(\mu)\right)P_L^{w}(\vk)+P_\mathrm{1-loop}^{w}(\vk)\right]
\eeq
\citep{2018JCAP...07..053I}, where `w' and `nw' refer to the oscillatory and broadband parts of the power spectrum, and the velocity dispersion $\Sigma^2(\mu)$ depends weakly on cosmology. An analogous procedure is possible for the power spectrum of biased tracers. Whilst a full treatment of IR resummation for $M(\vk)$ is beyond the scope of this work, for the purpose of comparing theory to data it is useful to include the effect in an approximate manner. Our approximation is in two parts: firstly, for the term involving $C_{\delta_M}^2(k)P_{NL}(\vk)$ (see Appendix \ref{appen: integral-simp}), we replace $P_\mathrm{NL}$ with its IR-resummed form $P^\mathrm{IR-res}_\mathrm{NLO}(\vk)$, and secondly, we replace any additional linear power spectra $P_L(\vk)$ with $P^\mathrm{IR-res}_\mathrm{LO}(\vk)$. Whilst this is not a full treatment, we expect the majority of the oscillatory behavior to be sourced by terms directly proportional to $P_{NL}(\vk)$ and $P_{L}(\vk)$ which our procedure correctly treats, thus the scheme is expected to be adequate.}\footnote{\resub{Empirically, this is demonstrated in Sec.\,\ref{sec: data}, as we do not notice clear residual wiggles between data and theory, though these are apparent if IR resummation is not included.}}

\begin{figure}
    \centering
    \includegraphics[width=\textwidth]{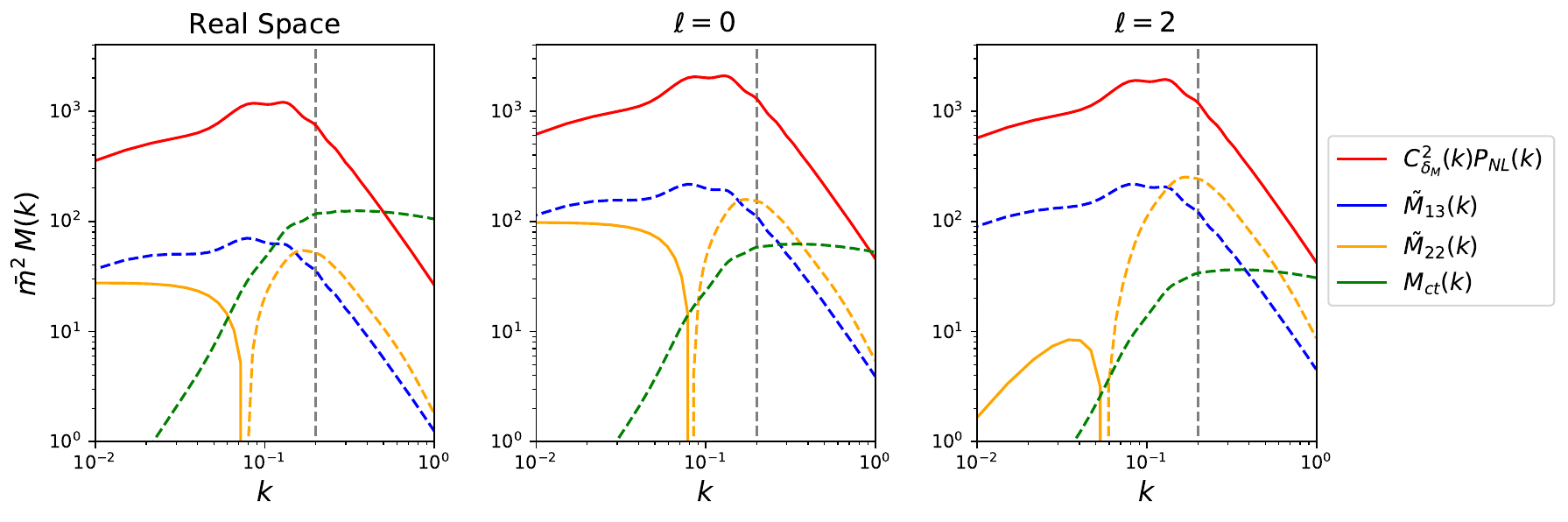}
    \caption{Components of the marked power spectrum of matter at one-loop order. This uses the mark parameters $\{p = 1, \delta_s = 0.25, R = 15\Mpch\}$ at $z = 1$, and the sum of all components (with $M_{ct}$ weighted by a free counterterm) is equal to the full model for $M(\vk)$. The first term, $C_{\delta_M}^2(k)P_\mathrm{NL}(k)$ includes both the linear `11' term and the contributions to the `22' and `13' components that are proportional to those in the unmarked power spectrum. $\tilde{M}_{22}$ and $\tilde{M}_{33}$ represent the additional contributions to the `22' and `13' terms; these remain important in the $k\rightarrow0$ limit. The left panel gives the results in real-space, with redshift-space monopole (quadrupole) being shown in the central (right) panel. $k$ and $M(k)$ are in $\hMpc$ and $h^{-3}\mathrm{Mpc}^3$ units respectively, and dashed lines indicate negative contributions.}
    \label{fig: mk-components}
\end{figure}

In summary, the one-loop EFT model for the marked spectrum of biased tracers in redshift-space has the following form, expressed in multipoles,
\beq\label{eq: Mk-EFT-model}
    \bar{m}^2M_\ell(k) = M_{11,\ell}(k) + 2M_{13,\ell}(k) + M_{22,\ell}(k) + M_{ct,\ell}(k) + M_{\mathrm{stoch},\ell}(k),
\eeq
which, assuming $\ell_\mathrm{max} = 2$, carries the following seven nuisance parameters;
\beq\label{eq: free-params}
    \{b_1,b_2,b_{\mathcal{G}_2},c_0^2,c_2^2,P_\mathrm{shot},\tilde{c}\}.
\eeq
These should parallel those in $P_\ell(k)$ (indeed only $b_1, b_2, b_{\mathcal{G}_2}$ enter into the $M(\vk)$ terms which are not proportional to $P(\vk)$ at one-loop order) though may be renormalized by higher-order contributions (as discussed below). In real-space, we do not require $c_2^2$ or $\tilde{c}$ (and can set $\mu = f = 0$), whilst for unbiased tracers, we can set $b_1 = 1$, $b_2 = b_{\mathcal{G}_2} = P_\mathrm{shot} = 0$. The one-loop components for matter in real- and redshift-space are plotted in Fig.\,\ref{fig: mk-components}.

%\alej{This $\tilde{c} \, k^4$ counterterm is new to me.  Now I see how can you go up to $k \sim 0.3$. Chen, Vlah \& White 2020 do not consider it.}\oliver{This is quite useful to capture the next-to-leading order FoG effects for BOSS since its FoG is quite strong. Its not always used, \textit{e.g.} the blinded EFT challenge paper didn't use it (and was limited to lower-$k$, due to the much higher precision data meaning that two-loops becomes important quite early on relative to the statistical errors.). As an aside, we don't go up to $k = 0.3$ for BOSS; only to $k = 0.25$ which is only possible due to the big error bars.}

\section{Reorganizing the Marked Spectrum}\label{sec: reorg}

The one-loop theory for $M(\vk)$ exhibits a curious property; many of the second- and third-order terms in $\delta_M(\vx)$ are not parametrically smaller than those in $\dptM{1}$ at low-$k$. More precisely, the low-$k$ limit of $M^{1-\mathrm{loop}}(\vk)$ contains terms proportional to $P_L(k)$, as in the `13'-type contributions in Fig.\,\ref{fig: mk-components}. This is in contrast to $P^\mathrm{1-loop}(\vk)$, where such terms are proportional to $\left(k/k_\mathrm{NL}\right)^2P_L(k)$. Further, there exist (UV-convergent) loop terms which tend to a constant at low-$k$ in $M^\mathrm{1-loop}(\vk)$, as in the `22' part of Fig.\,\ref{fig: mk-components}. The reason for this is straightforward; the marked density field cannot be well approximated by its first-order Taylor expansion even at large scales. %\footnote{\alej{Maybe is worthy to say something like: The reason for this is simple, the marked density field cannot be approximated as linear even at large scales.}\oliver{Agreed, though I think the exact point is that $\delta_M$ cannot be well represented by the first term in its Taylor expansion, rather than whether the underlying theory is linear. e.g. in a Universe without gravitational non-linearities, $\delta$ is linear, yet we still need higher order terms in $\delta_M$.}}
This complicates the theory, since there is no longer a well-defined radius of convergence: all loops contribute non-trivially to low-$k$ (though their amplitudes are diminishing, assuming that the convergence condition \eqref{eq: taylor-condition} is met). In this section, we discuss this subtlety, and show how the theory can, at least formally, be reorganized into a form in which loop contributions are manifestly small on large scales.

\subsection{Loop Renormalization}\label{subsec: reorg-motiv}
The additional complexity of the marked power spectrum can be related to the existence of \textit{contact terms} in the theory; products of operators evaluated at the same physical location, e.g., $\delta_g(\vx)\delta_{g,R}(\vx)$. A similar effect is present in the usual EFT bias expansions \eqref{eq: bias-expan}; for instance, the tracer overdensity $\delta_g(\vx)$ contains a second-order term proportional to $\delta^2(\vx)$. In this case, the contact terms are not UV-safe and must be corrected via some renormalization scheme \citep{2014JCAP...08..056A,2018PhRvD..98h3541A}, which effectively leads to rescaling of the bias operators and stochastic terms, as discussed in Sec.\,\ref{subsec: theory-uv}. As an example, the cubic bias $\delta^3$ term requires renormalization, via $\delta^3\rightarrow \delta^3 - 3\sigma^2\delta$ to avoid spurious (UV-unsafe) $\sigma^2$ factors appearing in the density correlators. This renormalization can be realized simply through a rescaling $b_1\rightarrow b_1^R\equiv b_1 + \frac{1}{2}b_3\sigma^2$, and thenceforth ignored \citep{2006PhRvD..74j3512M}. %Further, \alej{...?}%\footnote{\alej{As a self-promotion: as far as I know, the contact terms were not systematic addressed until my paper \href{https://arxiv.org/abs/1805.05304}{https://arxiv.org/abs/1805.05304} (section IV), and the UV sensitivity is absorbed by stochastic parameters. Assasi et al. only mention this problem and ignore these UV sensitive terms. There are other papers starting with McDonald 2016, but my approach is more general, so I would cite my paper.}\oliver{A good point. I think that the Assassi formalism is a systematic way of accounting for the UV sensitivity from all terms in the bias expansion (e.g. $b_2\delta^2$) at any given order in PT, though their expansions are only correct up to stochasticity-induced pieces, e.g. the low-k constant. I'll make sure your paper is cited for this.}}

For marked statistics, the situation is somewhat different. As previously noted, none of the new contact terms encounter UV divergences, since all are of the form $\delta_g(\vx)\delta^n_{g,R}(\vx)$ or $\delta^n_{g,R}(\vx)$ for integer $n$, thus, due to the presence of smoothing windows, none strictly require renormalization, making the assumption that $R > k_\mathrm{NL}^{-1}$ for non-linear scale $k_\mathrm{NL}$.\footnote{The presence of this window makes the UV limits more complex, and dependent on the form of $W_R(k)$; this is elaborated upon in Refs.\,\citep{2020PhRvD.102d3516P,2020JCAP...01..006A} and becomes crucial in configuration space.} However, these UV (or equivalently low-$k$) limits give significant contributions to the one-loop power spectrum, and this is present at all loop order. As an example, consider the term $\delta^3_{g,R}(\vx)$ with each density field evaluated according to linear theory. The term is of one-loop order (since it involves three linear density fields), yet in correlators we will find contributions of the form $\av{\dptgR{1}(\vx)\dptgR{1}(\vx)}\dptgR{1}(\vx)\equiv \sigma^2_{g,RR}\,\dptgR{1}(\vx)$, which is just the linear term suppressed by the variance of the biased smoothed field, $\sigma^2_{g,RR}$. For the Taylor expansion to be valid, this is necessarily smaller than unity, yet it contributes on all scales; unlike familiar one-loop terms, it remains important down to $k = 0$. 

\begin{figure}%
    \centering
    \subfloat[Contributions sourcing $P(k)$-like terms]{{\includegraphics[width=\textwidth]{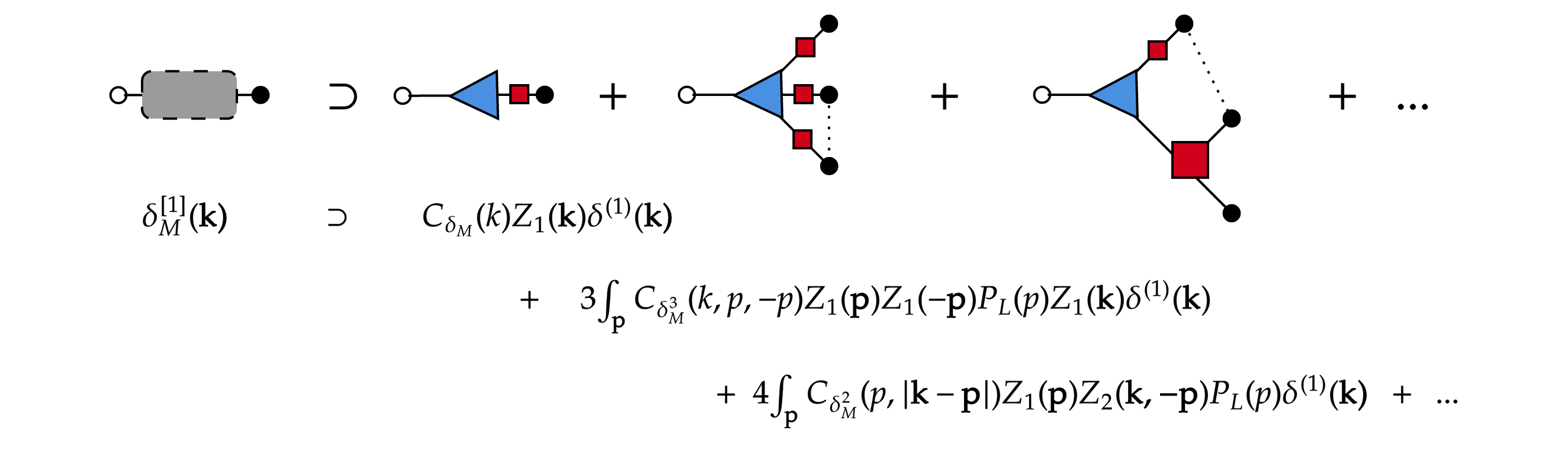} }}%
    \\
    \subfloat[Contributions sourcing constant terms]{{\includegraphics[width=\textwidth]{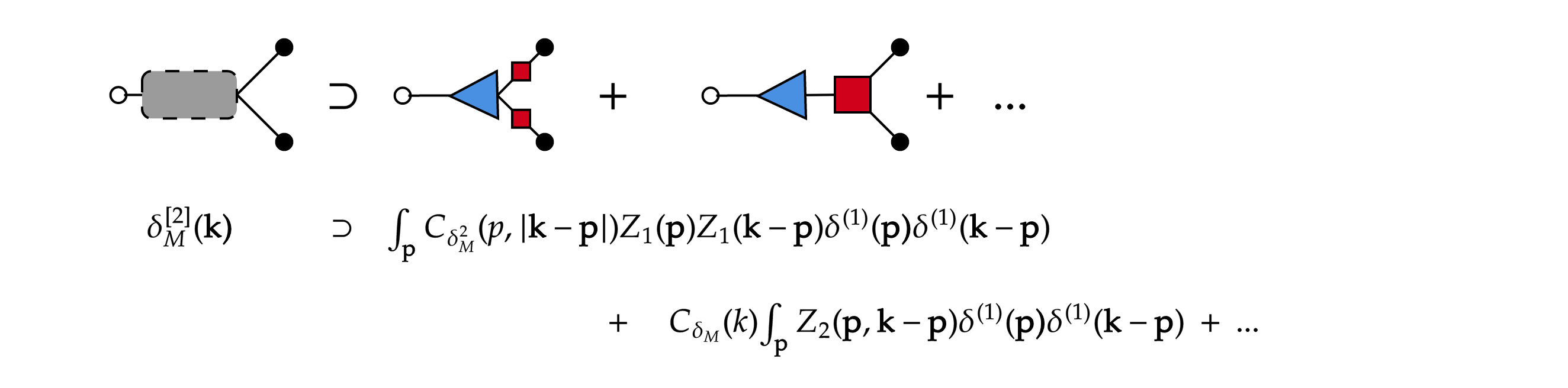} }}%
    \caption{Diagrammatic representation of the various terms contributing to the low-$k$ limit of the marked power spectrum and their corresponding loop integrals. In particular, we list all terms up to one-loop order in $\delta_M(\vk)$ that give contributions to $M(\vk)$ which, in the low-$k$ limit, are proportional to $P_L(k)$ (top panel) or a constant (bottom panel). As shown in the leftmost diagram, these all have (a) one external $\dpt{1}$ leg (black filled circle), or (b) two external legs and some intermediary contractions, depending on the loop order. Here, red squares with $n$ legs represent $Z_n$, blue triangles with $n$ legs represent $C_{\delta_M^n}$ and dotted lines represent the linear power spectrum $P_L(k)$. The full low-$k$ theory for $M(\vk)$ is obtained by contracting two such sets of diagrams.}%
    \label{fig: feyn-reorg}%
\end{figure}

The above behavior is not limited to one-loop order, nor to the linear pieces of $\delta_{g,R}$ and $\delta_g$: all $\dptM{n}(\vx)$ pieces with odd $n$ contribute terms linearly proportional to $\dpt{1}(\vx)$.\footnote{This is trivially true via Wick's theorem: $\dptM{n}$ contains $n$ copies of $\dpt{1}$, thus contracting $n-1$ internal fields gives a term proportional to $\dpt{1}$, provided $n-1$ is even.} Fig.\,\ref{fig: feyn-reorg}a gives a diagrammatic representation of the contributions to $\delta_M(\vk)$ which lead to $M(\vk)$ terms with low-$k$ limits proportional to $P_L(k)$ at zero- and one-loop order. In general, these are sourced by any diagrams with a single external $\dpt{1}$ leg (visualized here by a black circle), with all other internal density fields contracted. Note that we ignore the $\dptg{3}$ correlator proportional to $\int_{\vp} Z_3(\vk,\vp,-\vp)P_L(p)$; due to the low-$k$ scalings of the kernel, this scales as $(k/k_\mathrm{NL})^2P_L(k)$ (with an additional UV-unsafe part for biased tracers being partially absorbed by $P_\mathrm{shot}$, as described above). This highlights another important property; the terms important at low-$k$ necessarily involve zero-lag correlators between $\delta_{g}$ and/or $\delta_{g,R}$ fields. %\footnote{\alej{I have a problem with this terminology: contact operators are terms like $\xi(\vx)\xi(\vx)$ that in Fourier space gives convolutions $[P_L * P_L](\vk)$ which are UV sensitive. Instead, you are referring to the zero-lag correlators as contact terms.}\oliver{For me, contact operators are simply pairs of operators at the same spatial location, \textit{i.e.} $\mathcal{O}_1(\vx)\mathcal{O}_2(\vx)$, \textit{i.e.} correlation functions with zero separation. Maybe this is unconventional. In this example, I meant zero-lag correlator however and have changed it.}}

Dealing with such terms is non-trivial. One approach would be to replace the Taylor series with an expansion in terms of renormalized operators, e.g.,
\beq
    \bar{m}\,\delta_M(\vx) = \tilde{C}_0^0 + \tilde{C}_0^1\delta_{g}(\vx)+\tilde{C}_1^0\delta_{g,R}(\vx)+\tilde{C}_1^1[\delta_{g}\delta_{g,R}](\vx) + \tilde{C}_2^0[\delta_{g,R}^2](\vx) + ...,
\eeq
where the square brackets indicate some renormalization scheme akin to that of Ref.\,\citep{2014JCAP...08..056A}, \textit{i.e.} $[\delta_X^n]$ is equal to $\delta_X^n$ but with all internal correlators up to a given order subtracted off. In principle, this means the theory can be evaluated simply by evaluating non-one particle irreducible (non-1PI) diagrams, since all 1PI diagrams (such as those of Fig.\,\ref{fig: feyn-reorg}) have been removed by construction. The difficulty with this approach lies in the effective coefficients $\tilde{C}_i^j$. Unlike for the bias expansion, the unrenormalized coefficients $C_n$ are fixed by the mark parameters, thus we cannot simply absorb constant pieces into them via redefinitions. Furthermore, due to the $W_R$ smoothing window, the renormalized coefficients will inherit non-trivial $k$ dependence. To properly evaluate such a theory, one would need to compute all possible 1PI diagrams, then compute their low-$k$ limit, vanquishing the gains from renormalization.

An alternative scheme would be to consider some kind of propagator formalism, explicitly expanding $\delta_M(\vx)$ as a series in the number of \textit{external} legs $\dpt{1}$, \textit{i.e.}
\beq
    \bar{m}\,\delta_M(\vk) = \sum_{n = 0}^\infty \delta_M^{[n]}(\vk),
\eeq
where $\delta_M^{[n]}$ contains all diagrams with $n$ external legs and any number of internal contractions. This is the premise of \textit{renormalized} perturbation theory \citep{2006PhRvD..73f3519C,2006PhRvD..73f3520C,2008PhRvD..78j3521B}, used for the evaluation of the log-normal density field in Ref.\,\citep{2011ApJ...735...32W}. Here, $\delta_M^{[n]}$ is given as an integral over $n$ linear density fields;
\beq
    \delta_M^{[n]}(\vk) &=& \frac{\bar{m}}{n!}\int_{\vk_1...\vk_n}\left\langle\frac{\partial^n \delta_M(\vec k)}{\partial\dpt{1}(\vk_1)...\partial\dpt{1}(\vk_n)}\right\rangle\dpt{1}(\vk_1)...\dpt{1}(\vk_n)\\\nonumber
    &\equiv& \int_{\vk_1...\vk_n}\Gamma^{[n]}(\vk_1,...,\vk_n)\delD(\vk_1+...+\vk_n-\vk)\dpt{1}(\vk_1)...\dpt{1}(\vk_n),
\eeq
defining the $(n+1)$-point propagator $\Gamma^{[n]}$, which can be related to the original $\delta_M$ expansion \eqref{eq: Hdef} using Wick's theorem;
\beq
    \delD(\vk_1+...+\vk_n-\vk)\Gamma^{[n]}(\vk_1,...,\vk_n) &\equiv& \frac{1}{n!}\sum_{m=1}^\infty \left\langle\frac{\partial^n\dptM{m}(\vk)}{\partial\dpt{1}(\vk_1)...\partial\dpt{1}(\vk_n)}\right\rangle\nonumber
\eeq
\beq\label{eq: propagator-def}
    \Gamma^{[n]}(\vk_1,...,\vk_n)&=& \sum_{m=n}^\infty\frac{(-1)^{n+m}+1}{2}\binom{m}{n}(m-n-1)!!\\\nonumber
    &&\,\times\int_{\vp_1..\vp_{q_{nm}}}H_m(\vk_1,...,\vk_n,\vp_1,-\vp_1,...,\vp_{q_{nm}},-\vp_{q_{nm}})P_L(p_1)...P_L(p_{q_{nm}}),
\eeq
where $q_{nm} = (m-n)/2$, $H_m$ are the marked kernels defined in \eqref{eq: Hdef} and we have dropped zero-lag terms. In this formalism, the $M(\vk)$ spectrum is given by
\beq
    \bar{m}^2M(\vk) = \sum_{n=1}^\infty M_{[nn]}(\vec k) = \sum_{n=1}^\infty n!\int_{\vk_1...\vk_n}\left|\Gamma^{[n]}(\vk_1,...,\vk_n)\right|^2P_L(k_1)...P_L(k_n),
\eeq
which is simply a reorganization of the Eulerian PT result. The principal benefit of this is the separation of terms proportional to $P_L(k)$; by construction, these appear \textit{only} in $M_{[11]}$, with amplitudes given by
\beq
    \frac{M_{[11]}(\vk)}{P_L(k)} = \left|\Gamma^{[1]}(\vk)\right|^2.
\eeq
Calculation of the low-$k$ $P_L(k)$ dependence thus reduces to computing the low-$k$ limit of $\Gamma^{[1]}$.\footnote{Note that $\Gamma^{[1]}$ contains also terms suppressed by powers of $(k/k_\mathrm{NL})^2$, e.g., those from $M_{13}(\vk)$.} The pudding is not yet proved however, since $\Gamma^{[n]}$ technically contains terms of all loop orders; computation of this at one-loop order and its infinite-order form will be discussed in Secs.\,\ref{subsec: theory-reorg-1-loop}\,\&\,\ref{subsec: theory-reorg-inf} respectively. A word of caution is needed when interpreting the above result, since the EFT counterterms have been ignored in the above definitions of $\Gamma^{[n]}$. Strictly speaking, they should be included in the definitions of the $H_n$ kernels, order-by-order, though, at one-loop, this can be ignored, since the relevant counterterms are suppressed by $(k/k_\mathrm{NL})^2$.\footnote{This does not hold at all loops: as an example, consider the 2-loop contribution involving $Z_1(\vk)Z_1(\vp)Z_3(\vec q,-\vec q,-\vp)$. The $\vec q$ integral is not UV-safe (since smoothing windows only occur for $W_R(p)$ and $W_R(k)$) and thus the counterterm is needed, even though the low-$k$ limit is not parametrically suppressed.}

Whilst $M_{[11]}$ encapsulates all terms scaling as $P_L(k)$ at low-$k$, it is not the full $k\rightarrow0$ limit; additional contributions occur which tend to a constant on the largest scales. These appear from all $M_{[nn]}$ terms with $n>1$, for example:
\beq\label{eq: m[22],m[33]-def}
    \lim_{k\rightarrow0}M_{[22]}(\vk) &=& \int_{\vp}\left|\lim_{k\ll p}\Gamma^{[2]}(\vp,\vk-\vp)\right|^2P_L^2(p)\\\nonumber
    \lim_{k\rightarrow0}M_{[33]}(\vk) &=& \int_{\vp,\vec q}\left|\lim_{k\ll p,q}\Gamma^{[3]}(\vp,\vq,\vk-\vp-\vq)\right|^2P_L(p)P_L(q)P_L(|\vp+\vq|),
\eeq
both of which are $k$-independent at $k = 0$ though inherit $k$ dependence with characteristic scale $k\sim 1/R$ from the window functions in $C_{\delta_M^n}$. For unmarked matter all diagrams vanish  (as e.g., $F_2(\vp,-\vp) = G_2(\vp,-\vp) = 0$), though they are non-zero for unmarked biased tracers, (e.g. with $Z_2(\vp,-\vp) = b_2/2$), yet degenerate with the shot-noise. For marked statistics, they are in general non-zero, and cannot be exactly captured by the shot-noise (if present), since the exact low-$k$ dependence is non-trivial.

\subsection{Reorganized Linear Theory}
Consideration of these low-$k$ limits motivates an alternative expansion, hereby dubbed the \textit{reorganized} theory, whereupon we collect together all terms of the same order in $(k/k_\mathrm{NL})^{2}$;
\beq
    \bar{m}^2M^\mathrm{reorg}(\vk) = M^{r,0}(\vk) + M^{r,1}(\vk) + ...,
\eeq
where $M^{r,n}$ scales as $(k/k_\mathrm{NL})^{2n}P_L(k)$ in the low-$k$ limit, and $M^{r,0}$ additionally includes the term tending towards a constant as $k\rightarrow0$. Practically, this work considers only one-loop theory explicitly, which contributes to both $M^{r,0}$ and $M^{r,1}$. In this context,
\beq\label{eq: Mreo1-loop}
    \left.M^{r,0}(\vk)\right|_{1-\mathrm{loop}} &=& M_{11}(\vk)+ \left(M_{22}^{r,0}(\vk)+2M_{13}^{r,0}(\vk)\right)+M_\mathrm{stoch}(\vk)\\\nonumber
    \left.M^{r,1}(\vk)\right|_{1-\mathrm{loop}} &=& M_{22}^{r,1}(\vk)+2M_{13}^{r,1}(\vk)+M_{ct}(\vk),
\eeq
in terms of the EFT contributions discussed in Sec.\,\ref{sec: theory} (assuming that the stochasticity power spectrum contains only the lowest order term in an expansion in powers of $(k/k_\mathrm{NL})^2$). This uses the definitions
\beq \label{ec: M13M22r0}
    M_{13}^{r,0}(\vk) &=& H_1(\vk)P_L(k)\lim_{k\rightarrow0}\left[\frac{M_{13}(\vk)}{H_1(\vk)P_L(k)}\right]\\\nonumber
    M_{22}^{r,0}(\vk) &=& \lim_{k\rightarrow0} M_{22}(\vk),
\eeq
with $M_{ab}^{r,1}(\vk) = M_{ab}(\vk) - M_{ab}^{r,0}(\vk)$ at one-loop order, noting that $M_{13}(\vk)$ ($M_{22}(\vk)$) contributes only to the $P_L(k)$-like (constant) low-$k$ limit. The explicit form of $M^{r,0}(\vk)$ will be discussed in the following subsection. 
%\footnote{\alej{I like a lot how you discuss the theory until here. However the two above equations are not entirely correct, I know you know. For %example: $\lim_{k\rightarrow0}\left(M_{22}(\vk)+2M_{13}(\vk)\right)$ is just a constant. I would write something like: For a kernel $m_{22}(\vk_1,\vk_2)$, such that  $M_{22}(\vk) = \int_{\vp} m_{22}(\vk-\vp,\vp)$
%\begin{align}
%  M_{22}(\vk) %&= \int_{\vp} m_{22}(\vk-\vp,\vp) \nonumber\\
%              &=   \int_{\vp} m_{22}(-\vp,\vp) + \int_{\vp} (m_{22}(\vk-\vp,\vp)-m_{22}(-\vp,\vp))  = M_{22}^{r,0}(\vk) + M_{22}^{r,1}(\vk) \\
%  M_{13}^{r,0}(\vk) &= H(\vk) P_L(\vk) \lim_{k\rightarrow0} \left[ \frac{M_{13}(\vk)}{ H(\vk) P_L(\vk)} \right], \qquad   M_{13}^{r,1}(\vk) =  M_{13}(\vk)- M_{13}^{r,0}(\vk)
%\end{align}
%and 
%\beq
%    \left.M^{r,0}(\vk)\right|_{1-\mathrm{loop}} &=& M_{11}(\vk)+ M_{22}^{r,0}(\vk) + M_{13}^{r,0}(\vk), \\\nonumber
%    \left.M^{r,1}(\vk)\right|_{1-\mathrm{loop}} &=& M_{22}^{r,1}(\vk) + M_{13}^{r,1}(\vk) +M_{ct}(\vk).
%\eeq
%I am trying to be general, and not only focusing on either real or redshift-space
%}\oliver{I agree with this, and have made the change above. I don't think its necessary to introduce the $m_{22}$ function however, since $\int_{\vp}m_{22}(\vp,-\vp)$ is just the $k\rightarrow0$ limit.}}

Of course, this expansion is just a reorganization of known terms with no change to the underlying theory. Its benefit however is that is allows for a well-defined `linear' theory, $M^{r,0}(\vk)$, with higher order terms being parametrically suppressed by powers of $(k/k_\mathrm{NL})^2$. Any exact calculation is limited to finite order in loops however, thus this decomposition is strictly only formal; \textit{i.e.} $M^{r,0}$ calculated from one-loop EFT does not contain $P_L(k)$ contributions from higher loops. Further, the number of contributing diagrams grows quickly with loop order, with eight non-trivial $P_L(k)$-like diagrams at two-loop, all of which contain six-dimensional integrals (and up to fourth-order bias parameters). In practice however, Sec.\,\ref{sec: data} shows the reorganized linear term at one-loop order to provide a far better match to simulations than linear theory, with the corrections from the full one-loop EFT only becoming important at higher $k$.

\subsubsection{One-Loop Order}\label{subsec: theory-reorg-1-loop}
We now demonstrate the character of the low-$k$ limit of $M(\vk)$ for the one-loop theory of biased tracers in redshift-space, and thus the one-loop contributions to the reorganized linear theory, $M^{r,0}(\vk)$. \textit{Viz.} the above discussion, two types of terms are required: (a) those proportional to $P_L(k)$, and (b) those that tend to a constant as $k\rightarrow0$. The derivation of both parts are sketched in Appendix \ref{appen: low-k-limits}; in combination, we obtain the low-$k$ limit, and hence reorganized linear theory:
\beq \label{eq: Mr01loop}
    \left.M^{r,0}(\vk)\right|_{1-\mathrm{loop}} &=& \lim_{k\rightarrow0}M_{[11]}(\vk)+\lim_{k\rightarrow0}M_{[22]}(\vk)+M_\mathrm{stoch}(\vk)\\\nonumber
    &=& C_{\delta_M}^2(k) (b_1 + f\mu^2 )^2 P_L(k)\\\nonumber
   &&\,+\,2C_{\delta_M}(k) (b_1 + f\mu^2 )^2 P_L(k) \left(b_1^2+\frac{2}{3}b_1f+\frac{1}{5}f^2\right)\\\nonumber
   &&\qquad\,\times\left[(C_2-3 C_3 W_R(k))\sigma^2_{RR}  +2 C_2 W_R(k) \sigma^2_{R}\right] \\\nonumber
   &&\,+\,2  C_{\delta_M}(k)  (b_1 + f\mu^2 ) P_L(k) \left[ \mathcal{A}_0(k) + \mathcal{A}_2(k) \mu^2 \right]\\\nonumber
   &&\,+\,\mathcal{B}(k)+M_\mathrm{stoch}(\vk),
\eeq
%\alej{To match eq.~\eqref{ec: limk0M11}, the second line of the last equality needs an extra factor of 2}
where terms in the third through fifth lines give the loop corrections to the $M_{11}$ piece, and the final line encodes the stochastic part of the spectrum and the constant terms in the $k\rightarrow0$ limit. Setting $f = \mu = 0$ straightforwardly gives the real-space counterpart to this.\footnote{For unmarked statistics, we can simply set $\mu = 0$ to recover the real-space prediction. Here, we must additionally set $f = 0$ since the angular parts are mixed up by the mark function, \textit{i.e.} $\mu = 0$ terms in $M(\vk)$ are not solely sourced by $\mu = 0$ terms in $P(\vk)$. This occurs because the mark transformation is applied \textit{after} redshift-space distortions; for a simple example consider the terms involving $\av{\delta_R(\vx)\delta_R(\vx)}$. These are integrated over angle, and thus contain $f$ factors but no explicit $\mu$. Setting $f = 0$ removes the anisotropy of the underlying unmarked field, giving the correct result.} The functions $\mathcal{A}_n(k)$ %\footnote{\alej{I assume that these functions coincide with my functions $2\times J_n^{LS}$ in the notes. Note the factor of 2.}\oliver{Yes. The 2 comes from $M_{[22]}\sim 2\left|\Gamma^{[2]}\right|^2$}} 
and $\mathcal{B}(k)$ depend on the second-order biases and have weak $k$ dependence through $W_R(k)$; these are given in \eqref{eq: calA-def}\,\&\,\eqref{eq: calB-def} in Appendix \ref{appen: low-k-limits}. Furthermore, this uses the variance definitions
\beq\label{eq: sig2R-defs}
    \sigma^2_{R} = \int_{\vp}W_R(p)P_L(p)\,,\qquad \sigma^2_{RR} = \int_{\vp}W^2_R(p)P_L(p).
\eeq

\begin{figure}
    \centering
    \includegraphics[width=\textwidth]{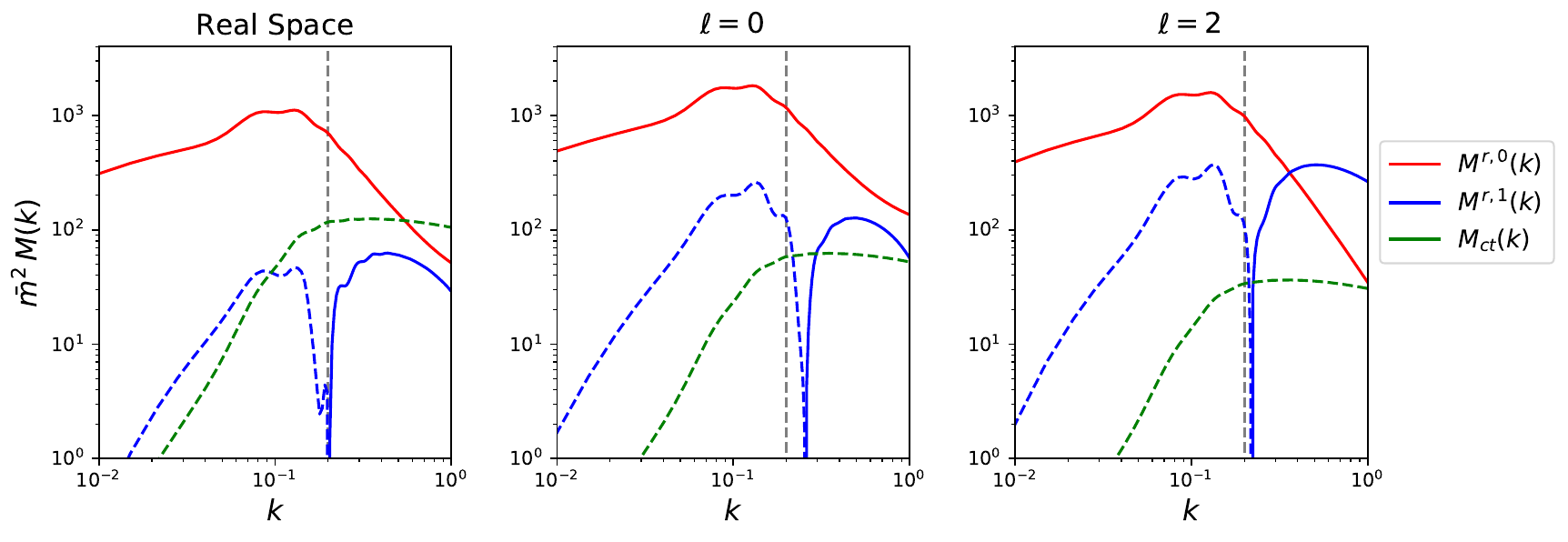}
    \caption{Components of the marked power spectrum of matter at one-loop order, following the reorganization scheme of Sec.\,\ref{sec: reorg}. This is analogous to Fig.\,\ref{fig: mk-components} but regroups the terms such that the lowest-order (reorganized linear, $M^{r,0}$) term contains all pieces relevant in the $k\rightarrow0$ limit and the first correction ($M^{r,1}$) is suppressed by a factor of $(k/k_\mathrm{NL})^2$ on large scales, allowing for a (formally) well-defined theory. Note that the counterterm piece must be multiplied by a free parameter fitted to simulations or data.}
    \label{fig: mk-components-reorg}
\end{figure}

For the redshift-space case, the above expression may be similarly written in terms of multipoles; in that case, we note that only $\ell =0,2,4$ are non-zero and the low-$k$ constant piece contributes only to the monopole. Importantly, the reorganized linear theory at one-loop order depends only on the free parameters $\{b_1,b_2,b_{\mathcal{G}_2},P_\mathrm{shot}\}$; if third order biases were present, they would not enter since this contribution is suppressed in the low-$k$ limit, as for the free counterterms. We plot the components of the reorganized theory of matter in real- and redshift-space in Fig.\,\ref{fig: mk-components-reorg} (which is simply a reshuffling of the terms present in Fig.\,\ref{fig: mk-components}). As expected, the lowest-order piece contains \textit{all} terms relevant on the largest scales, with the correction term, $M^{r,1}$, being parametrically suppressed.

% Alternatively, in terms of multipoles, we obtain
% \beq \label{eq: reorg-Mell-1loop}
%     \left.M^{r,0}_\ell(k)\right|_{1-\mathrm{loop}} &=& \left[C_0-C_1W_R(k)\right]^2P_{11,\ell}(k) + 2\left[C_0-C_1W_R(k)\right]P_{11,\ell}(k)\mathcal{A}_\ell^{1-\mathrm{loop}}(k)\\\nonumber
%     && + \delta^K_{\ell0}\mathcal{B}^{1-\mathrm{loop}}(k)+M_\mathrm{shot,\ell}(k),
% \eeq
% where $P_{11,\ell}$ are the multipoles of the linear theory (Kaiser) power spectrum, and we introduce the loop-correction functions
% \beq \label{AellandB}
%     \mathcal{A}_\ell^{1-\mathrm{loop}}(k) &=& \left[2C_2W_R(k)\mathcal{S}_R + \left(C_2-3C_3W_R(k)\right)\mathcal{S}_{RR}^2\right]\\\nonumber
%     &&\,%+b_2\sigma^2\left[\left(C_2-\frac{1}{2}C_1\right)W_R(k)-\frac{C_1}{2}\right]
%     + \frac{1}{K_\ell(k)}\left[4\left(C_2A_{RR,\ell}^{(1)}-C_1A_{R,\ell}^{(1)}\right)-\left(2C_2A_{RR,\ell}^{(2)}-C_1A_{R,\ell}^{(2)}\right)\right]\\\nonumber
%     \mathcal{B}^{1-\mathrm{loop}}(k) &=& \frac{b_2^2}{2}\left[C_0-C_1W_R(k)\right]^2\int\frac{p^2dp}{2\pi^2}P_L^2(p)\\\nonumber
%     &&\,- 2b_2\left[C_0-C_1W_R(k)\right]\int \frac{p^2dp}{2\pi^2}\left[C_1-C_2W_R(p)\right]\left(b_1^2+\frac{2}{3}f(p)b_1+\frac{1}{5}f^2(p)\right)W_R(p)P_L^2(p)\\\nonumber
%     &&\,+2\int\frac{p^2dp}{2\pi^2}\left[C_1-C_2W_R(p)\right]^2\left(b_1^4+\frac{4}{3}b_1^3f(p)+\frac{6}{5}b_1^2f^2(p)+\frac{4}{7}b_1f^3(p)+\frac{1}{9}f^4(p)\right)W_R^2(p)P_L^2(p).
% \eeq
% Note that the low-$k$ constant piece contributes only to the monopole.

\subsubsection{Infinite-Loop Order}\label{subsec: theory-reorg-inf}

\begin{figure}
    \centering
    \includegraphics[width=\textwidth]{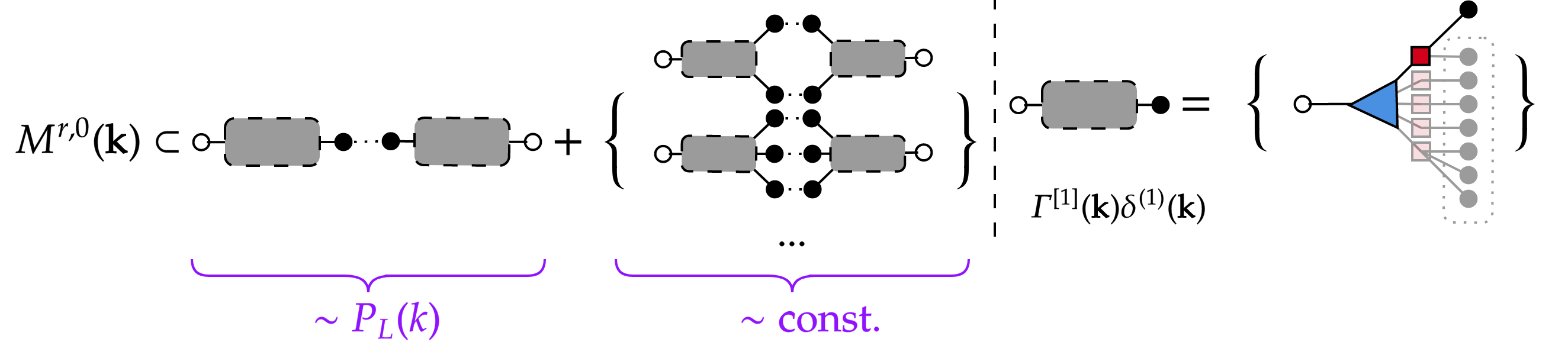}
    \caption{Diagrammatic form of terms entering the reorganized linear theory ($M^{r,0}(\vk)$) for the marked power spectrum $M(\vk)$ at infinite order in loops. On the left side, we show the terms proportional to $P_L(k)$ (from the two-point propagator) and a constant (from all higher point propagators), whilst the right side displays the general form of contributions to the two-point propagator $\Gamma^{[1]}$, noting that the piece of $M^{r,0}(\vk)$ proportional to $P_L(k)$ is just $\left|\Gamma^{[1]}(\vk)\right|^2P_L(k)$. As in Fig.\,\ref{fig: feyn-reorg}, grey rectangles connected to $n$ black circles represent the $(n+1)$-point propagator, \textit{i.e.} the sum of all diagrams with $n$ external legs. In general, $\Gamma^{[1]}$ is composed of any diagrams with a single $\dpt{1}$ vertex and a complete set of contractions between the other (even) number of fields, as indicated by the dashed lines. Note that this also contains terms suppressed by powers of $(k/k_\mathrm{NL})^2$, which may be dropped in the low-$k$ limit.}
    \label{fig: feyn-reorg-inf}
\end{figure}

Whilst the above subsection provides a useful separation of the one-loop theory into terms important at low-$k$ and those parametrically suppressed, to obtain the full expression for the reorganized linear piece it is necessary to work to infinite-loop order, to capture all terms important on large scales. Though the contributions necessarily decay with the number of loops (since an $\ell$-loop diagram contains $(\ell-1)$ powers of $b_1^2\sigma^2_{RR}$, which is below unity if the Taylor expansion \eqref{eq: taylor-condition} is convergent), this convergence is quite slow, especially for the mark parameter regimes (low $R$ and $\delta_s$) found favorable for cosmological parameter estimation in Ref.\,\citep{2020arXiv200111024M}. Further, the expansion is slower to converge if $\delta_g$ is non-linear \citep{2020PhRvD.102d3516P}. For studies using marked statistics to constrain modified gravity models, alternative parameters have been used \cite{2016JCAP...11..057W,2020JCAP...01..006A}, which are more suited for the one-loop reorganization proposed above. 
Further, whilst it may be useful to evaluate the low-$k$ theory to a higher loop order than the other contributions, this becomes intractable beyond a few-loop calculation (and, for biased tracers, unuseful, due to the large number of bias coefficients). It is thus useful to understand the contributions to $M^{r,0}$ from an arbitrary loop theory, such that we can arrive at an \textit{ansatz} for the functional form of the reorganized linear piece at infinite order.

As discussed above, the low-$k$ limit is the sum of two contributions;
\beq\label{eq: inf-Mr0-form}
    M^{r,0}(\vk) = \lim_{k\rightarrow0} M_{[11]}(\vk) + \lim_{k\rightarrow0}\sum_{n>1}M_{[nn]}(\vk),
\eeq
where the first term includes diagrams with one contraction between the two $\delta_M$ fields (which are proportional to $P_L(k)$) and the second those with any higher number of external contractions (giving a low-$k$ constant). These are illustrated schematically in the leftmost panel of Fig.\,\ref{fig: feyn-reorg-inf}, with the grey boxes indicating that the diagrams can contain any number of internal contractions, providing they have the required number of external $\dpt{1}$ legs. The right panel of the figure shows the form of the $\delta_M$ diagrams that contribute to the $P_L(k)$-like limit of $M^{r,0}$. As shown, there can be an arbitrary number of $\delta_g$ fields (red and pink squares), each of which can contain an arbitrary number of density fields (filled black and gray circles), but all bar one must be connected in some fashion. There are some limitations on these diagrams however: there can be no self-connections on the external density field  (\textit{i.e.} two black circles ($\delta$) contracted from the same red square ($\delta_g$)), else the contribution is suppressed in the low-$k$ limit or removed by bias renormalization. Furthermore, a diagram containing self-connections on \textit{any} galaxy density field is not UV-safe and will be affected by counterterm contributions.

We now consider the contribution of a general diagram to the large-scale limit of $M_{[11]}$ by first discussing the form of the $\Gamma^{[1]}$ propagator \eqref{eq: propagator-def}, recalling that $M_{[11]}(\vk) = |\Gamma^{[1]}(\vk)|^2P_L(k)$. %\elena{is this the $M_{[11]}$ in eq. 3.13? For me (probably because I am not an EFT expert) it is hard to navigate between the notation in "reorganized theory" with $M^{r,0}$ and in "renormalized theory" with gammas. Similar in Figure 4, I was confused on why there would be the gamma1 there.}. 
From Fig.\,\ref{fig: feyn-reorg-inf}, we see that $\vk$-dependence occurs only in two places: (1) the coupling $C_{\delta_M^n}$ (blue triangle) has one argument involving $k$ (from the condition that all paths out of the coupling must sum to $\vk$), and (2) the $Z_m$ kernel on the output leg (dark red square) contains a single $\vk$ argument. Schematically, the dependence of the two-point propagator is thus
\beq
    \lim_{k\rightarrow 0}\left.\Gamma^{[1]}(\vk)\right|_{\infty-\mathrm{loop}} = \sum_{\mathrm{diagrams}}\int_{...}C_{\delta_M^n}(...,|\vk-...|)Z_m(\vk,...)...,
\eeq
where ellipses indicatet contributions that are $k$-independent on large scales. Since the absolute orientation of $\vk$ arises only in a single argument of $Z_m$, we expect the $\mu$-dependence to arise only as a first-order polynomial in $\mu^2$, just as in the one-loop case. In terms of $k$ dependence, we consider only terms in $Z_m$ which are $k$-independent in the large-scale limit; contributions arise however from window functions in the $C_{\delta_M^n}$ coupling.\footnote{Whilst $W_R(k)\rightarrow 1$ as $k\rightarrow 0$, taking this limit is dangerous for finite $k$, since $W_R(k)$ has characteristic scale $\sim 1/R$, which can be far less than the non-linear scale $k_\mathrm{NL}$, which parametrizes when higher order terms in the reorganized expansion become important.} Since only one $\delta_g$ leg depends on $\vk$, we have at most one $W_R(k)$ function in $\Gamma^{[1]}(\vk)$.

The above discussion motivates the following \textit{ansatz} for the two-point propagator:
\beq
    \lim_{k\rightarrow 0}\left.\Gamma^{[1]}(\vk)\right|_{\infty-\mathrm{loop}} = \left(a_0+a_1W_R(k)\right)+\left(a_2+a_3W_R(k)\right)\mu^2,
\eeq
which has the correct functional form and is consistent with the structure of the one-loop result \eqref{eq: Gamma1-1-loop}. In real-space, we may simply set $a_2=a_3 = 0$. The $M_{[11]}$ result follows trivially;
\beq
    \lim_{k\rightarrow 0}\left.M_{[11]}(\vk)\right|_{\infty-\mathrm{loop}} = \left[\left(a_0+a_1W_R(k)\right)+\left(a_2+a_3W_R(k)\right)\mu^2\right]^2P_L(k),
\eeq
noting that no additional $\mu^4$ components are needed due to symmetry. In general, the coefficients $\{a_i\}$ depend on all higher loops and nuisance parameters and are not known; %\footnote{\alej{I still think there is a way, not in the approach we have followed, but with a different expansion from the beginning. But OK, this is out of reach by now.}\oliver{I think they can be resummed analytically for Gaussian fields. For non-Gaussian fields I don't think a general solution is possible, without requiring the Taylor expansion parameters to be free.}};
however, they may be treated as free nuisance parameters. For biased tracers, this form contains fewer parameters than an explicit evaluation of the low-$k$ limit, since there is not a proliferation of bias parameters (though these have important roles at larger $k$). 

We further require an \textit{ansatz} for the higher-point propagators. Motivated by the one-loop result, we expect these to have the functional form
\beq
    \lim_{k\rightarrow 0}\left.M_{[nn]}(\vk)\right|_{\infty-\mathrm{loop}} = \left[c_0+c_1W_R(k)\right]^2
\eeq
for $n>1$. As above, the $W_R$ dependence appears through the $C_{\delta_M^n}$ functions in $\delta_M$, with at most one contribution per field. As for the one-loop case, we do not expect this to have $\mu$ dependence. 

In summary, we make the following prediction for the reorganized marked spectrum at infinite loop:
\beq\label{eq: Mr0-inf-loop}
    \left.M^{r,0}(\vk)\right|_{\infty-\mathrm{loop}} &=& \left[\left(a_0+a_1W_R(k)\right)+\left(a_2+a_3W_R(k)\right)\mu^2\right]^2P_L(k)\\\nonumber
    &&\,+ \left[c_0+c_1W_R(k)\right]^2,
\eeq
depending on six free parameters (or four in real-space), one of which is partially degenerate with the shot-noise, $P_\mathrm{shot}$, if present.\footnote{Note that we do not simply absorb the leading order $k^2$ correction of $W_R(k)$ into the $c_s^2$ counterterm; this would lead to the counterterm-like contributions being important at very low $k$, due to the assumption that $R^{-1}\ll k_\mathrm{NL}, \Lambda$ for EFT smoothing scale $\Lambda$.} Whilst this is a fairly significant number of free parameters, it is important to note that we expect it to \textit{fully} encapsulate the theory until corrections of order $(k/k_\mathrm{NL})^2P_L(k)$ become important. Additionally, this number can likely be significantly reduced in practice; an approximate form which we expect to capture most of the smoothing function dependence is given by
\beq\label{eq: Mr0-inf-loop-approx}
    \left.M^{r,0}(\vk)\right|_{\infty-\mathrm{loop}} \approx \left[C_0-C_1W_R(k)\right]^2\left\{(\tilde{a}_0+\tilde{a}_1\mu^2)P_L(k) + \tilde{c}_0\right\}.
\eeq
Furthermore, the $\mu^2$ coefficients are unnecessary if only real-space or the redshift-space monopole is used. In this approximation, we have just three additional parameters; two for the monopole (or real-space) and one for the quadrupole, or, to good approximation, only one per multipole if the shot-noise is already free.%\footnote{\alej{very nice section. I am worried of the need of 6 free parameters for linear theory. I get that all are independent and this is what theory motivates. However, for practical purposes may be sufficient with 3? $a_0$, $a_2$ and $b0$? We already had this discussion: $W=1+c R^2 k^2 + ...$, the terms propto $R^2 k^2$ are degenerated with EFT parameters. I get $R$ is very large so $k^4$ appears rapidly. But c'mon, 6 free parameters will spoil any advantage for an optimal mark function for parameter estimation.}\oliver{I think that the 6 free parameters control the full functional form, and combining the $k^2$ part of the window function Taylor expansion with the counterterm is dangerous since they have quite different scale lengths. If we assume a known $k$ dependence as in Eq.\,\ref{eq: Mr0-inf-loop-approx} we only have 3 free parameters, one of which is degenerate with shot-noise, if present. Then we only have one free parameter per multipole.}}

\section{Comparison to Data}\label{sec: data}
\subsection{Quijote Simulations}

To test the model for $M(\vk)$ developed above, we compute the statistic on density fields from the \texttt{Quijote} suite \citep{2020ApJS..250....2V}, a collection of over $40,000$ $N$-body simulations spanning a wide variety of cosmologies. In this work, we use 50 `fiducial' simulations with cosmology $\{\Omega_m = 0.3175, \Omega_b = 0.049, h = 0.6711, n_s = 0.9624, \sigma_8 = 0.834, M_\nu = 0\,\mathrm{eV}, w = -1\}$ at redshift $z = 1$ (matching the peak sensitivity of upcoming spectroscopic surveys), each of which contains $512^3$ particles, followed from second-order Lagrangian perturbation theory (2LPT) initial conditions at $z = 127$. To test the impact of massive neutrino approximations, we also make use of the `M$\nu\texttt{+}$' % and `M$\nu\texttt{+++}$'
simulations, featuring three degenerate neutrinos with $\sum m_\nu = 0.1\,$eV, initialized using first-order Zel`dovich displacements. Analysis using the massive neutrino simulations is shown in Appendix \ref{appen: massive-nu}. For each simulation snapshot, particles are optionally displaced along the $z$-axis by their peculiar velocities, and real- and redshift-space marked spectrum computed as in Ref.\,\citep{2020arXiv200111024M}. We perform an analogous procedure for the biased tracer spectra, but using instead the density field of halos identified by the friends-of-friends algorithm \citep[e.g.][]{1982ApJ...257..423H} with linking length $b = 0.2$, \resub{and at least 20 dark matter particles}.\footnote{\resub{The below analysis was also performed on a sample restricting to halos with $M>3.1\times10^{13}h^{-1}M_\odot$, finding analogous results, but with increased shot-noise.}} We consider two sets of mark parameters; a default set of $\{p = 1, \delta_s = 0.25, R = 15\Mpch\}$ matching Ref.\,\citep{2020PhRvD.102d3516P}, and an alternative set $\{p=4,\delta_s = 10, R = 15\Mpch\}$, shown to have utility in measuring modified gravity parameters in Ref.\,\citep{2020JCAP...01..006A} (and similar to those of Ref.\,\citep{2016JCAP...11..057W}). 
% \alej{The parameters of \citep{2016JCAP...11..057W} are similar, but never tested with MG models, more closer to those parameters were used in my paper \cite{2020JCAP...01..006A}}.
\resub{In all cases we will compare the theory model to the mean of 50 simulations, and fit parameters using a Gaussian $M(\vk)$ likelihood, by simply performing a numerical $\chi^2$-minimization. For the covariance matrix, we adopt a diagonal approximation measured from those same simulations, motivated by Ref.\,\citep{2020arXiv200111024M}, which showed this to be an excellent approximation for the marked power spectrum of matter in real-space. Furthermore, since our interest here is only in finding a best-fit theory model, rather than performing parameter inference via MCMC, the exact choice of covariance is not of great importance.}

\subsection{Matter Spectra}

\begin{figure}
    \centering
    \includegraphics[width=\textwidth]{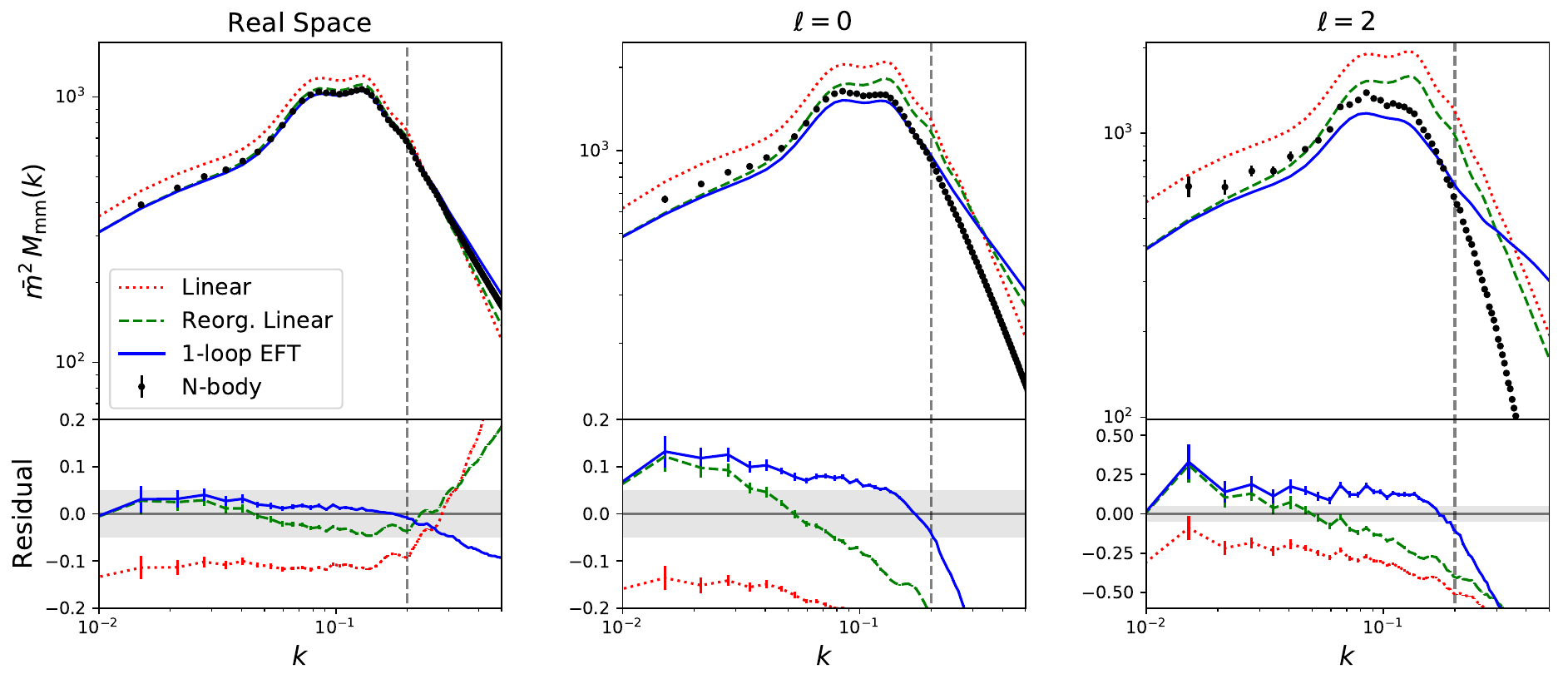}
    \caption{Comparison of the theory models for the marked power spectrum for matter, $M_{mm}(\vk)$, with data from $N$-body simulations. This uses the mark parameters $\{p = 1, \delta_s = 0.25, R = 15\Mpch\}$ in the fiducial \texttt{Quijote} cosmology at $z = 1$ (without massive neutrinos). For the theory models, we plot linear theory (dotted red, simply the $M_{11}$ term of Sec.\,\ref{sec: theory}), the reorganized linear term of Sec.\,\ref{subsec: theory-reorg-1-loop} (dashed green) and the full one-loop EFT (full blue), with the free EFT counterterms fitted using data up to $k_\mathrm{max} = 0.2\hMpc$, indicated by the vertical dashed line. We show results in real- and redshift-space, using the first two Legendre multipoles in the latter case. The bottom panels show the residual of observations from theory (defined as $\left(M_\mathrm{obs}-M_\mathrm{theory}\right)/M_\mathrm{theory}$), with the grey shaded band indicating 5\% deviations. In all cases, error bars are obtained from the mean of 50 simulations, \resub{and free parameters are fitted using $\chi^2$-minimization, assuming a diagonal covariance}.}
    \label{fig: model_comparison_matter}
\end{figure}

We begin by presenting results for the marked power spectrum of matter in real- and redshift-space, using the one-loop EFT model defined in Sec.\,\ref{sec: theory}. This is shown in Fig.\,\ref{fig: model_comparison_matter} for the fiducial mark parameters. Three models are compared: linear theory (including only terms up to $\mathcal{O}(P_L)$), one-loop EFT (up to $\mathcal{O}(P_L^2)$), and the `reorganized' linear theory of Sec.\,\ref{subsec: theory-reorg-1-loop}, \textit{i.e.} linear theory but incorporating the low-$k$ corrections of the one-loop terms. Only the full EFT model carries free parameters (in this example, just a counterterm scaling as $k^2P_L$ per multipole), which is set by likelihood minimization, % minimization of a Gaussian likelihood (assuming a diagonal $M(\vk)$ covariance), %\alej{I thought the fitting were made by eye}\oliver{no I do it properly}, 
with a fitting range of $k\in[0,0.2]\hMpc$. From \eqref{eq: taylor-condition}\,\&\,\eqref{eq: taylor-condition-rsd} we require $(1+\delta_s)^{-1}\sigma_{RR}(z)<1$ and $\left(1+\frac{2}{3}f(z)+\frac{1}{5}f^2(z)\right)^{1/2}(1+\delta_s)^{-1}\sigma_{RR}(z)<1$ for convergent Taylor series in real- and redshift-space; these evaluate to 0.13 and 0.17 respectively.

\begin{figure}
    \centering
    \includegraphics[width=\textwidth]{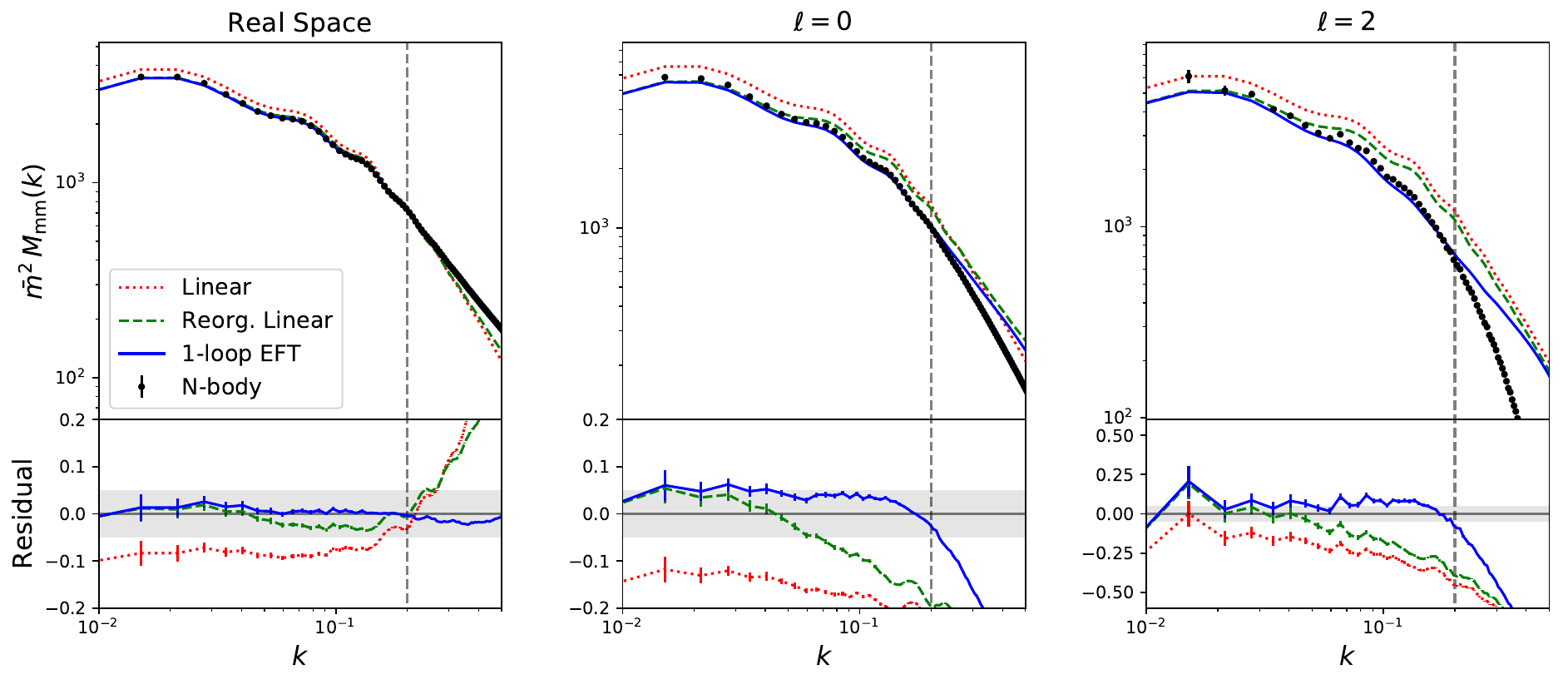}
    \caption{As Fig.\,\ref{fig: model_comparison_matter} but for the parameter set $\{p = 4, \delta_s = 10, R = 15\Mpch\}$, relevant to modified gravity studies. The one-loop fit is significantly better than for the default parameter set, as expected from the improved convergence of the underlying Taylor expansion.}
    \label{fig: model_comparison_matter_MG}
\end{figure}

Considering first the real-space case (which matches Ref.\,\citep{2020PhRvD.102d3516P}), we see that linear theory provides a poor model for $M(\vk)$ on all scales, with a $\sim 10\%$ error extending down to low-$k$. Whilst this is significantly different from the results familiar for the unmarked power spectrum, it is justified by the discussion of Sec.\,\ref{sec: reorg}, \textit{i.e.} that there are non-negligible terms that contribute at low-$k$ from higher loop orders. Indeed, the reorganized linear theory (incorporating the one-loop low-$k$ corrections) is seen to produce a much improved fit to the data on large scales, though underestimates the simulation data at the few percent level on the largest scales. The one-loop result performs better still at mildly non-linear scales (and is equal to the reorganized curve at low-$k$ by construction), since it also accounts for non-linearities in the matter density field and the backreaction of small-scale physics onto the large-scale modes.

A similar narrative is observed in redshift-space; linear theory strongly overpredicts $M(\vk)$, whilst the one-loop results are closer to the simulated data, though still an underestimate. The reorganized model is also a good approximation \resub{to} the one-loop result (and significantly simpler to compute since it does not require evaluation of convolution integrals), with deviations becoming important at $k\approx 0.08\hMpc$. That the theory model performs worse in redshift-space is well-understood; the Taylor expansion is more slowly convergent due to the greater (angle-averaged) density field, thus the neglected higher-loop contributions are more important in the low-$k$ limit. Furthermore, the matter quadrupole is notoriously difficult to fit in redshift-space, due to significant FoG suppression at relatively large scales that cannot be well modeled by a simple damping function \citep{2020arXiv200910724C}. We thus conclude that although the one-loop theory is a significant improvement over linear theory, it still struggles to provide a good match to the data in redshift-space. This can be ameliorated by using a broader smoothing scale $R$ though this is likely to reduce the information content of $M(\vk)$ \citep{2020arXiv200111024M}.

Results obtained using the second set of marked parameters are shown in Fig.\,\ref{fig: model_comparison_matter_MG}. In this case the magnitude of the Taylor series expansion parameter is significantly smaller; 0.014 (0.019) in real- (redshift-) space, thus we expect better convergence. Indeed, this is observed: whilst linear theory still provides a $\sim 10\%$ inaccurate model for $M(\vk)$, in real-space, the one-loop corrections give a model with $\sim 1\%$ accuracy up to $k\approx 0.1\hMpc$ (for the reorganized linear theory) or close to $k\approx 0.5\hMpc$ (for the full one-loop theory). In redshift-space, the one-loop and reorganized fits are accurate at the $\sim 5\%$ level (due to the greater magnitude of beyond-linear terms), which is again a significant improvement to the results for the fiducial parameters. From the convergence condition one might na\"ively expect the low-$k$ modeling of the redshift-space multipoles to perform better; in reality, this is more complex, since the low-$k$ contributions arise not only from higher order terms in the mark expansion (e.g., higher powers of $\delta_R$), but the non-linear contributions to the density field itself. As shown in Ref.\,\citep{2020PhRvD.102d3516P}, these (involving higher-order gravitational kernels $Z_n$) have non-trivial impact on the low-$k$ limit.

Given the above results, it is evident that our modelling can be improved by greater understanding of the low-$k$ limits of the theory. As discussed in Sec.\,\ref{subsec: theory-reorg-inf}, the low-$k$ limit has a well-defined form, which may be used as an \textit{ansatz} for the full reorganized theory. Here, we consider the model of \eqref{eq: Mr0-inf-loop-approx}, which gives a contribution proportional to $\left[C_0-C_1W_R(k)\right]^2P_L(k)$ for each multipole, and an additional term scaling as $\left[C_0-C_1W_R(k)\right]^2$ for the monopole (or real-space spectrum), with two (three) free parameters in real-space (redshift-space monopole and quadrupole). To ensure that we still capture all effects present in our theory model at a given order we simply add this correction term onto the usual model rather than substituting for the reorganized linear piece, \textit{i.e.} for the EFT model $M_\mathrm{EFT}(\vk)$ (equal to $\left.M^{r,0}(\vk)\right|_{1-\mathrm{loop}}+\left.M^{r,1}(\vk)\right|_{1-\mathrm{loop}}$), we use
\beq\label{eq: EFT+inf-loop}
    \left.M(\vk)\right|_{\mathrm{low-}k + \mathrm{EFT}} = \left.M^{r,0}(\vk)\right|_{\infty-\mathrm{loop}} + M_\mathrm{EFT}(\vk).
\eeq
In this case, the parameters of the infinite-loop $M^{r,0}$ piece correspond to the \textit{difference} between the true theory and our explicitly calculated $n$-loop model. This preseves the non-trivial $k$-dependencies in the low-$k$ limit of $M_\mathrm{EFT}(\vk)$. A similar model is possible for the combination of the low-$k$ effects with linear theory.

\begin{figure}
    \centering
    \includegraphics[width=\textwidth]{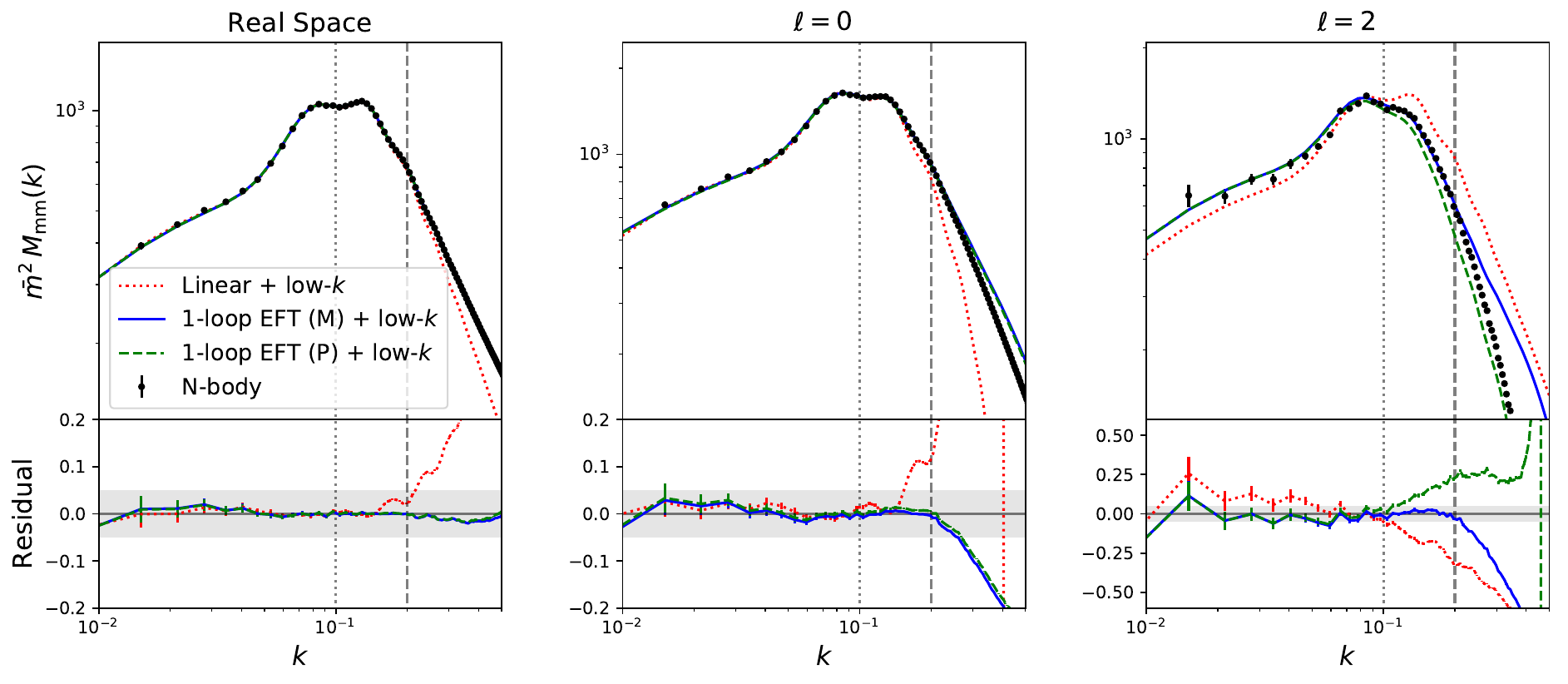}
    \caption{As Fig.\,\ref{fig: model_comparison_matter}, but adding the low-$k$ correction term of Sec.\,\ref{subsec: theory-reorg-inf}. This adds \resub{two (three)} free parameters to capture the (hard to model) behavior of large scales in \resub{real- (redshift-)space}, motivated by the infinite-loop form of the $k\rightarrow0$ theory \eqref{eq: Mr0-inf-loop-approx}. Three models are shown, all including the correction term; linear theory (dotted red), one-loop EFT with free parameters calibrated from $M(\vk)$ (M; solid blue) and one-loop EFT with free parameters fixed from modeling $P(\vk)$ (P; dashed green). We fit up to $k_\mathrm{max} = 0.2\hMpc$ (vertical dashed line) for the EFT model, and $k_\mathrm{max} = 0.1\hMpc$ (vertical dotted line) when fitting only low-$k$ parameters. Notably, we obtain good agreement at low-$k$ for all multipoles, which extends to $\sim 0.2\hMpc$ in the one-loop EFT model. When fixing the counterterm parameters to $P(\vk)$ we obtain good agreement for the monopole, but some evidence of deviations in the quadrupole, suggesting that higher loops are influencing these parameters.}
    \label{fig: model_comparison_matter_lowk}
\end{figure}

The results are shown in Fig.\,\ref{fig: model_comparison_matter_lowk} for the default mark parameter set, comparing results from linear theory and one-loop EFT, both supplemented with the above large-scale correction term. We fit the \resub{two (three) real-space (redshift-space)} free low-$k$ parameters \resub{via $\chi^2$-minimization} up to $k_\mathrm{max} = 0.1\hMpc$ (linear theory) or $0.2\hMpc$ (one-loop EFT). In all cases, we observe excellent agreement between the model and theory on large-scales, justifying our infinite-loop prediction and providing a flexible approach by which to fit the data. This is strongly dependent on the assumed $k$-dependence of the reorganized linear term however (\textit{i.e.} $M^{r,0}\propto [C_0-C_1W_R(k)]^2$); ignoring this and including only the $k = 0$ result significantly degrades the fit. We further note that the constant term $\tilde{c}_0$ in \eqref{eq: Mr0-inf-loop-approx} is insignificant in the real-space case.
%\footnote{\alej{Is this correct?  $c_0$ is the parameter that gives the LS constant shift}\oliver{yes, it's numerically found to be small in real-space}}. 
In the mildly non-linear regime, we find good agreement up to $k = 0.2\hMpc$, particularly in real-space, with the inclusion of one-loop reorganized terms being vital for the quadrupole term.

Finally, we test whether this approach allows for accurate models which can fit the power spectrum and marked spectrum simultaneously; an important check for overfitting. For this, we consider the one-loop EFT model as in \eqref{eq: EFT+inf-loop}, but fit the counterterm parameters, $\{c_0^2, c_2^2\}$ to the \textit{unmarked} power spectrum (up to $k_\mathrm{max} = 0.2\hMpc$), then the extra low-$k$ parameters directly to the large-scale $M(\vk)$ modes (up to $k_\mathrm{max} = 0.1\hMpc$). Without the low-$k$ correction, this gives a poor model for $M(\vk)$ (indicating that the free counterterms are also absorbing higher-order effects), though, as seen in Fig.\,\ref{fig: model_comparison_matter_lowk}, their inclusion allows for accurate fitting up to $k\sim 0.2\hMpc$ for real-space or the redshift-space monopole, and $k\approx 0.1\hMpc$ for the quadrupole. Thus, at the price of a slightly reduced radius of convergence, one may perform joint analyses of $P(\vk)$ and $M(\vk)$. We note that the above conclusions hold also for the alternative set of mark parameters and models including massive neutrinos. Discussion of the latter can be found in Appendix \ref{appen: massive-nu}.

\subsection{Biased Tracer Spectra}
\begin{figure}
    \centering
    \includegraphics[width=\textwidth]{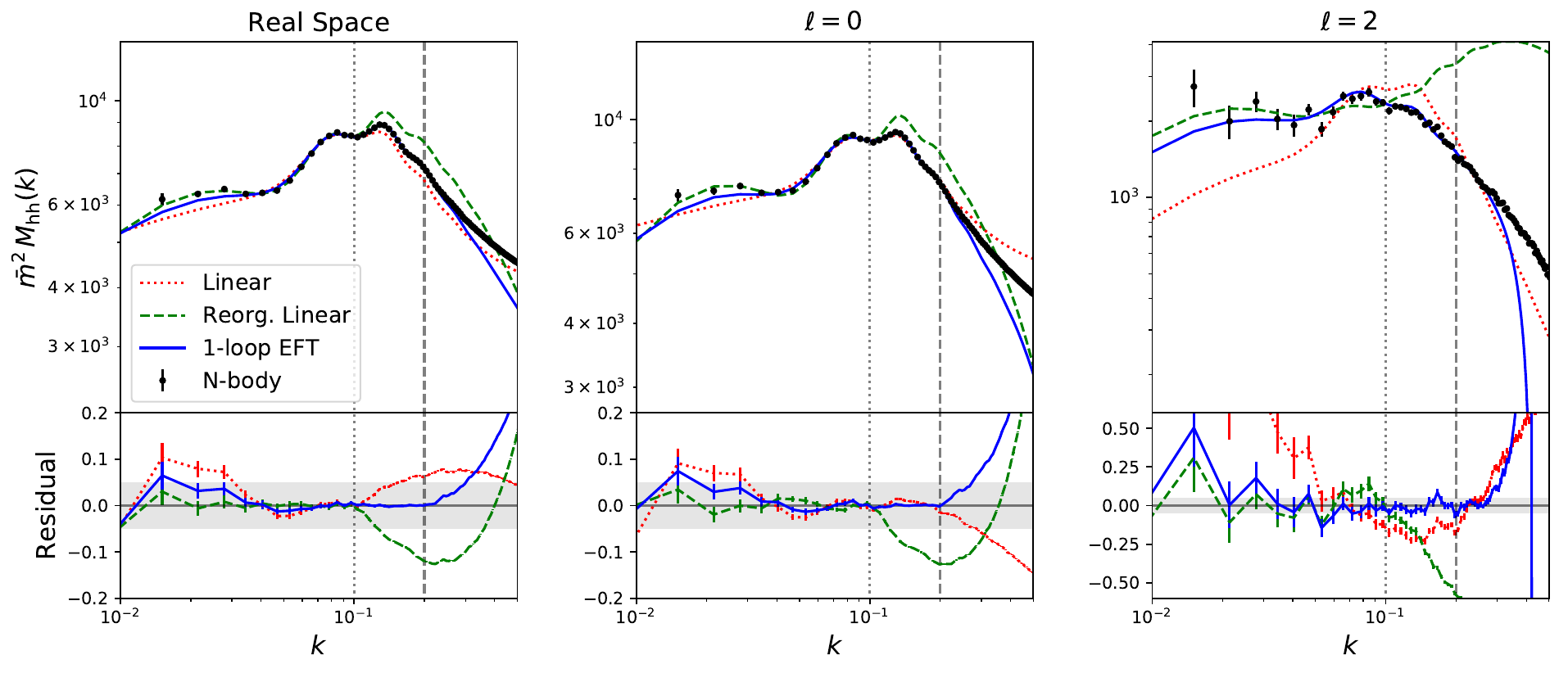}
    \caption{As Fig.\,\ref{fig: model_comparison_matter} but for the marked power spectrum of halos, $M_{hh}(\vk)$. These are generated from the \texttt{Quijote} halo catalogs, with halos identified via the Friends-of-Friends algorithm. In this case, free parameters are fitted up to $k = 0.1\hMpc$ (for linear and reorganized linear models) or $k = 0.2\hMpc$ (one-loop EFT). A total of two, four, and seven free parameters are needed to fully specify the three models for biased tracers in redshift-space (reducing to five for the real-space EFT). The fits are significantly better than those for unbiased matter (Fig.\,\ref{fig: model_comparison_matter}) indicating that the free bias parameters are absorbing higher loop effects.}
    \label{fig: model_comparison_halos}
\end{figure}

We now turn to the marked spectrum of halos. \textit{A priori}, one may expect the corresponding model fits to be more accurate than those of matter since (a) the free bias parameters can absorb some higher-loop effects, and (b) redshift-space distortions are, in general, weaker.
%\elena{An halo field shouldn't have FoG, isn't it?}.\oliver{depends on how you classify halos} 
Considering first the approaches without the infinite-loop reorganized \textit{ansatz}, a comparison between models is shown in Fig.\,\ref{fig: model_comparison_halos}. Each model carries a number of free parameters: $\{b_1, P_\mathrm{shot}\}$ in the linear case, $\{b_1,b_2,b_{\mathcal{G}_2},P_\mathrm{shot}\}$ for the linear reorganized model at one-loop order and $\{b_1,b_2,b_{\mathcal{G}_2},c_0^2,c_2^2,\tilde{c},P_\mathrm{shot}\}$ for the full one-loop EFT. These are fit to $M(\vk)$ up to $k_\mathrm{max} = 0.1\hMpc$ (linear and reorganized theory) or $0.2\hMpc$ (one-loop EFT).

In all cases, the biased tracer model outperforms that presented in the previous section for matter, even though the convergence condition \eqref{eq: taylor-condition} becomes more difficult to satisfy due to halo bias greater than unity. This is attributed to the linear and one-loop bias parameters absorbing higher-order corrections. \resub{As an example, we find that the linear bias, $b_1$, shifts from $2.11$ to $1.47$ when fitting the redshift-space marked spectrum $M(\vk)$ with EFT instead of the power spectrum. Similar shifts are seen for other bias parameters.} 

Whilst linear theory is aided by the free $b_1$ and $P_\mathrm{shot}$ parameters and provides a modest fit to the real-space and redshift-space monopole up to $k \approx 0.1\hMpc$, it clearly fails for the quadrupole; even though $b_1$ and $P_\mathrm{shot}$ can absorb higher-order terms, the linear model does not contain sufficient freedom to capture the true ratio of monopole to quadrupole. Including the large-scale contributions from one-loop terms in the reorganized linear model gives a significantly better fit, with the low-$k$ limit now being set by four parameters. Given that we already have significant freedom in the bias parameters, it may seem somewhat unusual that the reorganized linear theory helps the fit at low-$k$ compared to the purely linear case (particularly in real-space). This is attributed to the non-trivial $k$-dependencies present in the low-$k$ limit of these terms (Sec.\,\ref{subsec: theory-reorg-1-loop}) which cannot be captured in the linear model (whose terms scale as $\left[C_0-C_1W_R(k)\right]^2$). Considering the EFT model, we obtain small ($\lesssim 5\%$) residuals up to $k_\mathrm{max} = 0.2\hMpc$ for all multipoles, with a clear gain from including the full loop corrections on mildly non-linear scales.

\begin{figure}
    \centering
    \includegraphics[width=\textwidth]{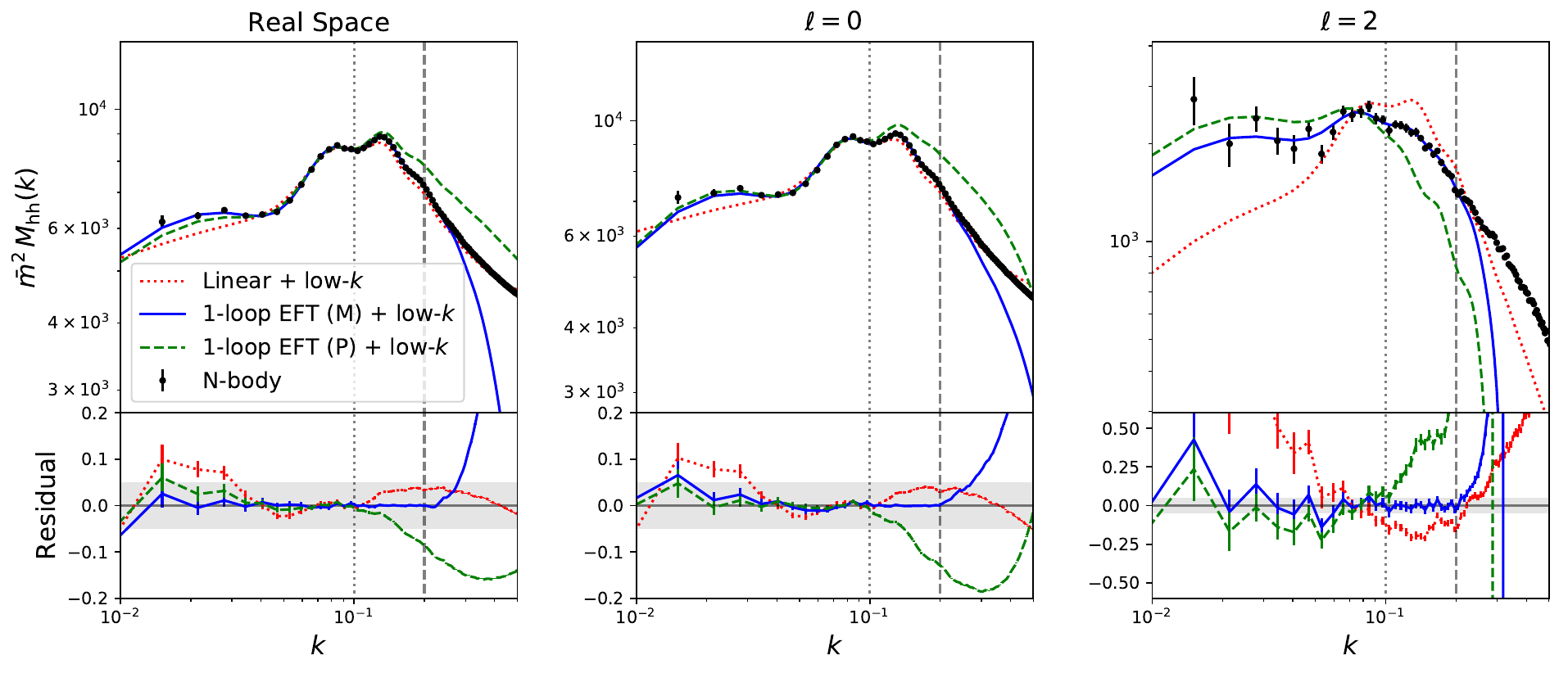}
    \caption{As Fig.\,\ref{fig: model_comparison_halos} but including the infinite-loop low-$k$ corrections of Sec.\,\ref{subsec: theory-reorg-inf} (analogous to Fig.\,\ref{fig: model_comparison_matter_lowk}). At the expense of three additional parameters, this improves convergence of the theory somewhat, allowing for the same bias parameters to be used for $P(\vk)$ and $M(\vk)$ up to $k = 0.1\hMpc$ (dashed green curve).}
    \label{fig: model_comparison_halos_lowk}
\end{figure}

We may also consider the fits involving the low-$k$ correction term derived from the infinite-loop reorganized linear theory, as in the previous subsection. Having such an approach is important for biased tracers, and results in a significantly more applicable model than extending the theory calculation to higher-order. This arises since the higher-loop contributions (which impact the large-scale theory) are dependent on higher-order bias parameters, the number of which quickly become very large. As an example, when explicitly computed at one-loop order, the reorganized linear theory contains three biases (plus shot-noise), all of which are important for the low-$k$ limit, in contrast to the zero-loop theory, with only $b_1$. Our \textit{ansatz} thus provides a tractable model of low-$k$, without requiring a prohibitively large number of free parameters. This is shown in Fig.\,\ref{fig: model_comparison_halos_lowk}, fitting parameters up to $k_\mathrm{max} = 0.2\hMpc$ (EFT) or $k_\mathrm{max} = 0.1\hMpc$ (linear and low-$k$), as before. \resub{Note that several of these parameters (e.g., $b_1$ and $\tilde{a}_0$ or $P_\mathrm{shot}$ and $\tilde{c}_0$) are highly degenerate.} Several points are of note: firstly, we find that the correction does not improve the fit of the linear model to simulations. This is as expected, since the correction terms are almost fully degenerate with the bias and shot-noise. To improve the large-scale fit in this case we need more complex $k$ dependencies (for example, utilizing the full six-parameter large-scale \textit{ansatz} of \eqref{eq: Mr0-inf-loop}). This is particularly evident for the quadrupole, which is poorly fit by the $M_2(k)\propto [C_0-C_1W_R(k)]^2P_L(k)$ model. 

We find little motivation for including the low-$k$ correction when EFT is included (comparing Figs.\,\ref{fig: model_comparison_halos}\,\&\,\ref{fig: model_comparison_halos_lowk}), indicating that the higher-order effects can already be encapsulated by the free biases. Of greater interest is the fit when the free parameters ($\{b_1,b_2,b_{\mathcal{G}_2},c_0^2,c_2^2,\tilde{c}\}$) are fixed from fitting to $P(\vk)$.\footnote{Note that we re-fit the shot-noise of $M(\vk)$, since it differs significantly between $M(\vk)$ and $P(\vk)$.} Without the low-$k$ correction, the fit is poor (and not shown in the figure), but including it gives a good model up to $k_\mathrm{max} = 0.1\hMpc$. \resub{As above, we note that the bias parameters are different between the two models (fitting the low-$k$ correction to $M(\vk)$ or $P(\vk)$, due to the absorption of higher-loop effects and parameter degeneracies.} Beyond this limit, the fit degrades, implying that higher-order effects also impact quasi-linear scales. As for the unbiased case, we conclude that inclusion of the low-$k$ correction terms allows for joint modeling of $P(\vk)$ and $M(\vk)$ (and hence bias parameters equal to their `physical' values), at the expense of slightly more modest scale limitations and additional free parameters. An extension would be the joint fit of $P(\vk)$ and \textit{multiple} $M(\vk)$ spectra with different mark parameters (shown to be optimal for parameter extraction in Ref.\,\citep{2020arXiv200111024M}); the bias and counterterm parameters would remain the same in all cases, but the three low-$k$ parameters would take differing values dependent on the choice of mark.

% \begin{figure}
%     \centering
%     \includegraphics[]{}
%     \caption{As Fig.\,\ref{fig: model_comparison_halos} but for massive neutrinos.}
%     \label{fig: model_comparison_halos_Mnu}
% \end{figure}

% \begin{figure}
%     \centering
%     \includegraphics{gfx/mk_real_halo_R15.0_p1.0_ds0.25_z1_M0.0.pdf}
%     \caption{Halos in real space, as above.}
%     \label{fig:my_label}
% \end{figure}

% \begin{figure}
%     \centering
%     \includegraphics{gfx/mk_RSD_halo_R15.0_p1.0_ds0.25_z1_M0.0.pdf}
%     \caption{Halos in redshift space, as above.}
%     \label{fig:my_label}
% \end{figure}

\section{Summary and Outlook}\label{sec: summary}
The marked power spectrum is a promising new statistic capable of placing strong constraints on a range of cosmological parameters. In this work, we have considered its modeling in the context of the Effective Field Theory of Large Scale Structure (EFTofLSS), building upon the framework of Ref.\,\citep{2020PhRvD.102d3516P}. The theory has been extended by including a self-consistent description of biased tracers and redshift-space distortions, following similar work performed for unmarked statistics. At their core, such modifications are straightforward, just requiring a redefinition of the perturbative kernels for the underlying density field; in practice, significant care is needed to formulate the theory in a manner that can be accurately and swiftly computed. The theory developed herein satisfies those goals, we have discussed its ultra-violet sensitivity and free parameters, which encode large scale tracer bias, stochasticity, Fingers-of-God (FoG) effects and the backreaction of small-scale physics on large-scale modes.

As noted in previous work, modeling the marked spectrum is difficult since the theory cannot be well-described as linear on any scale, with higher order terms (both Gaussian and non-Gaussian) being important even on the largest scales. Indeed, this problem is exacerbated in redshift-space (where the relevant expansions are slower to converge), and for biased tracers, giving the undesirable quality that, to fully describe the large-scale modes, one must consider contributions from all orders in perturbation theory, each of which may carry unknown parameters. Much of this work has been devoted to understanding and ameliorating this problem. In particular, we have introduced a `reorganized' formalism for the theory, which explicitly regroups terms into those that are important in the low-$k$ limit and those that are parametrically suppressed by powers of $(k/k_\mathrm{NL})^2$. Whilst the first (`reorganized linear') contribution technically depends on all orders in perturbation theory, we have provided its form in the one-loop case, and given an \textit{ansatz} for its infinite-loop structure that can be used for modeling, given a small number of additional free parameters.

Using $N$-body simulations from the \texttt{Quijote} suite, the theory has been rigorously tested for both biased and unbiased tracers, in real- and redshift-space. As in previous work, we find that one-loop EFT provides an accurate model, which is, by construction, equal to the reorganized linear theory (evaluated at one-loop order) on the largest scales. For matter, the agreement is worse in redshift-space, particularly for the quadrupole (which suffers from significant FoG effects), though inclusion of the infinite-loop corrections via three free parameters (two in real-space) allows for accurate predictions up to mildly non-linear scales, even when the usual free parameters are fit from external data. The fit is additionally improved using the mark parameters appropriate for modified gravity studies. For biased tracers, the fit is significantly improved, though there is evidence that the higher-loop effects are being absorbed into the free parameters of the theory. This is again reduced with the low-$k$ infinite-loop \textit{ansatz}. Theory models have also been derived for massive neutrino cosmologies, obtaining a comparable fit to that of the massless case, with the impact of post-EdS corrections to the kernels found to be negligible for realistic neutrino masses.

Whilst this work represents a significant step forward in our modeling of the marked power spectrum, the story is not yet concluded. Considering the pure EFT model, there are a number of aspects that must be discussed before the theory can be applied to data: a \resub{complete} infra-red resummation scheme to \resub{fully} capture non-perturbative long-wavelength modes that impact the BAO wiggles; the impact of the Alcock-Paczynski effect from the conversion of redshifts and angles to Cartesian co-ordinates; consideration of the survey window function; and a proper understanding of the shot-noise. Unlike for the matter power spectrum, adding these ingredients does not complete the recipe; as in the above discussion, the one-loop EFT is clearly incomplete, since it does not account for the renormalization of terms via the higher-loop contributions. Whilst this work provides a model for the inclusion of such effects on the large-scale theory, complications will arise on mildly non-linear scales, since those terms, normally sourced by the one-loop corrections, are again impacted by higher perturbative orders. Ideally, one would develop some form of renormalization scheme such that the loop expansion is formally convergent, akin to the `reorganized linear' scheme discussed herein. It remains to be seen whether such a theory can be developed which predicts the contributions directly, rather than through free coefficients.

It is additionally uncertain whether the gains in cosmological parameters reported in Ref.\,\citep{2020arXiv200111024M} will translate to the more realistic cases of biased tracers and redshift-space. For the unmarked power spectrum, redshift-space induces a significant change to the information content (e.g., breaking the $b_1-\sigma_8$ degeneracy through redshift-space distortions), thus one might expect analogous changes for the marked spectrum. It is thus crucial to perform tests such as simulation-based Fisher forecasts to understand whether the marked spectrum retains utility in this context. Even if this holds true, it is not guaranteed that the theory model for $M(\vk)$ can capture the requisite information, due to the necessary addition of free parameters to accurately model the large-scale limit. Whilst these will generically degrade the information content on amplitude parameters, \resub{in particular $\sigma_8$ for the monopole, and the growth-rate $f(z)$ for the quadrupole} (analogous to the effect of the power spectrum bias parameters), information arising from \textit{features} in the marked spectra is expected to remain, since it is not fully degenerate. Furthermore, we expect this to be partially ameliorated by combining multiple spectra, both marked and unmarked, which use the same bias parameters and counterterms. Such effects are simply explored via Fisher forecasts; if it transpires that the theory model cannot recover the cosmological information encoded in the mark, it may indicate the necessity for alternative methods for information extraction, such as simulation-based inference.

\acknowledgments
%We thank Simon Foreman for useful discussions that motivated in part this work. 
We thank Simon Foreman and Henrique Rubira for comments on a draft of this work, \resub{and are additionally grateful to the anonymous referee for providing an insightful report.} OHEP acknowledges funding from the WFIRST program through grants NNG26PJ30C and NNN12AA01C and thanks the %Institute for Advanced Study and 
Max Planck Institute for Astrophysics for hospitality, where part of this work was carried out. AA acknowledges partial support from Conacyt Grant No. 283151 \resub{and Conacyt Ciencia de Frontera grant No.~102958}.

\appendix

\section{Simplifying the One-Loop Integrals}\label{appen: integral-simp}
Below, we consider simplifications of the one-loop integrals in the most general case: biased tracers in redshift-space. Much of this parallels Ref.\,\citep{2020PhRvD.102d3516P}, though additional complication is added by the redshift-space dependence on the LoS. Below, we consider the $M_{11}$, $M_{13}$ and $M_{22}$ contributions in turn, and note that the real-space and unbiased cases are obtained by setting $\{f = \mu = 0\}$ and $\{b_1=1, b_2=b_{\mathcal{G}_2} = 0\}$ respectively.

\subsection{Linear Piece}
Starting from \eqref{eq: Mexpan}, we may write
\beq
    M_{11}(k,\mu) = C_{\delta_M}^2(k)Z_1^2(\vk)P_L(k) = \left[C_0-C_1W_R(k)\right]^2\left(b_1+f\mu\right)^2P_L(k),
\eeq
where we have inserted the definitions of $C_{\delta_M}(k)$ and $Z_1(\vk)\equiv b_1+f\mu^2$. Note that we assume the growth-factor $f$ to be scale-independent here; the generalization (relevant for massive neutrino cosmologies) is given in Appendix \ref{appen: massive-nu}. This yields the standard Kaiser multipoles \citep{1984ApJ...284L...9K};
\beq\label{eq: M11simp}
    M_{11,\ell}(k) = \left[C_0-C_1W_R(k)\right]^2P_L(k)\times \begin{cases}b_1^2+\frac{2}{3}b_1f+\frac{1}{5}f^2 & \ell = 0\\\nonumber \frac{4}{3}b_1^2+\frac{4}{7}b_1f & \ell = 2 \\\nonumber \frac{7}{35}f^2 & \ell = 4.\end{cases}
\eeq

\subsection{One-Loop Terms: 13-Piece}\label{appen: 13-simp}
The $M_{13}$ integral is given by
\beq
    M_{13}(k,\mu) &=& 3C_{\delta_M}(k)Z_1(\vec k)P_L(k)\int_{\vec p}P_L(p)\Bigl\{^{}_{}C_{\delta_M}(k)Z_3(\vec p,-\vec p,\vec k)\Bigr.\\\nonumber
    &&\,+\frac{2}{3}C_{\delta_M^2}(k,0)Z_1(\vec k)Z_2(\vec p,-\vec p)+\frac{2}{3}C_{\delta_M^2}(p,|\vec k-\vec p|)Z_1(\vec p)Z_2(\vec k,-\vec p)\\\nonumber
    &&\,\Bigl.+\frac{2}{3}C_{\delta_M^2}(p,|\vec k+\vec p|)Z_1(-\vec p)Z_2(\vec k,\vec p)+C_{\delta_M^3}(p,p,k)Z_1(\vec p)Z_1(-\vec p)Z_1(\vec k)\Bigr\}
\eeq
\eqref{eq: Mexpan}\,\&\,\eqref{eq: Hdef}. This can be split into three pieces involving $C_{\delta_M}$, $C_{\delta_M^2}$ and $C_{\delta_M^3}$, the first is simply 
\beq
    M_{13}^A(k,\mu) = 3C_{\delta_M}^2(k)Z_1(\vec k)P_L(k)\int_{\vec p}Z_3(\vec p,-\vec p,\vec k)P_L(p) \equiv C_{\delta_M}^2(k)P_{13}(k,\mu),
\eeq
where $P_{13}$ is the unmarked spectrum. The third is also straightforward;
\beq\label{eq: M13C}
     M_{13}^C(k,\mu) &=& 3C_{\delta_M}(k)Z_1^2(\vec k)P_L(k)\int_{\vec p}C_{\delta_M^3}(p,p,k)Z_1(\vec p)Z_1(-\vec p)P_L(p)\\\nonumber
     &=& \left[C_0-C_1W_R(k)\right]P_L(k)(b_1+f\mu^2)^2\\\nonumber
     &&\,\times \left(b_1^2+\frac{2b_1f}{3}+\frac{f^2}{5}\right)\int_{\vec p} \left[2C_2W_R(k)W_R(p)+(C_2-3C_3W_R(k))W_R^2(p)\right]P_L(p),
\eeq
inserting $Z_1(\vk)$ and $C_{\delta_M^3}$ coefficients \eqref{eq: CdeltaMdef}. Defining 
\beq\label{eq: calS-def}
    \mathcal{S}_R = \left(b_1^2+\frac{2b_1f}{3}+\frac{f^2}{5}\right)\int_{\vec p}W_R(p)P_L(p),
\eeq
and analogously for $\mathcal{S}_{RR}$,\footnote{Note that these are simply the Kaiser monopole prefactor \eqref{eq: M11simp} multiplying $\sigma_R^2$ and $\sigma_{RR}^2$} \eqref{eq: sig2R-defs}, $M_{13}^C$ can be written
\beq
    M_{13}^C(k,\mu) &=& \left[C_0-C_1W_R(k)\right]P_L(k)(b_1+f\mu^2)^2\\\nonumber
    &&\,\times \left[2C_2W_R(k)\mathcal{S}_R+(C_2-3C_3W_R(k))\mathcal{S}_{RR}\right].
\eeq

The second set of contributions to $M_{13}$ are less trivial. For unbiased tracers, $Z_2(\vec p,-\vec p) = 0$, but for biased tracers, $Z_2(\vec p,-\vec p) = b_2/2$, giving
\beq
    M_{13}^B(k,\mu) &=& 2C_{\delta_M}(k)C_{\delta^2_M}(k,0)Z_1^2(\vec k)P_L(k)\int_{\vec p}\frac{b_2}{2}P_L(p)\\\nonumber
    &&+2C_{\delta_M}(k)Z_1(\vec k)P_L(k)\left[\int_{\vec p}C_{\delta_M^2}(p,|\vec k-\vec p|)Z_1(\vec p)Z_2(\vec k,-\vec p)P_L(p) + (\vec p\leftrightarrow -\vec p)\right].
\eeq
The first integral is simply $b_2\sigma^2/2$, which appears UV divergent, but is exactly cancelled by the bias renormalization contribution of \eqref{eq: bias-renorm-cont}. Furthermore, we can transform $\vp \rightarrow -\vp$ in the second integral, giving the renormalized form
\beq \label{ec: M13B}
    M_{13}^B(k,\mu) &=& 4C_{\delta_M}(k)Z_1(\vec k)P_L(k)\int_{\vec p}C_{\delta_M^2}(p,|\vec k-\vec p|)Z_1(\vec p)Z_2(\vec k,-\vec p)P_L(p),
\eeq
which is simplified by noting that it is always possible to expand the $Z_n$ kernels as polynomials in $\hat{\vec p}\cdot\hat{\vec n}$;
\beq
    Z_1(\vec p)Z_2(\vec k,-\vec p) = \sum_n (\hat{\vec p}\cdot\hat{\vec n})^n z_n(p,|\vec k-\vec p|,k,\mu).
\eeq
Using the integral relation of Appendix \ref{appen: rot-integ}, this can be simplified;
\beq
    \frac{M_{13}^{B}(k,\mu)}{4C_{\delta_M}(k)Z_1(\vec k)P_L(k)} &=& \sum_n\int_{\vec p}(\hat{\vec p}\cdot\hat{\vec n})^nC_{\delta_M^2}(p,|\vec k-\vec p|)z_n(p,|\vec k-\vec p|,k,\mu)P_L(p)\\\nonumber
    &=& 2\sum_n\sum_{m\leq n}\mu^m\int_0^\infty\frac{p^2dp}{2\pi}\int_{-1}^1\frac{dx}{2}C_{\delta_M^2}(p,k,x)z_n(k,p,\mu,x)G_{nm}(x)P_L(p),
\eeq
where $x = \hat{\vec p}\cdot\hat{\vec k}$ and the $n$-th order polynomials $G_{nm}$ are defined in \eqref{eq: Gnm-def}.\footnote{Our integrand strictly breaks rotational symmetry since it depends on $\mu$ as well as $\hat{\vec p}\cdot\hat{\vec k}$; since it can be expanded as a (finite) polynomial in $\mu$, Eq.\,\ref{eq: rotational-integrand} still applies.} Whilst the angular integration is, in principle, analytic, the resulting expressions are prohibitively lengthy, thus we opt instead to perform it numerically, first expanding the integrand as a (closed) power series up to $\mu^8$. The full expression is lengthy, and thus omitted from this work.

\subsection{One-Loop Terms: 22-Piece}\label{appen: 22-simp}
The $M_{22}$ terms can be similarly split into three pieces;
\beq
    \frac{1}{2}M_{22}^A(k,\mu) &=& C_{\delta_M}^2(k)\int_{\vp} \left[Z_2(\vp,\vk-\vp)\right]^2P_L(p)P_L(|\vp-\vk|) \equiv \frac{1}{2}C_{\delta_M}^2(k) P_{22}(k,\mu)\\\nonumber
    \frac{1}{2}M_{22}^B(k,\mu) &=& 2C_{\delta_M}(k)\int_{\vp}\left[C_{\delta_M^2}(p,|\vk-\vp|)Z_2(\vp,\vk-\vp)Z_1(\vp)Z_1(\vk-\vp)\right]P_L(p)P_L(|\vk-\vp|)\\\nonumber
    \frac{1}{2}M_{22}^C(k,\mu) &=& \int_{\vp} \left[C_{\delta_M^2}(p,|\vk-\vp|)Z_1(\vp)Z_1(\vk-\vp)\right]^2P_L(p)P_L(|\vk-\vp|).
\eeq
The first is just a rescaling of $P_{22}(k,\mu)$ and requires no special treatment. For the third, we notice that this is just a convolution of two $P_{11}(\vk)\equiv Z_1^2(\vk)P_L(k)$ functions, which, inserting $C_{\delta_M^2}$ \eqref{eq: CdeltaMdef}, is given by
\beq
    \frac{1}{2}M_{22}^C(k,\mu) &=& C_2^2 \ast[W_R^2P_{11},W_R^2P_{11}](\vk)+\frac{1}{2}C_1^2 \ast[W_RP_{11},W_RP_{11}](\vec k)\\\nonumber
    &&\, + \frac{1}{2}C_1^2 \ast[W_R^2P_{11},P_{11}](\vec k)-2C_1C_2\ast[W_R^2P_{11},W_RP_{11}](\vec k),
\eeq
analogously to Ref.\,\citep{2020PhRvD.102d3516P}, where $\ast[X,Y]$ is a convolution, simply expressed as a Fourier-transformed real-space multiplication:
\beq
    \ast[X,Y](\vec k) = \mathcal{F}\left[X(\vec r)Y(\vec r)\right](\vec k).
\eeq
Since the convolvands have angular dependence, some care is needed for their evaluation; efficient evaluation is possible by first expressing $X, Y$ in terms of their multipoles, then computing the multipoles of $\left[XY\right](\vr)$ via the relation
\beq
    \left[XY\right]_L(r) &=& (2L+1)\sum_{\ell,\ell'} X_\ell(r)Y_{\ell'}(r)\begin{pmatrix}\ell & \ell'& L \\ 0 &0 &0 \end{pmatrix}^2
\eeq
\citep[Eq.\,34.3.1]{nist_dlmf}, where parentheses represent Wigner $3j$ symbols. Relation of Fourier- and configuration-space multipoles is achieved via FFTLog transforms \citep{2000MNRAS.312..257H}, using the formulae \beq\label{eq: mult-config-fourier}
    X_\ell(k) = 4\pi i^\ell \int r^2dr\,j_\ell(kr)X_\ell(r) \quad \Leftrightarrow \quad X_\ell(r) = (-i)^{\ell}\int \frac{k^2dk}{2\pi^2}j_\ell(kr)X_\ell(k),
\eeq
\citep[e.g.,][]{2020MNRAS.492.1214P}, making direct computation of the multipoles of $M_{22}^C$ straightforward.

The second piece can be written
\beq 
    \frac{M_{22}^B}{4C_{\delta_M}(k)} = C_2\ast_{Z_1Z_1Z_2}\left[W_RP_L,W_RP_L\right](\vec k)-C_1\ast_{Z_1Z_1Z_2}\left[W_RP_L,P_L\right](\vec k),
\eeq
where $\ast_Z[X,Y]$ is the convolution of $X$ and $Y$ with kernel $Z$, and $Z_1Z_1Z_2\equiv Z_1(\vec k_1)Z_1(\vec k_2)Z_2(\vec k_1,\vec k_2)$. Computation thus requires integrals of the form
\beq
    \mathcal{I}[X,Y]= \int_{\vec p} Z_1(\vec p)Z_1(\vec k-\vp)Z_2(\vp,\vk-\vp)X(p)Y(|\vk-\vp|).
\eeq
Whist this is technically possible with FFTLog approaches, it is difficult since the integrand is sixth order in the angle $(\hat{\vec p}\cdot\hat{\vec n})$. Simplification is possible through relation \eqref{eq: rotational-integrand} however, \textit{i.e.}
\beq
     \mathcal{I}[X,Y]&=&\ikk Z_1(\vec k_1)Z_1(\vec k_2)Z_2(\vk_1,\vk_2)X(k_1)Y(k_2)\\\nonumber
     &=& \sum_n \int_{\vec p} (\hat{\vec p}\cdot\hat{\vec n})^nz_n(k,\mu,p,x)X(p)Y(\sqrt{k^2+p^2-2kpx})\\\nonumber
     &=& \sum_n\sum_{m\leq m}\mu^m\int_0^\infty \frac{p^2dp}{2\pi^2}\int_{-1}^1\frac{dx}{2}z_n(k,p,\mu,x)G_{nm}(x)X(p)Y(\sqrt{k^2+p^2-2kpx}),
\eeq
and the resulting 2D integral can be performed numerically.

\section{Low-$k$ Limits of the One-Loop Terms}\label{appen: low-k-limits}
Here, we sketch the derivations of the low-$k$ limits stated in Sec.\,\ref{subsec: theory-reorg-1-loop}. To compute the low-$k$ limit of $M_{[11]}$, we first require the two-point propagator, $\Gamma^{[1]}$, given by \eqref{eq: propagator-def}; 
\beq
    \Gamma^{[1]}(\vk) &=& \sum_n \left\langle\frac{\partial\dptM{n}(\vk)}{\partial\dpt{1}(\vk)}\right\rangle\\\nonumber
    &=& \sum_{\mathrm{odd \,} n}n!!\int_{\vp_1..\vp_m}H_n(\vk,\vp_1,-\vp_1,...,\vp_m,-\vp_m)P_L(p_1)...P_L(p_m),
\eeq
where $m = (n-1)/2$. At one-loop, we obtain
\beq
    \left.\Gamma^{[1]}(\vk)\right|_{1-\mathrm{loop}} &= & C_{\delta_M}(k)Z_1(\vk) + 3C_{\delta_M}(k)\int_{\vp}Z_3(\vk,\vp,-\vp)P_L(p)\\\nonumber
    &&\,+4\int_{\vp}C_{\delta_M^2}(p,|\vk-\vp|)Z_1(\vp)Z_2(\vk,-\vp)P_L(p)\\\nonumber
    &&\,+3Z_1(\vk)\int_{\vp}C_{\delta_M^3}(k,p,p)Z_1(\vp)Z_1(-\vp)P_L(p),
\eeq
from the $H_n$ kernel definitions \eqref{eq: Hdef}, dropping a term involving $Z_2(\vp,-\vp)$ due to bias renormalizations (Sec.\,\ref{subsec: theory-uv}). At low-$k$, $Z_3(\vk,\vp,-\vp)\propto k^2/p^2$ thus this term may also be dropped.

Taking the low-$k$ limit for a Gaussian window $W_R$, we arrive at
\beq\label{eq: Gamma1-1-loop}
    \lim_{k\rightarrow 0}\left.\Gamma^{[1]}(\vk)\right|_{1-\mathrm{loop}} &=& Z_1(\vk)\Bigl[C_{\delta_M}(k)+\left(C_2-3C_3W_R(k)\right)\mathcal{S}_{RR}+2C_2W_R(k)\mathcal{S}_R\Bigr]\\\nonumber
    &&\,+\sum_{n \in \{0,2\}}\Bigl[A^{(1)}_{\mu^n}\left(2C_2W_R(k)\sigma^2_{RR} - C_1(1+W_R(k))\sigma_R^2\right)\Bigr.\\\nonumber
    &&\,\qquad\qquad\Bigl.-A^{(2)}_{\mu^n}\left(2C_2\overline{\sigma}^2_{RR}-C_1\overline{\sigma}^2_R\right)W_R(k)\Bigr]\mu^{n},
\eeq
where $\mathcal{S}_{R}$ and $\mathcal{S}_{RR}$ are defined in \eqref{eq: calS-def}.The second line contains terms from the $Z_2(\vk,-\vp)$ integral, which simplifies to a linear function of $\mu^2$ after angular integration.\footnote{At leading order, $Z_2(\vk,-\vp)$ is proportional to $p/k$ at leading order, which appears divergent. This contribution vanishes after angular integration over a symmetric domain.}  The associated $A^{(m)}_{\mu^{n}}$ coefficients are
\beq\label{eq: A-mu-def}
    A_{\mu^0}^{(1)} &=& \frac{b_1 f^2}{5}-\frac{8 f b_{G_2}}{15}+\frac{b_1^2 f}{3}+\frac{26 b_1 f}{35}+\frac{b_2
   f}{3}-\frac{4}{3} b_1 b_{G_2}+\frac{34 b_1^2}{21}+b_1 b_2+\frac{36 f^2}{245}\\\nonumber
   A_{\mu^0}^{(2)} &=& \frac{2 b_1 f}{15}+\frac{b_1^2}{3}+\frac{f^2}{35}\\\nonumber
   A_{\mu^2}^{(1)} &=& b_1 f^2+\frac{4 f b_{G_2}}{15}+b_1^2 f+\frac{22 b_1 f}{35}+\frac{2 f^3}{5}+\frac{74
   f^2}{245}\\\nonumber
   A_{\mu^2}^{(2)} &=& \frac{2 b_1 f^2}{5}+\frac{b_1^2 f}{3}+\frac{4 b_1 f}{15}+\frac{f^3}{7}+\frac{4 f^2}{35},
\eeq
and we define
\beq
    \overline{\sigma}_R^2 = \int_{\vp}(pR)^2W_R(p)P_L(p), \quad \overline{\sigma}_{RR}^2 = \int_{\vp}(pR)^2W^2_R(p)P_L(p).
\eeq
For convenience we denote
\beq\label{eq: calA-def}
    \mathcal{A}_n(k) &=& A^{(1)}_{\mu^n}\left(2C_2W_R(k)\sigma^2_{RR}- C_1(1+W_R(k))\sigma_R^2\right)\\\nonumber
    &&\,-A^{(2)}_{\mu^n}\left(2C_2\overline{\sigma}^2_{RR}-C_1\overline{\sigma}^2_R\right)W_R(k),
\eeq
such that the sum in \eqref{eq: Gamma1-1-loop} is just $\mathcal{A}_0(k)+\mathcal{A}_2(k)\mu^2$.

The contribution to the low-$k$ limit of $M_{[11]}$ is thus
\beq \label{ec: limk0M11}
    \left.\lim_{k\rightarrow 0}M_{[11]}(\vk)\right|_{1-\mathrm{loop}} &=& Z_1^2(\vk)P_L(k)C_{\delta_M^2}(k)\\\nonumber
    &&\,+Z_1^2(\vk)P_L(k)\left[C_{\delta_M^2}(k) + 2(C_2-3C_3W_R(k))\mathcal{S}_{RR}+4C_2W_R(k)\mathcal{S}_R\right]\\\nonumber
    &&\,+2Z_1(\vk)P_L(k)\left[\mathcal{A}_0(k)+\mathcal{A}_2(k)\mu^2\right],
\eeq
where we have truncated the quadratic expansion to avoid impartially including terms of order $\sigma_{RR}^4$. Note that these results can be alternatively derived by taking (twice) the $k\rightarrow 0$ limit of $M_{13}^B(\vk)$ and $M_{13}^C(\vk)$ in Appendix \ref{appen: 13-simp}.

For the low-$k$ constant at one-loop order, we require only the low-$k$ limit of $M_{[22]}(\vk)$, which can be derived from $\Gamma^{[2]}$ using \eqref{eq: m[22],m[33]-def}. This sources two one-loop contributions, arising from the diagrams shown in Fig.\,\ref{fig: feyn-reorg}b;
\beq
    \left.\lim_{k\ll p}\Gamma^{[2]}(\vp,\vk-\vp)\right|_{1-\mathrm{loop}} &=& C_{\delta_M^2}(p,|\vk-\vp|)Z_1(\vp)Z_1(\vk-\vp) + C_{\delta_M}(k)Z_2(\vp,\vk-\vp)\\\nonumber
    &\approx& C_{\delta_M^2}(p,|\vk-\vp|)Z_1(\vp)Z_1(-\vp) + C_{\delta_M}(k)\frac{b_2}{2},
\eeq
where we note that $Z_2(\vp,-\vp) = b_2/2 + \mathcal{O}(k^2/p^2)$ in the low-$k$ limit (with $b_2 = 0$ for matter). Following some computation, this leads to the following contribution to $M_{[22]}$, and hence the low-$k$ constant:
\beq\label{eq: M22lim}
    \left.\lim_{k\rightarrow 0}M_{[22]}(\vk)\right|_{1-\mathrm{loop}} &=& \frac{b_2^2}{2}C_{\delta_M}^2(k)\int_{\vp}P_L^2(p) + \mathcal{B}(k)
\eeq
with 
\beq\label{eq: calB-def}
    \mathcal{B}(k) &=& - 2b_2\left(b_1^2+\frac{2}{3}fb_1+\frac{1}{5}f^2\right)C_{\delta_M}(k)\int_{\vp}\left[C_1-C_2W_R(p)\right]W_R(p)P_L^2(p)\\\nonumber
    &&\,+2\left(b_1^4+\frac{4}{3}b_1^3f+\frac{6}{5}b_1^2f^2+\frac{4}{7}b_1f^3+\frac{1}{9}f^4\right)\int_{\vp}\left[C_1-C_2W_R(p)\right]^2W_R^2(p)P_L^2(p),
\eeq
where, for simplicity, we have assumed $W_R(|\vk-\vp|)\approx W_R(p)$.\footnote{Note that there is a hidden $\mu^2$ contribution which appears if one keeps the full $W_R(|\vk-\vp|)$ factor; this piece scales as $k^2$ at low k, and is thus shifted into the reorganized loop term by construction.}. Note that the first term in \eqref{eq: M22lim} does not contain window functions in the loop integral, and is thus not UV controlled. This is fully captured by the shot-noise counterterm, assuming it to have a form proportional to $C_{\delta_M}^2(k)$, and can hence be ignored. These expressions can be equivalently derived from the $k\rightarrow0$ limits of the $M_{22}^B$ and $M_{22}^C$ terms outlined in Appendix \ref{appen: 22-simp}.

% \section{One-loop Kernels in Redshift Space}
% For completeness, we here give the one-loop perturbation theory kernels $Z_n$ used in this work, for biased tracers in redshift-space. With the identifications $\{\mu_i\rightarrow0,f\rightarrow0\}$ and $\{b_1\rightarrow1, b_2\rightarrow0,b_{\mathcal{G}_2}\rightarrow0,b_{\Gamma_3}\rightarrow0\}$, these apply also to real-space and unbiased tracers respectively.\oliver{Actually worth including these?}

\section{Application to Massive Neutrino Cosmologies}\label{appen: massive-nu}

%\vspace{1cm}
\subsection{Theory Model}
In the presence of massive neutrinos, the perturbation theory kernels become significantly more complex than their EdS equivalents. For Lagrangian perturbation theory, these are given in Ref.\,\citep{2020JCAP...10..034A}, but can be mapped to the usual Eulerian $F_n$ and $G_n$ kernels to construct the full redshift-space $Z_n$ kernels. The main difficulty is that the free-streaming of neutrinos introduces an additional scale into the theory, such that linear growth factor becomes wavenumber-dependent, \textit{i.e.} $f\rightarrow f(k)$. The outcome of this is that, even in linear theory, velocity and density fields become non-locally related by $\theta^{(1)}(\vk) = (f(k)/f_0) \delta^{(1)}(\vk)$ with $f_0=f(k\rightarrow 0)$ and $\theta=-i \vk \cdot \vec v /(a f_0 H)$.. The linear density and velocity spectra become
\begin{equation}\label{PLs}
P_{\delta\delta}^L(k) = P_L(k), \qquad  P_{\delta\theta}^L(k) = \frac{f(k)}{f_0}P_L(k), \qquad  P_{\theta\theta}^L(k) = \left(\frac{f(k)}{f_0}\right)^2 P_L(k),    
\end{equation}
and the Kaiser boost is no-longer scale independent, but takes the form
\begin{equation}
Z_1(\vk) = b_1 + f(k) \mu^2,    
\end{equation}
such that the Kaiser power spectrum (equal to that of linear theory, ignoring infra-red resummation) is $P_{11}^{(s)}(\vk) = Z_1^2(\vk) P_L(k)$. At higher-order, the effects of neutrinos become increasingly complex. In this work, one must consider both the linear and second-order kernels to obtain the reorganized linear theory of Sec.\,\ref{subsec: theory-reorg-1-loop}; at second order in EPT, the density and velocity kernels become 
\begin{align}
F_2(\vk_1,\vk_2) &= \frac{1}{2} + \frac{3}{14}\A + \left( \frac{1}{2} - \frac{3}{14}\B  \right)   \frac{(\vk_1\cdot\vk_2)^2}{k_1^2 k_2^2}
        + \frac{\vk_1\cdot\vk_2}{2 k_1k_2} \left(\frac{k_2}{k_1} + \frac{k_1}{k_2} \right), \\\nonumber
G_2(\vk_1,\vk_2) &= \frac{3\A(f_1+f_2) + 3 \dot{\A}/H }{14 f_0} +
\left(\frac{f_1+f_2}{2 f_0} - \frac{3\B(f_1+f_2) + 3 \dot{\B}/H }{14 f_0}\right) \frac{(\vk_1\cdot\vk_2)^2}{k_1^2 k_2^2} \\\nonumber
&\quad + \frac{\vk_1\cdot\vk_2}{2 k_1k_2} \left( \frac{f_2}{f_0}\frac{k_2}{k_1} + \frac{f_1}{f_0}\frac{k_1}{k_2} \right),
\end{align}
where $f_{1}=f(\vk_{1})$ and $f_{2}=f(\vk_{2})$. Functions $\A,\B= \A(\vk_1,\vk_2,t), \B(\vk_1,\vk_2,t)$ depend on the wave-vectors of the two interacting plane waves, and are solutions to the second order differential equations given in Ref.\,\citep{2020JCAP...10..034A}. Utilizing these forms, the $Z_2$ kernel becomes
\begin{align}\label{Z2K}
Z_2(\vk_1,\vk_2) &=   b_1 F_2(\vk_1,\vk_2) +  f_0 \mu_{\vk_{12}}^2 G_2(\vk_1,\vk_2)  + \frac{b_2}{2} + b_{\mathcal{G}_2} \left[ \frac{(\vk_1 \cdot \vk_2)^2}{k_1^2 k_2^2} -1 \right] \\\nonumber
                 &\quad + \frac{f_0 \mu_{\vk_{12}} k_{12}}{2}\left[ \frac{\mu_{\vk_1}}{k_1} (b_1 + f(k_2) \mu_{\vk_2}^2 ) +  \frac{\mu_{\vk_2}}{k_2} (b_1 + f(k_1) \mu_{\vk_1}^2 ) \right].
\end{align}
where $\mu_{\vk_{12}}$ is the polar angle of $\vk_1+\vk_2$. Whilst a similar expression can be constructed for the $Z_3$ kernel, it appears only in the reorganized loop corrections (scaling as $k^2/k_\mathrm{NL}^2$ on large-scales) in the one-loop reorganized theory, and thus can be safely neglected here (noting also that any terms proportional to $k^2P_L(k)$ are absorbed by the EFT counterterms).
%by considering that the growth factor $k$-dependence is always either inside the SPT kernels or the Kaiser boosts. 
The EdS kernels can be simply recovered by setting $\A=\B=1$ and $f(k)=f_0$.

Given the above kernels, one may compute the reorganized linear theory, as in Appendix \ref{appen: low-k-limits}. For a general loop correction $I(\vk) = \int_{\vp} \mK(\vk,\vp)$, this works by writing $I(\vk) = \int_{\vp} \lim_{|\vk|\rightarrow0}\mK(\vk,\vp) + \int_{\vp} (\mK(\vk,\vp)- \lim_{|\vk|\rightarrow0}\mK(\vk,\vp) )   \equiv I^{r,0}(\vk) + I^{r,1}(\vk)$, with  
$ I^{r,1}(\vk)$ vanishing at small scales, indeed scaling as $ k^2  P_L(k)$. 
%Loop corrections to spectra have the general form $I(\vk) = \int_{\vp} \mK(\vk,\vp)$ for some kernel $\mK$. For the ``standard'' power spectrum these behave as $I(\vk) \propto k^2 P_L(k)$ at large scales. However, this does not happen for marked spectra; reorganization achieves it by splitting $I(\vk)$  in pieces that have considerable influence on the broadband large scales and terms that are subdominant. Schematically, it works by writing $I(\vk) = \int_{\vp} K(\vk\rightarrow 0,\vp) + \int_{\vp} (K(\vk,\vp)- K(\vk\rightarrow 0,\vp) )   \equiv I^{(0)}(\vk) + I^{(1)}(\vk)$, with $ I^{(1)}(\vk)$ vanishing at small scales, indeed scaling as $ k^2  P_L(k)$. 
Unfortunately, the exact neutrino kernels do not allow for straightforward expressions for the reorganized marked spectra, since the functions $\A$ and $\B$ are not analytic. To ameliorate this, we henceforth simplify the above kernels by setting these functions to unity, which gives only a subdominant error in the modeling of redshift-space spectra for realistic neutrino masses, %(only slighty larger than the error introduced by using EdS instead of $\Lambda$CDM kernels in massless neutrino cosmologies).  
since the main effects of the free-streaming scale enter through the linear power spectrum and the growth rates $f(k)$.%\footnote{\alej{Not sure if show plots for the standard power spectrum comparing the different Kernels. Or perhaps plots of the $\A$ and $\B$ functions. what do you think?}\oliver{Probably just easier to add a reference to a previous paper that does this.} \alej{there is no paper that does it. I should have done it in my paper with Arka on LPT.}} 

Following lengthy manipulations and repeated use of the identity proved in Appendix \ref{appen: rot-integ}, we obtain the reorganized linear theory explicitly calculated at one-loop order, analogous to \eqref{ec: M13M22r0}\,\&\,\eqref{eq: Mr01loop}:
\begin{align} \label{Mr0nu}
    \left.M^{r,0}(\vk)\right|_{1-\mathrm{loop}} &= C_{\delta_M}^2(k)\left(b_1+f(k)\mu^2\right)^2P_L(k) \\\nonumber
   &\quad +2  C_{\delta_M}(k) Z_1^2(\vk) P_L(k) \Big[ b_1^2 \mathcal{D}_{\delta\delta}(k)  + \frac{2b_1 f_0}{3}\mathcal{D}_{\delta\theta}(k) 
       + \frac{f_0^2}{5} \mathcal{D}_{\theta\theta}(k) \Big] \\\nonumber
       &\quad + 2  C_{\delta_M}(k)   Z_1(\vk) P_L(k) \left[ \tilde{\mA}_0(k) +\tilde{\mA}_2(k) \mu^2 \right]\\\nonumber
   &\quad+ \tilde{\mathcal{B}}(k) + M_\mathrm{stoch}(\vk) ,
\end{align}
where the generalized functions $\tilde{\mA}_{n}(k)$ are given by
\begin{align} \label{JmLS}
\frac{1}{4}\tilde{\mA}_0(k) &=  b_1^2 \left[\frac{17 \mC}{21}-\frac{\bar{\mC}}{12} \right] + b_1 f_0 \left[\frac{b_1 \mC}{6}-\frac{\bar{\mC}_{\delta\theta}}{30}+\frac{19\mC}{210}+\frac{59 \mC_{\delta\theta}}{210}\right]  \\\nonumber 
&\quad +  f_0^2 \left[\frac{b_1 \mC_{\delta\theta}}{10}-\frac{\bar{\mC}_{\theta\theta}}{140}+\frac{5\mC_{\delta\theta}}{98}+\frac{11 \mC_{\theta\theta}}{490}\right]  \\\nonumber 
&\quad + \frac{b_1 b_2 \mC}{2}-\frac{2 b_1 b_{\mathcal{G}_2} \mC}{3}+\frac{b_2 f_0 \mC_{\delta\theta}}{6}-\frac{4 b_{\mathcal{G}_2} f_0 \mC_{\delta\theta}}{15},\\\nonumber
\frac{1}{4}\tilde{\mA}_2(k) &=   b_1 f_0 \left[ b_1
   \left(\frac{\mC}{2}-\frac{\bar{\mC}}{12}\right)-\frac{\bar{\mC}_{\delta\theta}}{15}+\frac{4 \mC}{105}+\frac{29 \mC_{\delta\theta}}{105} \right] \\\nonumber
   &\quad + f_0^2 \left[ b_1 \left(-\frac{\bar{\mC}_{\delta\theta}}{10}+\frac{\mC}{6}+\frac{\mC_{\delta\theta}}{3}\right)-\frac{\bar{\mC}_{\theta\theta}}{35}+\frac{8 \mC_{\delta\theta}}{245}+\frac{29\mC_{\theta\theta}}{245} \right]   \\\nonumber
   &\quad +  f_0^3  \left[-\frac{\bar{\mC}_{\theta\theta}}{28}+\frac{\mC_{\delta\theta}}{10}+\frac{\mC_{\theta\theta}}{10}  \right]  + \frac{2}{15} f_0 b_{\mathcal{G}_2} \mC_{\delta\theta},  
\end{align}
with
\begin{align}
\mC = C_2 \sigma^2_{RR}  -C_1 \sigma^2_R, \qquad & \bar{\mC} =2 C_2\bar{\sigma}^2_{RR} - C_1 \bar{\sigma}^2_{R}, \\\nonumber
\mC_{\delta\theta} = C_2  \sigma^2_{\delta\theta,RR}  -C_1 \sigma^2_{\delta\theta,R}, \qquad & \bar{\mC}_{\delta\theta} =2 C_2  \bar{\sigma}^2_{\delta\theta,RR} - C_1 \bar{\sigma}^2_{\delta\theta,R}, \\\nonumber
\mC_{\theta\theta} = C_2 \sigma^2_{\theta\theta,RR}  -C_1 \sigma^2_{\theta\theta,R}, \qquad & \bar{\mC}_{\theta\theta} =2 C_2 \bar{\sigma}^2_{\theta\theta,RR} - C_1 \bar{\sigma}^2_{\theta\theta,R}, 
\end{align}
where we have assumed $W_R(k)\approx 1$. This makes use of the zero-lag correlators
\begin{align}
%\bar{\sigma}^2_{RR} &= \int_p W_R^2(p) P_L(p) \nonumber\\
\sigma^2_{RR,\delta\theta} = \int_p W_R^2(p) P_{\delta\theta}^L(p), \qquad &                \bar{\sigma}^2_{RR,\delta\theta} = \int_p W_R^2(p) p^2 R^2 P_{\delta\theta}^L(p), \\\nonumber
\sigma^2_{RR,\delta\theta} = \int_p W_R^2(p)P_{\theta\theta}^L(p), \qquad & \bar{\sigma}^2_{RR,\theta\theta} = \int_p W_R^2(p) p^2 R^2 P_{\theta\theta}^L(p),  
\end{align}
accounting for the fact that the linear power spectra, $P^L_{\delta\delta}$, $P^L_{\delta\theta}$ and $P^L_{\theta\theta}$ are not equal in massive neutrino cosmologies but are instead related by \eqref{PLs}. 
Strictly, the window functions $W_R(k)$ are important even at low-$k$ since $R$ is typically much larger than the non-linear scale $k_\mathrm{NL}^{-1}$. To restore them, one may simply substitute
\begin{align}
\mC= C_2\sigma_{RR}^2-C_1\sigma_{R}^2  &\longrightarrow C_2W_R(k) \sigma_{RR}^2-\frac{1}{2}C_1 \sigma_{R}^2 (1+ W_R(k)),    \\\nonumber
\bar{\mC} = 2 C_2\bar{\sigma}_{RR}^2-C_1\bar{\sigma}_{R}^2  &\longrightarrow  \big( 2 C_2  \bar{\sigma}_{RR}^2- C_1 \bar{\sigma}_{R}^2 \big) W_R(k),
\end{align}
and analogously for $\mC_{\delta\theta},\, \mC_{\theta\theta},\,\bar{\mC}_{\delta\theta}\,\bar{\mC}_{\theta\theta}$. Note that in the EdS limit, $\mC=\mC_{\delta\theta}=\mC_{\theta\theta}$ and $\bar{\mC}=\bar{\mC}_{\delta\theta}=\bar{\mC}_{\theta\theta}$. Additionally, \eqref{JmLS} further defines $\mathcal{D}$ functions; these are given by
\begin{align}
\mathcal{D}_{\delta\delta}(k)  &= (-3 C_3 W_R(k)+ C_2 )\sigma^2_{RR}  +2 C_2 W_R(k) \sigma^2_{R}, \\\nonumber
\mathcal{D}_{\delta\theta}(k)  &= (-3 C_3 W_R(k)+ C_2 )\sigma^2_{RR,\delta\theta}  +2 C_2 W_R(k) \sigma^2_{R,\delta\theta} , \\\nonumber
\mathcal{D}_{\theta\theta}(k)  &= (-3 C_3 W_R(k)+ C_2 )\sigma^2_{RR,\theta\theta}  +2 C_2 W_R(k) \sigma^2_{R,\theta\theta}, 
\end{align}
Finally, we note that the function $\tilde{\mB}(k)$ in \eqref{Mr0nu} is given by \eqref{eq: calB-def} but by replacing the growth functions with $f(p)$ and keeping them inside the integral. It is straightforward to show that for cosmologies with a scale independent growth factor, in which $f(k)=f_0=f$, \eqref{Mr0nu} reduces to \eqref{eq: Mr01loop}.

%Computing the Legendre multipoles and going to EdS kernels ($f(k) \rightarrow f$, scale independent), one recovers eq.~\eqref{Mell_LS}.
The first-order correction $M^{r,1}$ in massive neutrino cosmologies may be defined analogously to Sec.\,\ref{sec: reorg};
\begin{equation}
  \left.M^\text{r,1}(\vk)\right|_{1-\mathrm{loop}} = M^\text{1-loop}(\vk) - \left[M^{r,0}(\vk) - M_{11}(\vk)\right] , 
\end{equation}
which, at large scales, scales as $k^2P_L(k)$. This is  more difficult to compute, since it depends also on the third-order kernels $Z_3$.

\subsection{Application to Data}

\begin{figure}
    \centering
    \includegraphics[width=\textwidth]{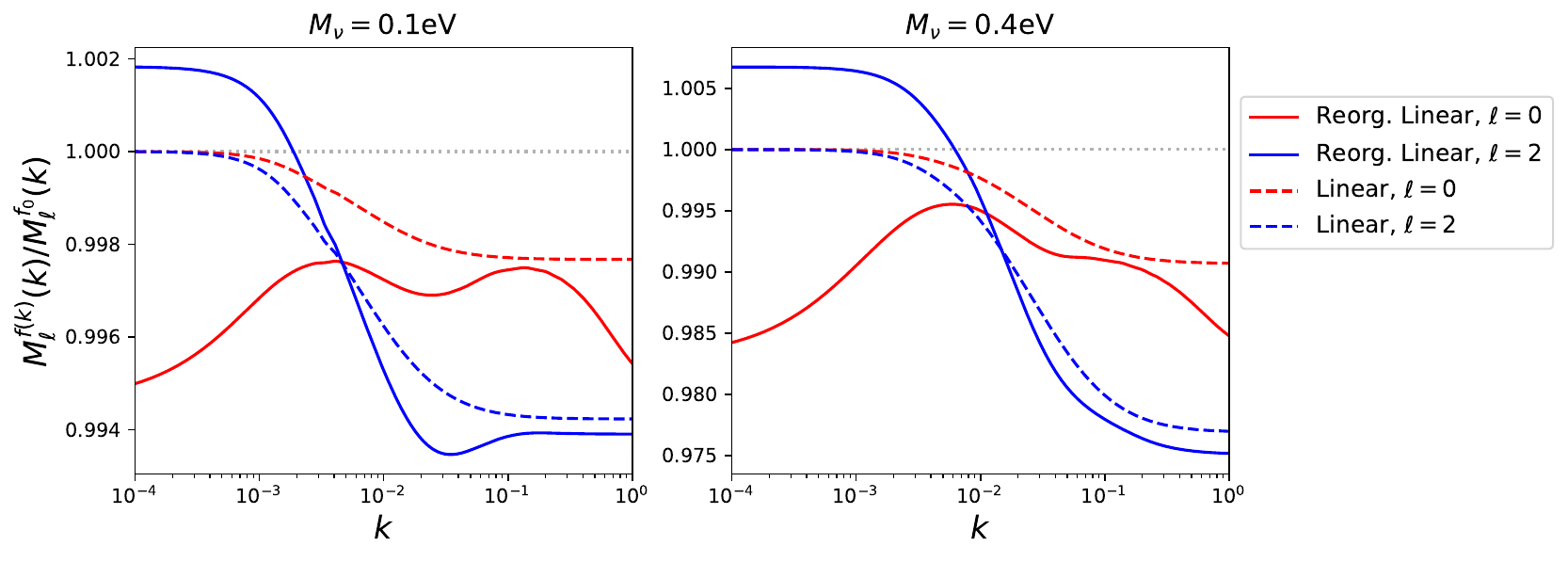}
    \caption{Comparison of models for the redshift-space marked power spectrum of CDM + baryons in the presence of massive neutrinos with total mass $0.1\,\mathrm{eV}$ (left) and $0.4\,\mathrm{eV}$ (right). For both linear (dashed lines) and one-loop reorganized linear (full lines) theories, we plot the ratio of the theory computed with a scale-dependent growth factor $f(k)$ (Appendix \ref{appen: massive-nu}) to that using the EdS approximation (Sec.\,\ref{sec: theory}) with $f(k) = f_0 = \mathrm{const.}$. The error induced by the EdS approximation is seen to be small and subdominant to the large-scale contributions from higher-loops in the theory.}
    \label{fig: fk_f0_ratio}
\end{figure}

Before comparing the massive neutrino theory models for $M(\vk)$ to data, we briefly consider the extent to which the scale-dependent growth factor $f(k)$ alters the model. This is shown in Fig.\,\ref{fig: fk_f0_ratio} for linear and reorganized linear theory, plotting the ratio of the full theory outlined in the above subsection to that assuming $f(k) = f_0$ (\textit{i.e.} the EdS approximation used in the rest of this work). We consider the (marked) power spectrum of CDM + baryons in all cases. The left panel shows the results for a total neutrino mass of $0.1\,\mathrm{eV}$ (comparable to the current observational limits); in this case, the error from assuming scale-independent $f$ is sub-percent on all scales considered for both the redshift-space monopole and quadrupole. For linear theory, the scale-dependent streaming creates a slight suppression of power on small scales, but the ratio asymptotes to unity at large scales, which is expected since $f_0$ is defined as the low-$k$ limit of $f(k)$. Notably, for the one-loop reorganized linear theory this limit is not recovered due to the mixing of scales induced by the mark, \textit{i.e.} the $k\rightarrow0$ limit depends on the density field at larger $k$ where $f(k)\neq f_0$. The exact low-$k$ limit will of course depend also on higher-loop contributions but is generally expected to be small. For a total neutrino mass of $0.4\,\mathrm{eV}$ (currently allowed in some non-minimal cosmological models), the deviations are larger, reaching $\sim 2\%$ by $k = 0.1$. However, as seen in Sec.\,\ref{sec: data}, such a deviation is not of interest in practice, since the theory model is \textit{not} capable of modeling $M(\vk)$ to such precision without using additional large-scale free parameters, that would be expected to absorb the bulk of these deviations (especially when coupled with the counterterms on smaller scales). Thus, the EdS approximation of $f(k) = f_0 = \mathrm{const.}$ is a valid assumption for physically reasonable scenarios.

\begin{figure}
    \centering
    \includegraphics[width=\textwidth]{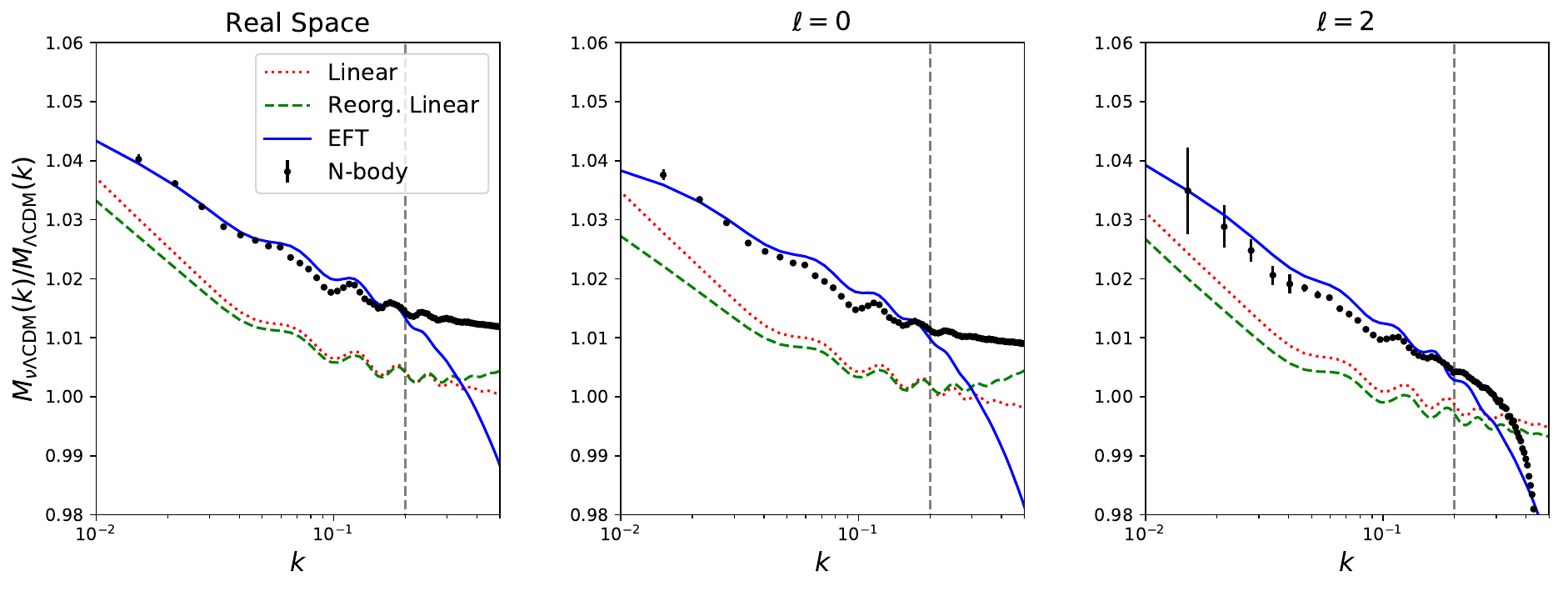}
    \caption{Ratio of marked spectra of CDM + baryons in massive and massless neutrino cosmologies in real- and redshift-space, for $M_\nu = 0.1\,\mathrm{eV}$. We include spectra from the \texttt{Quijote} simulations (black points) in addition to linear theory (Sec.\,\ref{subsec: theory-basics}, dotted red), one-loop reorganized linear theory (Sec.\,\ref{subsec: theory-reorg-1-loop}, dashed green) and one-loop EFT (Sec.\,\ref{subsec: theory-basics}, full blue). The free EFT counterterms are fitted up to $k_\mathrm{max} = 0.2\hMpc$ (indicated by the dashed line) and there are no additional free parameters. Functionally, there is a good agreement between the ratios, though the large-scale limits of the individual spectra do not match the data exactly, as in previous figures.}
    \label{fig: model_comparison_matter_Mnu}
\end{figure}

Fig.\,\ref{fig: model_comparison_matter_Mnu} compares theory and data for the marked power spectrum of CDM + baryons in real- and redshift-space, plotting the ratio of spectra with $M_\nu = 0.1\,\mathrm{eV}$ to those in the massless case. We assume EdS kernels in all cases, \textit{viz.} the above discussion, but fully include the effect of neutrinos in the linear power spectra from \texttt{CLASS}. Additionally, the two sets of simulations use the same initial phases of the density field, thus we are less susceptible to noise. Notably, the principal effect of massive neutrinos is to impart a few percent suppression in power which increases towards small-scales, as well as a slight modification to the oscillatory BAO feature; the functional form of this effect is well modeled by all variants of the theory, though the amplitude ratio is underestimated by the linear theories. With suitably chosen counterterms, the EFT model provides a reasonable fit to the marked power spectrum ratio up to $k\approx 0.2\hMpc$ in both real- and redshift-space, indeed, it significantly outperforms the reorganized linear theory even at $k\sim 0.01\hMpc$. Whilst this may seem paradoxical, since the one-loop EFT and reorganized linear theories agree by construction at low-$k$, this is due to one-loop contributions which are non-negligible even at $k\sim 10^{-2}\hMpc$, and we find good agreement between the theories by $k\sim 10^{-3}\hMpc$. Though EFT is thus shown to be a good model for the marked spectrum ratio, we caution that this does not imply accuracy for the individual spectra; rather we find that the theory performs similarly in massive and massless neutrino cosmologies, still with a large-scale error due to higher-loop contributions. \resub{Whether our model for marked spectra in the presence of massive neutrinos allows tighter constraints to be placed on cosmological parameters is uncertain, and we defer consideration to future work.}

\section{Rotational Integrand Formula}\label{appen: rot-integ}
A rotational scalar function $S(\vk,\vp)=S(k,p,x)$, with $x\equiv \hat{\vk}\cdot\hat{\vp}$, obeys the following relation 
\beq\label{eq: rotational-integrand}
    \int_{\vp}(\hat{\vp} \cdot \hat{\vn})^n S(\vk,\vp)  &=& \sum_{m=0}^n (\hat{\vk}\cdot\hat{\vn})^m \int_{\vp} G_{nm}(x)  S(k,p,x), 
\eeq
where
\beq\label{eq: Gnm-def}
    %G_{nm}(x) &=& \sum_{\ell=0}^n \frac{(1+(-1)^{\ell+n}) (2 \ell+1)}{2(1+\ell+n)}  \binom{\ell}{m} \binom{2 \ell}{\ell}  \binom{\frac{\ell+m-1}{2}}{\ell} \nonumber\\
    %&&\,\times
    %{}_3F_2\left(\frac{1-\ell}{2},-\frac{\ell}{2},\frac{1}{2} (-1-\ell-n);\frac{1}{2}-\ell,\frac{1}{2} (1-\ell-n); 1 \right)L_\ell(x)\\\nonumber
    G_{nm}(x) &=& \sum_{\ell = 0}^n\left(1+(-1)^{\ell+n}\right)(2\ell+1)\binom{\ell}{m}\binom{\frac{\ell+m-1}{2}}{l}\frac{ 2^{2\ell}\,n!\left(\frac{n+\ell}{2}+1\right)!}{\left(\frac{n-\ell}{2}\right)!(n+\ell+2)!}L_\ell(x)
.
\eeq
%for generalized ($p = 3, q = 2$) hypergeometric function ${}_3F_2(\vec a;\vec  b;z)$.

To prove this, first expand $(\hat{\vp}\cdot\hat{\vn})^n$ as a Legendre series in $\hat{\vp}\cdot\hat{\vn}$ and $S(k,p,x)$ in terms of a Legendre series in $x$;
\beq\label{eq: rot-int-tmp}
    \int_{\vp}(\hat{\vp} \cdot \hat{\vn})^n S(\vk,\vp) = \sum_{\ell_1=0}^nQ_{\ell_1}^{(n)}\sum_{\ell_2=0}^\infty\int_{\vp}S_{\ell_2}(k,p)L_{\ell_1}(\hat{\vp}\cdot\hat{\vn})L_{\ell_2}(\hat{\vk}\cdot\hat{\vp})
\eeq
where $Q^{(n)}_{\ell}$ are the Legendre multipoles of $(\hat{\vp}\cdot\hat{\vn})^n$, given by
\beq\label{eq: Qnell-def}
    Q^{(n)}_{\ell} &=& \left(1+(-1)^{\ell+n}\right)\frac{(2\ell+1)\sqrt{\pi}}{2^{n+2}}\frac{\Gamma(n+1)}{\Gamma(1+\frac{n}{2}-\frac{\ell}{2})\Gamma(\frac{n}{2}+\frac{\ell}{2}+\frac{3}{2})}\\\nonumber
    &=& \left(1+(-1)^{\ell+n}\right)2^{\ell}(2\ell+1)\frac{n!\left(\frac{n+\ell}{2}+1\right)!}{\left(\frac{n-\ell}{2}\right)!(n+\ell+2)!}
\eeq
\citep[Eq.\,7.126,][]{2007tisp.book.....G}, noting that $\ell,n$ are integers, and, via the first set of parentheses, $\ell-n$ must be even. The angular integral in \eqref{eq: rot-int-tmp} is straightforward;
\beq
    \int\frac{d\hat{\vec p}}{4\pi}L_{\ell_1}(\hat{\vp}\cdot\hat{\vn})L_{\ell_2}(\hat{\vk}\cdot\hat{\vn}) = \frac{1}{2\ell_1+1}\delta^K_{\ell_1\ell_2}L_{\ell_1}(\hat{\vk}\cdot\hat{\vn})
\eeq
\citep[Eq.\,14.17.6,][]{nist_dlmf}, giving
\beq
    \int_{\vp}(\hat{\vp} \cdot \hat{\vn})^n S(\vk,\vp) &=& \sum_{\ell=0}^n\frac{Q_{\ell}^{(n)}}{2\ell+1}L_\ell(\hat{\vk}\cdot\hat{\vn})\int_{\vp}S_{\ell}(k,p)\\\nonumber
    &=& \sum_{\ell=0}^nQ_\ell^{(n)}2^\ell\sum_{m=0}^\ell\binom{\ell}{m}\binom{\frac{\ell+m-1}{2}}{\ell}(\hat{\vk}\cdot\hat{\vn})^m\int_{\vp}S(\vk,\vp) L_{\ell}(\hat{\vk}\cdot\hat{\vp}),
\eeq
where, in the last line we have expressed $L_{\ell}(\hat{\vk}\cdot\hat{\vn})$ as its polynomial series and inserted the definition of $S_\ell$ as an angular integral. Inserting \eqref{eq: Qnell-def} into this and simplifying (noting that $\binom{\ell}{m} = 0$ for $m>\ell$) yields \eqref{eq: rotational-integrand}, completing the proof.
%\paragraph{Note added.} This is also a good position for notes added after the paper has been written.

\bibliographystyle{JHEP}
\bibliography{adslib,otherlib}%.lib}%,otherlib.bib.}

\providecommand{\href}[2]{#2}\begingroup\raggedright\begin{thebibliography}{10}

\bibitem{2001ApJ...553...14V}
L.~{Verde} and A.~F. {Heavens}, \emph{{On the Trispectrum as a Gaussian Test
  for Cosmology}}, \href{https://doi.org/10.1086/320656}{\emph{\apj} {\bfseries
  553} (2001) 14} [\href{https://arxiv.org/abs/astro-ph/0101143}{{\ttfamily
  astro-ph/0101143}}].

\bibitem{2016JCAP...06..052B}
D.~{Bertolini}, K.~{Schutz}, M.~P. {Solon} and K.~M. {Zurek}, \emph{{The
  trispectrum in the Effective Field Theory of Large Scale Structure}},
  \href{https://doi.org/10.1088/1475-7516/2016/06/052}{\emph{\jcap} {\bfseries
  2016} (2016) 052} [\href{https://arxiv.org/abs/1604.01770}{{\ttfamily
  1604.01770}}].

\bibitem{2020arXiv200902290G}
D.~{Gualdi}, S.~{Novell}, H.~{Gil-Mar{\'\i}n} and L.~{Verde}, \emph{{Matter
  trispectrum: theoretical modelling and comparison to N-body simulations}},
  {\emph{arXiv e-prints} (2020) arXiv:2009.02290}
  [\href{https://arxiv.org/abs/2009.02290}{{\ttfamily 2009.02290}}].

\bibitem{2001ApJ...546..652S}
R.~{Scoccimarro}, H.~A. {Feldman}, J.~N. {Fry} and J.~A. {Frieman}, \emph{{The
  Bispectrum of IRAS Redshift Catalogs}},
  \href{https://doi.org/10.1086/318284}{\emph{\apj} {\bfseries 546} (2001) 652}
  [\href{https://arxiv.org/abs/astro-ph/0004087}{{\ttfamily
  astro-ph/0004087}}].

\bibitem{2006PhRvD..74b3522S}
E.~{Sefusatti}, M.~{Crocce}, S.~{Pueblas} and R.~{Scoccimarro},
  \emph{{Cosmology and the bispectrum}},
  \href{https://doi.org/10.1103/PhysRevD.74.023522}{\emph{\prd} {\bfseries 74}
  (2006) 023522} [\href{https://arxiv.org/abs/astro-ph/0604505}{{\ttfamily
  astro-ph/0604505}}].

\bibitem{2020JCAP...03..040H}
C.~{Hahn}, F.~{Villaescusa-Navarro}, E.~{Castorina} and R.~{Scoccimarro},
  \emph{{Constraining M$_{{\ensuremath{\nu}}}$ with the bispectrum. Part I.
  Breaking parameter degeneracies}},
  \href{https://doi.org/10.1088/1475-7516/2020/03/040}{\emph{\jcap} {\bfseries
  2020} (2020) 040} [\href{https://arxiv.org/abs/1909.11107}{{\ttfamily
  1909.11107}}].

\bibitem{2017MNRAS.468.1070S}
Z.~{Slepian}, D.~J. {Eisenstein}, F.~{Beutler}, C.-H. {Chuang}, A.~J. {Cuesta},
  J.~{Ge} et~al., \emph{{The large-scale three-point correlation function of
  the SDSS BOSS DR12 CMASS galaxies}},
  \href{https://doi.org/10.1093/mnras/stw3234}{\emph{\mnras} {\bfseries 468}
  (2017) 1070}.

\bibitem{2015JCAP...05..007B}
T.~{Baldauf}, L.~{Mercolli}, M.~{Mirbabayi} and E.~{Pajer}, \emph{{The
  bispectrum in the Effective Field Theory of Large Scale Structure}},
  \href{https://doi.org/10.1088/1475-7516/2015/05/007}{\emph{\jcap} {\bfseries
  2015} (2015) 007} [\href{https://arxiv.org/abs/1406.4135}{{\ttfamily
  1406.4135}}].

\bibitem{2015MNRAS.454.4142S}
Z.~{Slepian} and D.~J. {Eisenstein}, \emph{{Computing the three-point
  correlation function of galaxies in O(N\^2) time}},
  \href{https://doi.org/10.1093/mnras/stv2119}{\emph{\mnras} {\bfseries 454}
  (2015) 4142} [\href{https://arxiv.org/abs/1506.02040}{{\ttfamily
  1506.02040}}].

\bibitem{2015PhRvD..91d3530S}
M.~{Schmittfull}, T.~{Baldauf} and U.~{Seljak}, \emph{{Near optimal bispectrum
  estimators for large-scale structure}},
  \href{https://doi.org/10.1103/PhysRevD.91.043530}{\emph{\prd} {\bfseries 91}
  (2015) 043530} [\href{https://arxiv.org/abs/1411.6595}{{\ttfamily
  1411.6595}}].

\bibitem{2017MNRAS.472.2436W}
C.~A. {Watkinson}, S.~{Majumdar}, J.~R. {Pritchard} and R.~{Mondal}, \emph{{A
  fast estimator for the bispectrum and beyond - a practical method for
  measuring non-Gaussianity in 21-cm maps}},
  \href{https://doi.org/10.1093/mnras/stx2130}{\emph{\mnras} {\bfseries 472}
  (2017) 2436} [\href{https://arxiv.org/abs/1705.06284}{{\ttfamily
  1705.06284}}].

\bibitem{2020MNRAS.492.1214P}
O.~H.~E. {Philcox} and D.~J. {Eisenstein}, \emph{{Computing the small-scale
  galaxy power spectrum and bispectrum in configuration space}},
  \href{https://doi.org/10.1093/mnras/stz3335}{\emph{\mnras} {\bfseries 492}
  (2020) 1214} [\href{https://arxiv.org/abs/1912.01010}{{\ttfamily
  1912.01010}}].

\bibitem{2020arXiv200501739P}
O.~H.~E. {Philcox}, \emph{{A Faster Fourier Transform? Computing Small-Scale
  Power Spectra and Bispectra for Cosmological Simulations in
  $\mathcal{O}(N^2)$ Time}}, {\emph{arXiv e-prints} (2020) arXiv:2005.01739}
  [\href{https://arxiv.org/abs/2005.01739}{{\ttfamily 2005.01739}}].

\bibitem{2020arXiv200903311P}
O.~H.~E. {Philcox}, M.~M. {Ivanov}, M.~{Zaldarriaga}, M.~{Simonovic} and
  M.~{Schmittfull}, \emph{{Fewer Mocks and Less Noise: Reducing the
  Dimensionality of Cosmological Observables with Subspace Projections}},
  {\emph{arXiv e-prints} (2020) arXiv:2009.03311}
  [\href{https://arxiv.org/abs/2009.03311}{{\ttfamily 2009.03311}}].

\bibitem{2017MNRAS.465.1757G}
H.~{Gil-Mar{\'\i}n}, W.~J. {Percival}, L.~{Verde}, J.~R. {Brownstein}, C.-H.
  {Chuang}, F.-S. {Kitaura} et~al., \emph{{The clustering of galaxies in the
  SDSS-III Baryon Oscillation Spectroscopic Survey: RSD measurement from the
  power spectrum and bispectrum of the DR12 BOSS galaxies}},
  \href{https://doi.org/10.1093/mnras/stw2679}{\emph{\mnras} {\bfseries 465}
  (2017) 1757} [\href{https://arxiv.org/abs/1606.00439}{{\ttfamily
  1606.00439}}].

\bibitem{2018MNRAS.478.4500P}
D.~W. {Pearson} and L.~{Samushia}, \emph{{A Detection of the Baryon Acoustic
  Oscillation features in the SDSS BOSS DR12 Galaxy Bispectrum}},
  \href{https://doi.org/10.1093/mnras/sty1266}{\emph{\mnras} {\bfseries 478}
  (2018) 4500} [\href{https://arxiv.org/abs/1712.04970}{{\ttfamily
  1712.04970}}].

\bibitem{2017MNRAS.469.1738S}
Z.~{Slepian}, D.~J. {Eisenstein}, J.~R. {Brownstein}, C.-H. {Chuang},
  H.~{Gil-Mar{\'\i}n}, S.~{Ho} et~al., \emph{{Detection of baryon acoustic
  oscillation features in the large-scale three-point correlation function of
  SDSS BOSS DR12 CMASS galaxies}},
  \href{https://doi.org/10.1093/mnras/stx488}{\emph{\mnras} {\bfseries 469}
  (2017) 1738} [\href{https://arxiv.org/abs/1607.06097}{{\ttfamily
  1607.06097}}].

\bibitem{1980lssu.book.....P}
P.~J.~E. {Peebles}, \emph{{The large-scale structure of the universe}}. 1980.

\bibitem{2019BAAS...51c..40P}
A.~{Pisani}, E.~{Massara}, D.~N. {Spergel}, D.~{Alonso}, T.~{Baker}, Y.-C.
  {Cai} et~al., \emph{{Cosmic voids: a novel probe to shed light on our
  Universe}}, {\emph{\baas} {\bfseries 51} (2019) 40}
  [\href{https://arxiv.org/abs/1903.05161}{{\ttfamily 1903.05161}}].

\bibitem{2019PhRvD.100d3514S}
M.~{Schmittfull}, M.~{Simonovi{\'c}}, V.~{Assassi} and M.~{Zaldarriaga},
  \emph{{Modeling biased tracers at the field level}},
  \href{https://doi.org/10.1103/PhysRevD.100.043514}{\emph{\prd} {\bfseries
  100} (2019) 043514} [\href{https://arxiv.org/abs/1811.10640}{{\ttfamily
  1811.10640}}].

\bibitem{2020JCAP...04..042C}
G.~{Cabass} and F.~{Schmidt}, \emph{{The EFT likelihood for large-scale
  structure}},
  \href{https://doi.org/10.1088/1475-7516/2020/04/042}{\emph{\jcap} {\bfseries
  2020} (2020) 042} [\href{https://arxiv.org/abs/1909.04022}{{\ttfamily
  1909.04022}}].

\bibitem{2020arXiv200714988C}
G.~{Cabass}, \emph{{The EFT Likelihood for Large-Scale Structure in Redshift
  Space}}, {\emph{arXiv e-prints} (2020) arXiv:2007.14988}
  [\href{https://arxiv.org/abs/2007.14988}{{\ttfamily 2007.14988}}].

\bibitem{2020JCAP...11..008S}
F.~{Schmidt}, G.~{Cabass}, J.~{Jasche} and G.~{Lavaux}, \emph{{Unbiased
  cosmology inference from biased tracers using the EFT likelihood}},
  \href{https://doi.org/10.1088/1475-7516/2020/11/008}{\emph{\jcap} {\bfseries
  2020} (2020) 008} [\href{https://arxiv.org/abs/2004.06707}{{\ttfamily
  2004.06707}}].

\bibitem{1992MNRAS.254..315W}
D.~H. {Weinberg}, \emph{{Reconstructing primordial density fluctuations. I -
  Method}}, \href{https://doi.org/10.1093/mnras/254.2.315}{\emph{\mnras}
  {\bfseries 254} (1992) 315}.

\bibitem{2011ApJ...731..116N}
M.~C. {Neyrinck}, I.~{Szapudi} and A.~S. {Szalay}, \emph{{Rejuvenating Power
  Spectra. II. The Gaussianized Galaxy Density Field}},
  \href{https://doi.org/10.1088/0004-637X/731/2/116}{\emph{\apj} {\bfseries
  731} (2011) 116} [\href{https://arxiv.org/abs/1009.5680}{{\ttfamily
  1009.5680}}].

\bibitem{2011ApJ...742...91N}
M.~C. {Neyrinck}, \emph{{Rejuvenating the Matter Power Spectrum. III. The
  Cosmology Sensitivity of Gaussianized Power Spectra}},
  \href{https://doi.org/10.1088/0004-637X/742/2/91}{\emph{\apj} {\bfseries 742}
  (2011) 91} [\href{https://arxiv.org/abs/1105.2955}{{\ttfamily 1105.2955}}].

\bibitem{2009ApJ...698L..90N}
M.~C. {Neyrinck}, I.~{Szapudi} and A.~S. {Szalay}, \emph{{Rejuvenating the
  Matter Power Spectrum: Restoring Information with a Logarithmic Density
  Mapping}}, \href{https://doi.org/10.1088/0004-637X/698/2/L90}{\emph{\apjl}
  {\bfseries 698} (2009) L90}
  [\href{https://arxiv.org/abs/0903.4693}{{\ttfamily 0903.4693}}].

\bibitem{2011ApJ...735...32W}
X.~{Wang}, M.~{Neyrinck}, I.~{Szapudi}, A.~{Szalay}, X.~{Chen},
  J.~{Lesgourgues} et~al., \emph{{Perturbation Theory of the Cosmological
  Log-density Field}},
  \href{https://doi.org/10.1088/0004-637X/735/1/32}{\emph{\apj} {\bfseries 735}
  (2011) 32} [\href{https://arxiv.org/abs/1103.2166}{{\ttfamily 1103.2166}}].

\bibitem{2007ApJ...664..675E}
D.~J. {Eisenstein}, H.-J. {Seo}, E.~{Sirko} and D.~N. {Spergel},
  \emph{{Improving Cosmological Distance Measurements by Reconstruction of the
  Baryon Acoustic Peak}}, \href{https://doi.org/10.1086/518712}{\emph{\apj}
  {\bfseries 664} (2007) 675}
  [\href{https://arxiv.org/abs/astro-ph/0604362}{{\ttfamily
  astro-ph/0604362}}].

\bibitem{2017MNRAS.464.3409B}
F.~{Beutler}, H.-J. {Seo}, A.~J. {Ross}, P.~{McDonald}, S.~{Saito}, A.~S.
  {Bolton} et~al., \emph{{The clustering of galaxies in the completed SDSS-III
  Baryon Oscillation Spectroscopic Survey: baryon acoustic oscillations in the
  Fourier space}}, \href{https://doi.org/10.1093/mnras/stw2373}{\emph{\mnras}
  {\bfseries 464} (2017) 3409}
  [\href{https://arxiv.org/abs/1607.03149}{{\ttfamily 1607.03149}}].

\bibitem{2020JCAP...05..032P}
O.~H.~E. {Philcox}, M.~M. {Ivanov}, M.~{Simonovi{\'c}} and M.~{Zaldarriaga},
  \emph{{Combining full-shape and BAO analyses of galaxy power spectra: a 1.6\%
  CMB-independent constraint on H$_{0}$}},
  \href{https://doi.org/10.1088/1475-7516/2020/05/032}{\emph{\jcap} {\bfseries
  2020} (2020) 032} [\href{https://arxiv.org/abs/2002.04035}{{\ttfamily
  2002.04035}}].

\bibitem{2020MNRAS.498.2492G}
H.~{Gil-Mar{\'\i}n}, J.~E. {Bautista}, R.~{Paviot}, M.~{Vargas-Maga{\~n}a},
  S.~{de la Torre}, S.~{Fromenteau} et~al., \emph{{The Completed SDSS-IV
  extended Baryon Oscillation Spectroscopic Survey: measurement of the BAO and
  growth rate of structure of the luminous red galaxy sample from the
  anisotropic power spectrum between redshifts 0.6 and 1.0}},
  \href{https://doi.org/10.1093/mnras/staa2455}{\emph{\mnras} {\bfseries 498}
  (2020) 2492} [\href{https://arxiv.org/abs/2007.08994}{{\ttfamily
  2007.08994}}].

\bibitem{doi:10.1002/mana.19841160115}
D.~{Stoyan}, \emph{On correlations of marked point processes},
  \href{https://doi.org/10.1002/mana.19841160115}{\emph{Mathematische
  Nachrichten} {\bfseries 116} (1984) 197}.

\bibitem{2005MNRAS.364..796S}
R.~K. {Sheth}, \emph{{The halo-model description of marked statistics}},
  \href{https://doi.org/10.1111/j.1365-2966.2005.09609.x}{\emph{\mnras}
  {\bfseries 364} (2005) 796}
  [\href{https://arxiv.org/abs/astro-ph/0511772}{{\ttfamily
  astro-ph/0511772}}].

\bibitem{2005astro.ph.11773S}
R.~K. {Sheth}, A.~J. {Connolly} and R.~{Skibba}, \emph{{Marked correlations in
  galaxy formation models}}, {\emph{arXiv e-prints} (2005) astro}
  [\href{https://arxiv.org/abs/astro-ph/0511773}{{\ttfamily
  astro-ph/0511773}}].

\bibitem{2006MNRAS.369...68S}
R.~{Skibba}, R.~K. {Sheth}, A.~J. {Connolly} and R.~{Scranton}, \emph{{The
  luminosity-weighted or `marked' correlation function}},
  \href{https://doi.org/10.1111/j.1365-2966.2006.10196.x}{\emph{\mnras}
  {\bfseries 369} (2006) 68}
  [\href{https://arxiv.org/abs/astro-ph/0512463}{{\ttfamily
  astro-ph/0512463}}].

\bibitem{2000ApJ...545....6B}
C.~{Beisbart} and M.~{Kerscher}, \emph{{Luminosity- and Morphology-dependent
  Clustering of Galaxies}}, \href{https://doi.org/10.1086/317788}{\emph{\apj}
  {\bfseries 545} (2000) 6}
  [\href{https://arxiv.org/abs/astro-ph/0003358}{{\ttfamily
  astro-ph/0003358}}].

\bibitem{2002A&A...387..778G}
S.~{Gottl{\"o}ber}, M.~{Kerscher}, A.~V. {Kravtsov}, A.~{Faltenbacher},
  A.~{Klypin} and V.~{M{\"u}ller}, \emph{{Spatial distribution of galactic
  halos and their merger histories}},
  \href{https://doi.org/10.1051/0004-6361:20020339}{\emph{\aap} {\bfseries 387}
  (2002) 778} [\href{https://arxiv.org/abs/astro-ph/0203148}{{\ttfamily
  astro-ph/0203148}}].

\bibitem{2009MNRAS.395.2381W}
M.~{White} and N.~{Padmanabhan}, \emph{{Breaking halo occupation degeneracies
  with marked statistics}},
  \href{https://doi.org/10.1111/j.1365-2966.2009.14732.x}{\emph{\mnras}
  {\bfseries 395} (2009) 2381}
  [\href{https://arxiv.org/abs/0812.4288}{{\ttfamily 0812.4288}}].

\bibitem{2016JCAP...11..057W}
M.~{White}, \emph{{A marked correlation function for constraining modified
  gravity models}},
  \href{https://doi.org/10.1088/1475-7516/2016/11/057}{\emph{\jcap} {\bfseries
  2016} (2016) 057} [\href{https://arxiv.org/abs/1609.08632}{{\ttfamily
  1609.08632}}].

\bibitem{2020arXiv200111024M}
E.~{Massara}, F.~{Villaescusa-Navarro}, S.~{Ho}, N.~{Dalal} and D.~N.
  {Spergel}, \emph{{Using the Marked Power Spectrum to Detect the Signature of
  Neutrinos in Large-Scale Structure}}, {\emph{arXiv e-prints} (2020)
  arXiv:2001.11024} [\href{https://arxiv.org/abs/2001.11024}{{\ttfamily
  2001.11024}}].

\bibitem{2018PhRvD..97b3535V}
G.~{Valogiannis} and R.~{Bean}, \emph{{Beyond {\ensuremath{\delta}} : Tailoring
  marked statistics to reveal modified gravity}},
  \href{https://doi.org/10.1103/PhysRevD.97.023535}{\emph{\prd} {\bfseries 97}
  (2018) 023535} [\href{https://arxiv.org/abs/1708.05652}{{\ttfamily
  1708.05652}}].

\bibitem{2018MNRAS.478.3627A}
J.~{Armijo}, Y.-C. {Cai}, N.~{Padilla}, B.~{Li} and J.~A. {Peacock},
  \emph{{Testing modified gravity using a marked correlation function}},
  \href{https://doi.org/10.1093/mnras/sty1335}{\emph{\mnras} {\bfseries 478}
  (2018) 3627} [\href{https://arxiv.org/abs/1801.08975}{{\ttfamily
  1801.08975}}].

\bibitem{2018MNRAS.479.4824H}
C.~{Hern{\'a}ndez-Aguayo}, C.~M. {Baugh} and B.~{Li}, \emph{{Marked clustering
  statistics in f(R) gravity cosmologies}},
  \href{https://doi.org/10.1093/mnras/sty1822}{\emph{\mnras} {\bfseries 479}
  (2018) 4824} [\href{https://arxiv.org/abs/1801.08880}{{\ttfamily
  1801.08880}}].

\bibitem{2020JCAP...01..006A}
A.~{Aviles}, K.~{Koyama}, J.~L. {Cervantes-Cota}, H.~A. {Winther} and B.~{Li},
  \emph{{Marked correlation functions in perturbation theory}},
  \href{https://doi.org/10.1088/1475-7516/2020/01/006}{\emph{\jcap} {\bfseries
  2020} (2020) 006} [\href{https://arxiv.org/abs/1911.06362}{{\ttfamily
  1911.06362}}].

\bibitem{2020PhRvD.102d3516P}
O.~H.~E. {Philcox}, E.~{Massara} and D.~N. {Spergel}, \emph{{What does the
  marked power spectrum measure? Insights from perturbation theory}},
  \href{https://doi.org/10.1103/PhysRevD.102.043516}{\emph{\prd} {\bfseries
  102} (2020) 043516} [\href{https://arxiv.org/abs/2006.10055}{{\ttfamily
  2006.10055}}].

\bibitem{2012JHEP...09..082C}
J.~J.~M. {Carrasco}, M.~P. {Hertzberg} and L.~{Senatore}, \emph{{The effective
  field theory of cosmological large scale structures}},
  \href{https://doi.org/10.1007/JHEP09(2012)082}{\emph{Journal of High Energy
  Physics} {\bfseries 2012} (2012) 82}
  [\href{https://arxiv.org/abs/1206.2926}{{\ttfamily 1206.2926}}].

\bibitem{2012JCAP...07..051B}
D.~{Baumann}, A.~{Nicolis}, L.~{Senatore} and M.~{Zaldarriaga},
  \emph{{Cosmological non-linearities as an effective fluid}},
  \href{https://doi.org/10.1088/1475-7516/2012/07/051}{\emph{\jcap} {\bfseries
  2012} (2012) 051} [\href{https://arxiv.org/abs/1004.2488}{{\ttfamily
  1004.2488}}].

\bibitem{2014arXiv1409.1225S}
L.~{Senatore} and M.~{Zaldarriaga}, \emph{{Redshift Space Distortions in the
  Effective Field Theory of Large Scale Structures}}, {\emph{arXiv e-prints}
  (2014) arXiv:1409.1225} [\href{https://arxiv.org/abs/1409.1225}{{\ttfamily
  1409.1225}}].

\bibitem{2016arXiv161009321P}
A.~{Perko}, L.~{Senatore}, E.~{Jennings} and R.~H. {Wechsler}, \emph{{Biased
  Tracers in Redshift Space in the EFT of Large-Scale Structure}}, {\emph{arXiv
  e-prints} (2016) arXiv:1610.09321}
  [\href{https://arxiv.org/abs/1610.09321}{{\ttfamily 1610.09321}}].

\bibitem{2015JCAP...09..029A}
R.~{Angulo}, M.~{Fasiello}, L.~{Senatore} and Z.~{Vlah}, \emph{{On the
  statistics of biased tracers in the Effective Field Theory of Large Scale
  Structures}},
  \href{https://doi.org/10.1088/1475-7516/2015/09/029}{\emph{\jcap} {\bfseries
  2015} (2015) 029} [\href{https://arxiv.org/abs/1503.08826}{{\ttfamily
  1503.08826}}].

\bibitem{2020JCAP...10..034A}
A.~{Aviles} and A.~{Banerjee}, \emph{{A Lagrangian perturbation theory in the
  presence of massive neutrinos}},
  \href{https://doi.org/10.1088/1475-7516/2020/10/034}{\emph{\jcap} {\bfseries
  2020} (2020) 034} [\href{https://arxiv.org/abs/2007.06508}{{\ttfamily
  2007.06508}}].

\bibitem{2002PhR...367....1B}
F.~{Bernardeau}, S.~{Colombi}, E.~{Gazta{\~n}aga} and R.~{Scoccimarro},
  \emph{{Large-scale structure of the Universe and cosmological perturbation
  theory}},
  \href{https://doi.org/10.1016/S0370-1573(02)00135-7}{\emph{\physrep}
  {\bfseries 367} (2002) 1}
  [\href{https://arxiv.org/abs/astro-ph/0112551}{{\ttfamily
  astro-ph/0112551}}].

\bibitem{2020JCAP...05..042I}
M.~M. {Ivanov}, M.~{Simonovi{\'c}} and M.~{Zaldarriaga}, \emph{{Cosmological
  parameters from the BOSS galaxy power spectrum}},
  \href{https://doi.org/10.1088/1475-7516/2020/05/042}{\emph{\jcap} {\bfseries
  2020} (2020) 042} [\href{https://arxiv.org/abs/1909.05277}{{\ttfamily
  1909.05277}}].

\bibitem{2018JCAP...04..030S}
M.~{Simonovi{\'c}}, T.~{Baldauf}, M.~{Zaldarriaga}, J.~J. {Carrasco} and J.~A.
  {Kollmeier}, \emph{{Cosmological perturbation theory using the FFTLog:
  formalism and connection to QFT loop integrals}},
  \href{https://doi.org/10.1088/1475-7516/2018/04/030}{\emph{\jcap} {\bfseries
  2018} (2018) 030} [\href{https://arxiv.org/abs/1708.08130}{{\ttfamily
  1708.08130}}].

\bibitem{2020PhRvD.102f3533C}
A.~{Chudaykin}, M.~M. {Ivanov}, O.~H.~E. {Philcox} and M.~{Simonovi{\'c}},
  \emph{{Nonlinear perturbation theory extension of the Boltzmann code CLASS}},
  \href{https://doi.org/10.1103/PhysRevD.102.063533}{\emph{\prd} {\bfseries
  102} (2020) 063533} [\href{https://arxiv.org/abs/2004.10607}{{\ttfamily
  2004.10607}}].

\bibitem{2015JCAP...02..013S}
L.~{Senatore} and M.~{Zaldarriaga}, \emph{{The IR-resummed Effective Field
  Theory of Large Scale Structures}},
  \href{https://doi.org/10.1088/1475-7516/2015/02/013}{\emph{\jcap} {\bfseries
  2015} (2015) 013} [\href{https://arxiv.org/abs/1404.5954}{{\ttfamily
  1404.5954}}].

\bibitem{2007tisp.book.....G}
I.~S. {Gradshteyn}, I.~M. {Ryzhik}, A.~{Jeffrey} and D.~{Zwillinger},
  \emph{{Table of Integrals, Series, and Products}}. 2007.

\bibitem{2006PhRvD..74j3512M}
P.~{McDonald}, \emph{{Clustering of dark matter tracers: Renormalizing the bias
  parameters}}, \href{https://doi.org/10.1103/PhysRevD.74.103512}{\emph{\prd}
  {\bfseries 74} (2006) 103512}
  [\href{https://arxiv.org/abs/astro-ph/0609413}{{\ttfamily
  astro-ph/0609413}}].

\bibitem{2014JCAP...08..056A}
V.~{Assassi}, D.~{Baumann}, D.~{Green} and M.~{Zaldarriaga},
  \emph{{Renormalized halo bias}},
  \href{https://doi.org/10.1088/1475-7516/2014/08/056}{\emph{\jcap} {\bfseries
  2014} (2014) 056} [\href{https://arxiv.org/abs/1402.5916}{{\ttfamily
  1402.5916}}].

\bibitem{2018JCAP...07..053I}
M.~M. {Ivanov} and S.~{Sibiryakov}, \emph{{Infrared resummation for biased
  tracers in redshift space}},
  \href{https://doi.org/10.1088/1475-7516/2018/07/053}{\emph{\jcap} {\bfseries
  2018} (2018) 053} [\href{https://arxiv.org/abs/1804.05080}{{\ttfamily
  1804.05080}}].

\bibitem{2018PhRvD..98h3541A}
A.~{Aviles}, \emph{{Renormalization of Lagrangian bias via spectral
  parameters}}, \href{https://doi.org/10.1103/PhysRevD.98.083541}{\emph{\prd}
  {\bfseries 98} (2018) 083541}
  [\href{https://arxiv.org/abs/1805.05304}{{\ttfamily 1805.05304}}].

\bibitem{2006PhRvD..73f3519C}
M.~{Crocce} and R.~{Scoccimarro}, \emph{{Renormalized cosmological perturbation
  theory}}, \href{https://doi.org/10.1103/PhysRevD.73.063519}{\emph{\prd}
  {\bfseries 73} (2006) 063519}
  [\href{https://arxiv.org/abs/astro-ph/0509418}{{\ttfamily
  astro-ph/0509418}}].

\bibitem{2006PhRvD..73f3520C}
M.~{Crocce} and R.~{Scoccimarro}, \emph{{Memory of initial conditions in
  gravitational clustering}},
  \href{https://doi.org/10.1103/PhysRevD.73.063520}{\emph{\prd} {\bfseries 73}
  (2006) 063520} [\href{https://arxiv.org/abs/astro-ph/0509419}{{\ttfamily
  astro-ph/0509419}}].

\bibitem{2008PhRvD..78j3521B}
F.~{Bernardeau}, M.~{Crocce} and R.~{Scoccimarro}, \emph{{Multipoint
  propagators in cosmological gravitational instability}},
  \href{https://doi.org/10.1103/PhysRevD.78.103521}{\emph{\prd} {\bfseries 78}
  (2008) 103521} [\href{https://arxiv.org/abs/0806.2334}{{\ttfamily
  0806.2334}}].

\bibitem{2020ApJS..250....2V}
F.~{Villaescusa-Navarro}, C.~{Hahn}, E.~{Massara}, A.~{Banerjee}, A.~M.
  {Delgado}, D.~K. {Ramanah} et~al., \emph{{The Quijote Simulations}},
  \href{https://doi.org/10.3847/1538-4365/ab9d82}{\emph{\apjs} {\bfseries 250}
  (2020) 2} [\href{https://arxiv.org/abs/1909.05273}{{\ttfamily 1909.05273}}].

\bibitem{1982ApJ...257..423H}
J.~P. {Huchra} and M.~J. {Geller}, \emph{{Groups of Galaxies. I. Nearby
  groups}}, \href{https://doi.org/10.1086/160000}{\emph{\apj} {\bfseries 257}
  (1982) 423}.

\bibitem{2020arXiv200910724C}
A.~{Chudaykin}, M.~M. {Ivanov} and M.~{Simonovi{\'c}}, \emph{{Optimizing
  large-scale structure data analysis with the theoretical error likelihood}},
  {\emph{arXiv e-prints} (2020) arXiv:2009.10724}
  [\href{https://arxiv.org/abs/2009.10724}{{\ttfamily 2009.10724}}].

\bibitem{1984ApJ...284L...9K}
N.~{Kaiser}, \emph{{On the spatial correlations of Abell clusters.}},
  \href{https://doi.org/10.1086/184341}{\emph{\apjl} {\bfseries 284} (1984)
  L9}.

\bibitem{nist_dlmf}
{NIST}, \emph{Nist digital library of mathematical functions}, .

\bibitem{2000MNRAS.312..257H}
A.~J.~S. {Hamilton}, \emph{{Uncorrelated modes of the non-linear power
  spectrum}},
  \href{https://doi.org/10.1046/j.1365-8711.2000.03071.x}{\emph{\mnras}
  {\bfseries 312} (2000) 257}
  [\href{https://arxiv.org/abs/astro-ph/9905191}{{\ttfamily
  astro-ph/9905191}}].

\end{thebibliography}\endgroup

\end{document}